\documentclass[journal, 10pt, onecolumn]{IEEEtran}
 \pdfoutput=1

\usepackage{amssymb}
\usepackage{amsmath,mathtools}
\usepackage{amsthm}
\usepackage{latexsym}
\usepackage{graphicx}
\usepackage{wrapfig}
\usepackage{color}
\usepackage[font={footnotesize}]{caption}
\usepackage[font={footnotesize}]{subcaption}
\usepackage{csquotes}
\usepackage{hyperref}
\usepackage{url}
\usepackage{epsf}
\usepackage{dcolumn}
\usepackage{float}
\usepackage{lipsum}
\usepackage{widetext}
\usepackage{cuted} 

\usepackage{textcomp}
\usepackage{algorithm2e}
\usepackage[noend]{algpseudocode}
\usepackage{tcolorbox}
\usepackage{glossaries}
\usepackage{cite} 

\ifCLASSINFOpdf
\else
\fi
\usepackage[cmintegrals]{newtxmath}
\usepackage{array}
\theoremstyle{definition}
\newtheorem{thm}{Theorem}
\newtheorem{lem}[thm]{Lemma}
\newtheorem{prop}[thm]{Proposition}

\newenvironment{pf}{{\noindent\sc Proof. }}{\IEEEQED}

\theoremstyle{definition}
\newtheorem*{defn}{Definition}

\theoremstyle{remark}
\newtheorem*{rem}{Remark}

\newcommand{\R}{{\mathbb R}}

\newcommand{\0}{{\mathbf 0}}

\newcommand{\argmax}{arg~max_}
\definecolor{awesome}{rgb}{1.0, 0.13, 0.32}
\definecolor{ao}{rgb}{0.0, 0.0, 1.0}
\definecolor{ao(english)}{rgb}{0.0, 0.5, 0.0}
\definecolor{cadmiumgreen}{rgb}{0.0, 0.42, 0.24}
\definecolor{darkpastelpurple}{rgb}{0.59, 0.44, 0.84}
\definecolor{darkorchid}{rgb}{0.6, 0.2, 0.8}
\definecolor{alizarin}{rgb}{0.82, 0.1, 0.26}

\begin{document}
%
\title{Throughput Optimal Decentralized Scheduling with Single-bit State Feedback for a Class of Queueing Systems}

%
%
%

\author{Avinash~Mohan,~\IEEEmembership{Member,~IEEE,}
        Aditya Gopalan,~\IEEEmembership{Member,~IEEE,}
        and~Anurag~Kumar,~\IEEEmembership{Fellow,~IEEE.}
\thanks{Avinash Mohan is with the Technion Israel Institute of Technology, Haifa-3200003, Israel. Aditya Gopalan and Anurag Kumar are with the Indian Institute of Science, Bangalore-560012, KA, India.
e-mail: avinashmohan@campus.technion.ac.il, \{aditya, anurag\}@iisc.ac.in, respectively.
\emph{Avinash Mohan is the corresponding author.}}
\thanks{This work was presented, in part, at the 13th IEEE International Conference on Computer Networks (IEEE Infocom, 2018). This research was supported by the Ministry of Human Resource Development Govt. of India, via a graduate fellowship for the first author, by Microsoft Research India, by a travel grant for the first author, by the SERB grant EMR/2016/002503 and the IUSSTF WAQM 2017 program for the second author, and the Department of Science and Technology, via a J.C. Bose Fellowship awarded to the last author.
}
\thanks{\textbf{Please note that all appendices are provided in the Supplementary Material.}}
}

%
%

\markboth{mohan-etal19single-bit-stable-scheduling, 2019}
{Version controlled}
%



\maketitle

\begin{abstract}
Motivated by medium access control for resource-challenged wireless Internet of Things (IoT) networks, whose main purpose is data collection, we consider the problem of queue scheduling with 
reduced queue state information.
In particular, we consider a time-slotted scheduling model with $N$ sensor nodes,  with pair-wise dependence, such that  nodes  $i$ and $i+1$, $ 1 \leq i \leq N-1$ cannot transmit together. 
We develop new throughput-optimal scheduling policies requiring only the empty-nonempty state of each queue that we term \emph{Queue Nonemptiness-Based} (QNB) policies. We revisit previously proposed policies and rigorously establish their throughput and delay-optimality. We propose a \emph{Policy Splicing} technique to combine scheduling policies for small networks in order to construct throughput-optimal policies for larger networks, some of which also aim for low delay. 
For $N=3$, there exists a sum-queue length optimal scheduling policy that requires only the empty-nonempty state of each queue. We show, however, that for $N \geq 4$, there is no scheduling policy that uses only the empty-nonempty states of the queues and is sum-queue length optimal over all arrival rate vectors in the capacity region. 	

We then extend our results to a more general class of interference constraints that we call cluster-of-cliques (CoC) conflict graphs. We consider two types of CoC networks, namely, Linear Arrays of Cliques (LAoC) and Star-of-Cliques (SoC) networks. 
We develop QNB policies for these classes of networks, study their stability and delay properties, and propose and analyze techniques to reduce the amount of state information to be disseminated across the network for scheduling. In the SoC setting, we propose a throughput-optimal policy that only uses information that nodes in the network can glean by sensing activity (or lack thereof) on the channel.

Our throughput-optimality results rely on two new arguments: a Lyapunov drift lemma specially adapted to policies that are queue length-agnostic, and a priority queueing analysis for showing strong stability. Our study throws up some new insights for the above classes of networks: 
\begin{itemize}
\item knowledge of queue length information
is not necessary to achieve optimal throughput/delay performance for certain classes of interference constraints,
\item for networks in these classes, it is possible to perform throughput-optimal scheduling by merely knowing whether queues in the network are empty or not, and
\item it is also possible to
be throughput-optimal by \emph{not} always scheduling the maximum possible number of nonempty non-interfering queues. 
\end{itemize}

\end{abstract}


%
\IEEEpeerreviewmaketitle

\section{Introduction}
\IEEEPARstart{T}he Internet of Things (IoT) paradigm is expected to make possible applications where vast numbers of devices coexist on a communication network. A typical example is a large-scale wireless sensor network comprising low-cost sensors that forward measurements from their respective locations. Given the massive scale and ubiquitous nature of these wireless networks, for IoT solutions to be viable, the embedded IoT devices, or \emph{motes} (as they are often called) will naturally have to cost very little (less than \$1, according to some estimates \cite{honrubia17industrial-iot-booming-drop-sensor-prices}). Such devices will
\begin{itemize}
\item \textit{\color{black}have to consume very low power} since a sensor in an IoT network is typically expected to last for several years without replacement. Battery and replacement expenses can affect the cost of deployment and maintenance adversely. \cite[Table~III]{raza-etal17critical-analysis-research-potential-challenges-future-directions-industrial-sensor} gives quite an extensive list of battery lifetimes expected from motes in various IoT applications. 

\item \textit{\color{black}possess very limited communication capabilities}\color{black}. This is, in fact, the first of two important consequences of the low power consumption constraint. Frequent communication with a centralized coordinating entity (such as a Path Computation Entity, or PCE, in the 6TiSCH architecture \cite{dujovne-etal14ip-enabled-industrial-iot}) will place unnecessary burdens on an already energy limited device.
	\textit{\color{black}Decentralized} MAC protocols, wherein the motes autonomously take transmission decisions based on limited  information about the network state\footnote{Like, perhaps, only the information they can gather by sensing their immediate surroundings, i.e, one-hop neighborhood.} will, therefore, need to be used instead of those that employ centralized control of transmissions.

\item \textit{\color{black}possess very limited memory and computing power}\color{black}. This is the second consequence of the power consumption constraint. To keep cost-per-device and energy drain low, one cannot equip such a sensor with anything beyond the bare minimum memory and processing abilities required for transmitting small packets based on a simple-to-compute transmission schedule.
\end{itemize}

The above requirements paint a picture of a \textit{\color{black}highly resource-challenged network}\color{black}. Furthermore, these constraints are starkly different from 
those that are encountered in the transmission of traditional voice and packet data over wireline or cellular wireless networks. 
In cellular systems, for example, the preponderance of scheduling decisions comes from the base station and hence, control is centralized \cite[Chapters~6, 13]{dahlman-etal13lte-advanced-mobile-broadband}, but some Quality of Service (QoS) is expected -- low packet delay, for instance \cite{lohstroh-etal19enabling-technologies-internet-important-things,alasti10quality-of-service-wimax-lte}. In contention access systems such as WiFi, scheduling is distributed but the service is best effort \cite{kong-etal04performance-analysis-802-11e-content-channel-access} or, at best, differentiated -- such as with the \enquote{Enhanced Distributed Channel Access} \cite {wu-etal06edca-throughput-analysis-802-11e} and the \enquote{Enhanced Distributed Coordination Function} (E-DCF) mechanisms in the IEEE 802.11e standard \cite{zhu-etal05performance-analysis-802-11e-dcf-service-differentiation}. In many IoT applications, e.g., condition monitoring or predictive maintenance, there is a need for distributed scheduling (for the earlier reasons) while also providing QoS. 
Consequently, resource allocation techniques developed to handle packet and voice data are insufficient to address the aforementioned issues in resource challenged networks. Most existing medium access protocols and scheduling algorithms suffer from limitations 
such as requiring too much state information to compute the schedule in any time slot making them hard to decentralize, being computationally intensive and thus unsuited for implementation on low-cost IoT devices, etc. 
Our aim in the paper is to propose decentralized MAC protocols with a focus on low packet delay (i.e., latency) and reduced exchange of control information across the network. 
The traditional approach for dynamic resource allocation has been to use backlog or queue length information to opportunistically schedule transmissions. One of the seminal contributions to scheduling in constrained queueing systems is the work of Tassiulas and Ephremides \cite{tassiulas92stability}. This paper introduces the model of a wireless network as a network of queues with pair-wise scheduling constraints (corresponding to wireless interference, half-duplex operation, etc.), and several flows over the network, each with its ingress queue and egress queue. The pair-wise constraints are represented by a \emph{conflict graph} (also known as an \emph{interference graph}) with the queues as the nodes and the pair-wise scheduling constraints being the edges. With stochastic arrivals to each flow to be routed from their ingress to egress points, the authors develop MaxWeight, a \emph{centralized} scheduling algorithm which requires the queue lengths of all nodes, and show that it is throughput-optimal, i.e., it stochastically stabilizes all queues under any stabilizable arrival rate.

\indent Attempts to decentralize MaxWeight include approximations based on  message passing between nodes \cite{sanghavi-etal09message-passing-maximum-weight-set, modiano-etal06maximizing-throughput-gossiping}, or using queue lengths to modulate backoff parameters in CSMA and ALOHA \cite{jiang-walrand11approaching-throughput-optimality-csma-collisions,rajagopalan09network-adiabatic}. Both of these methods, while being throughput-optimal, suffer from poor delay performance.  
Another method to reduce the amount of information required for scheduling is proposed in \cite{tassiulas-ephremides94dynamic-scheduling-tandem-parallel}, where, for two classes of constrained queueing systems, algorithms relying only on the empty-nonempty state of queues is proposed and analysed for delay performance. Our interest lies in the second half of \cite{tassiulas-ephremides94dynamic-scheduling-tandem-parallel}, where a scheduling algorithm is proposed for a system of $N$ parallel queues in which adjacent queues cannot be served simultaneously. The authors give the delay optimal policy for $N=3.$ This has been extended to $N=4$ by Ji et al \cite{ji-srikant10optimal-scheduling-small-switches} where the heavy-traffic delay-optimality of the proposed policy was proved. One of the contributions of our work in this paper is to refine these earlier contributions, and then to provide a novel method (policy \emph{splicing}) to develop scheduling algorithms for larger numbers of nodes.
It is not yet clear if it is even possible to extend these algorithms to general wireless networks while preserving performance guarantees such as throughput-optimality.

\indent A third class of strategies has focused on completely uncoordinated medium access, in contrast to methods using network state information. Here the focus is on improving the saturation throughput of Abramson's ALOHA protocol beyond $(1/e),$ without any queue length knowledge. The main idea is to allow collisions to take place, but use physical layer techniques like successive interference cancellation to decode the garbled messages over multiple time slots \cite{taghavi-etal16design-universal-schemes-uncoordinated-multiple-access, narayanan-etal12iterative-collision-resolution-slotted-aloha}. 
These techniques, like Abramson's ALOHA, allow lightweight uncoordinated access, but are not throughput-optimal, in that they can hope to achieve only the saturation throughput (the maximum sum rate point in the stability region). In contrast, we are interested in contention-free, low coordination and throughput-optimal MAC schemes. 

One other important contribution of this paper comes from a purely theoretical standpoint. As opposed to collocated networks, for which it is well-known that the empty-nonempty statuses of queues is sufficient for stability and good delay performance, it is a common perception that queue length information is required to achieve throughput-optimality in non-collocated networks. The authors in \cite{mckeown-etal19achieving-100-percent-throughput-switch} provide an example which helps reinforce this point: let us consider the classical input-queued switch (shown below in Fig.~\ref{PreambleToPart2FigMcKeownsExampleAndIGraph}).
\begin{figure}[htb]
\begin{subfigure}{.5\textwidth}
\hspace{1.00cm}
\includegraphics[height=5.150cm, width=8.65cm]{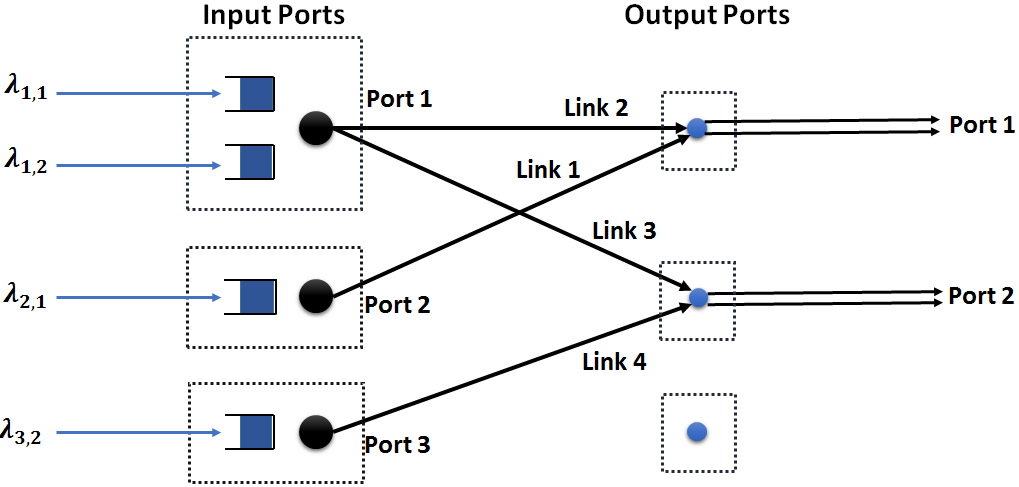}
\caption{A $3\times3$ input-queued switch with only $4$ active flows. $\lambda_{i,j}$ denotes the rate of arrival of packets to input port $i$ and headed for output port $j$.}
\label{PreambleToPart2FigInputQueuedSwitch}
\end{subfigure}
\begin{subfigure}{.5\textwidth}
\centering
\includegraphics[height=5.150cm, width=1.25cm]{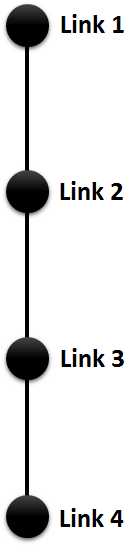}
\caption{The associated conflict (interference) graph is a path graph.}
\label{PreambleToPart2FigIGraphAssociatedWithMcKeownsIQSwitch}
\end{subfigure}
\caption{An input-queued switch and the associated interference graph.}
\label{PreambleToPart2FigMcKeownsExampleAndIGraph}
\end{figure}
As can be seen from the associated conflict graph in Fig.~\ref{PreambleToPart2FigMcKeownsExampleAndIGraph}, the arrival rate to the queue associated with Link~1 is $\lambda_{2,1}$, to Link~2 is $\lambda_{1,1}$, to Link~3 is $\lambda_{1,2}$ and  to Link~4 is $\lambda_{3,2}$. The backlog of the queue associated with Link~$i$ is denoted $Q_i(t)$, where we continue with the queue-length, arrival and departure embedding that we have been using until now. Activation constraints prevent scheduling of adjacent links, which means that adjacent queues cannot be served simultaneously and hence, 
\begin{eqnarray}
\lambda_{2,1}+\lambda_{1,1}&<&1,\\
\lambda_{1,1}+\lambda_{1,2}&<&1\text{, and}\\
\lambda_{1,2}+\lambda_{3,2}&<&1.
\label{PreambleToPart2EqnMcKeownExampleArrivalRateConstraints}
\end{eqnarray}
As discussed in Sec.~\ref{Chapter03SecSystemModel}, this is, in fact, the capacity region of the network in Fig.~\ref{PreambleToPart2FigMcKeownsExampleAndIGraph}, which means that for every rate vector $[\lambda_{1,1},\lambda_{1,2},\lambda_{2,1},\lambda_{3,2}]$ in the above region (i.e., satisfying \eqref{PreambleToPart2EqnMcKeownExampleArrivalRateConstraints}), there exists a scheduling policy that ensures stability. 
Now, consider a randomized scheduling policy that schedules the largest number of non interfering links in every slot and ties are broken with equal probability. In particular, this means that 
\begin{itemize}
\item In every slot, the policy chooses between the following three pairs: Queues 1 and 3, Queues 2 and 4 and Queues 1 and 4. Also,
\item If, for example, only Queues 1, 3 and 4 are nonempty, the policy serves Queues 1 and 3 w.p. $\frac{1}{2}$, and Queues 1 and 4 w.p. $\frac{1}{2}$.
\end{itemize}
Assume that the arrival process to each queue is IID across time and Bernoulli. 
Let $\lambda_{i,j}=0.5-\delta,$ where $\delta>0,$ i.e., arrival rates are all equal, and note that this satisfies the constraints in \eqref{PreambleToPart2EqnMcKeownExampleArrivalRateConstraints}. Now consider input port $1$ which, in the interference graph, is the combination of Queues 2 and 3. The total arrival rate to this pair = $\lambda_{1,1}+\lambda_{1,2}=1-2\delta.$ In any slot with $Q_2(t)>0\text{, and }Q_3(t)>0,$ since arrivals to Queues 1 and 4 are Bernoulli, they will both be non empty w.p. $(0.5-\delta)^2,$ which means that the $Q_2+Q_3$ pair will receive service w.p. $\frac{2}{3}.$ Therefore, the per-slot probability that this pair receives service is \emph{upper bounded by}
\begin{eqnarray}
&&\frac{2}{3}\times\left(0.5-\delta\right)^2+1\times\underbrace{\left(1-\left(0.5-\delta\right)^2\right)}_{\text{At least one of Queue 1 or 4 received 0 arrivals}}\\
&&<1-2\delta=\lambda_{1,1}+\lambda_{1,2},~\forall~\delta<0.0358,
\end{eqnarray}
which proves that this policy clearly renders input port $1,$ unstable. Of particular note is the fact that this policy does not attempt of break ties in favor of the two inner queues that are more constrained (Queue 2 cannot be served if either Queue 1 or Queue 3 is being served) than the two outer queues. As we will show in the sequel, prioritizing inner queues will have a strong impact on both stability and delay. Any policy that schedules the largest number of non-empty non-interfering queues in every slot is called a \emph{Maximum Size Matching} (MSM) policy and the one analysed above is an example of this class. 
One obvious candidate for a throughput-optimal scheduling policy is MaxWeight. We will, however, show that the MaxWeight policy is \emph{not desirable} in this case due to {two} reasons
\begin{itemize}
\item The first being that it violates our requirement of low state information exchange for taking scheduling decisions.
\item The second being that MaxWeight is provably \emph{non delay optimal} over these types of interference graphs.
\end{itemize}
Another obvious candidate is the aforementioned class of QCSMA, i.e., queue length-modulated CSMA, policies (\cite{jiang-walrand11approaching-throughput-optimality-csma-collisions,rajagopalan09network-adiabatic}), which, although stabilizing and amenable to distributed implementation shows unacceptable delay performance, as discussed before. These two observations combined with the instability of the MSM policy bring forth an important question: is it truly possible to stabilize non-collocated networks and achieve low queueing delay without using queue lengths? In the sequel, we will answer this question in the affirmative.

\subsection{Our Contributions and Organization}\label{secOurContributionsAndOrganization}%
It is well known that collocated networks admit stable and delay-optimal scheduling policies that require only the empty-non empty states of constituent queues. In this work, we first develop \emph{centralized} throughput-optimal and low-delay\footnote{By low delay, we mean a low sum queue length across the system.} scheduling policies that rely only on reduced state information, namely the empty-nonempty states of queues. Thereafter, we use these policies to construct reduced-state \emph{decentralized} scheduling protocols for multiple classes of networks. Our specific contributions are as follows.
\begin{itemize}
	\item 
	We begin by studying scheduling of transmissions on \enquote{path-graph interference networks.} We restrict ourselves to the subclass of Maximum Size Matching policies (MSM) mentioned above, that additionally require only the empty-nonempty statuses of the queues therein (introduced in Sections~\ref{secMSMPolicies} and \ref{secPrimerQLengthAgnostic}) and will provide a complete characterization of the set of such policies for the case with $N=3$ queues (Sec.~\ref{secTOSchedulingWith3Queues}). The fact that  the policies we discuss do not require any information about the queues except their empty-nonempty status helps satisfy our reduced state information requirement. We establish several interesting results about (in)stability and delay optimality, specifically, that MaxWeight is \emph{not} delay-optimal in such networks in a stochastic ordering sense. 
	
	\item 
	Continuing with path-graph networks, we propose a \enquote{policy splicing} technique (see Figures \ref{Chapter04FigTheGeneralSplicingProcessToGiveNonMSMPolicy} and \ref{Chapter04FigTheGeneralSplicingProcessStepsProjectionAndInnerQueuePriority}) to combine policies for small networks to give rise to policies for larger networks (Sections \ref{Chapter04SecQLenAgnosticTOSplicing} and \ref{Chapter04SecConstructAgnosticSplicingSubecTDandBUFor4Qs},). We use this technique to propose MSM scheduling policies for several such networks. We also provide an in-depth analysis of delay with MSM policies (Sec.~\ref{chapter04SecDelayAnalysisAgnostic}) culminating in a result that shows that there do not exist delay optimal MSM policies for such networks with $N\geq4$ queues (Thm.~\ref{Chapter04ThmNonexistenceOfDOPoliciesForNGreaterThan4}). 

	\item
	We then extend our theory of MSM policies to schedule transmissions over a much more general class of networks that we call \enquote{Cluster-of-Cliques Constraint Networks,} such as the ones in Figures~\ref{Chapter05FigStarOfCliquesSensorNetwork} and \ref{Chapter05FigClusterOfCollocatedNetworks} (Sec.~\ref{Chapter05SecTOSchedulingPolicies}). We will also see multiple methods to further reduce the amount of state information (empty-nonempty statuses of the queues) that has to be exchanged across the network to make these protocols amenable to distributed implementation. We finally use this theory to propose a throughput-optimal protocol, akin to the QZMAC protocol \cite{mohan-etal16hybrid-macsMASSversion}, wherein scheduling decisions are taken using only the information about activity on the channel (or lack thereof) that can be \emph{sensed} by the nodes and will study its performance in detail (Sec.~\ref{Chapter05SecRemarksDecenralized}). 

\begin{figure}[tb]
\centering
\includegraphics[height=5.0cm, width=8.00cm]{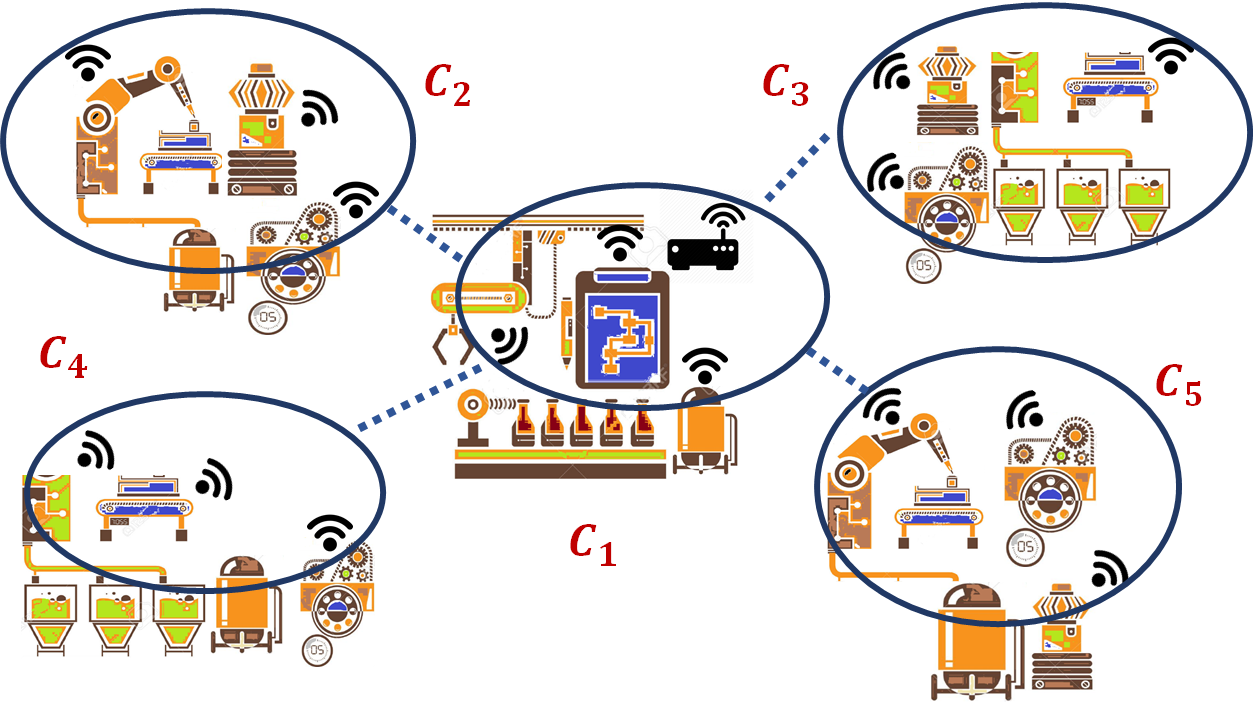}
\caption{A wireless sensor network on a factory floor showing multiple workstations. Sensors within a workstation form a clique or a collocated network. The network in the figure comprises $N=5$ cliques. Fig.~\ref{Chapter05FigExtendingPi3PoliciesToOtherInterferenceGraphs} shows the associated conflict graph.}
\label{Chapter05FigStarOfCliquesSensorNetwork}
\end{figure}

\begin{figure}[tb]
\hspace{4.50cm}
\includegraphics[height=4.0cm, width=12.0cm]{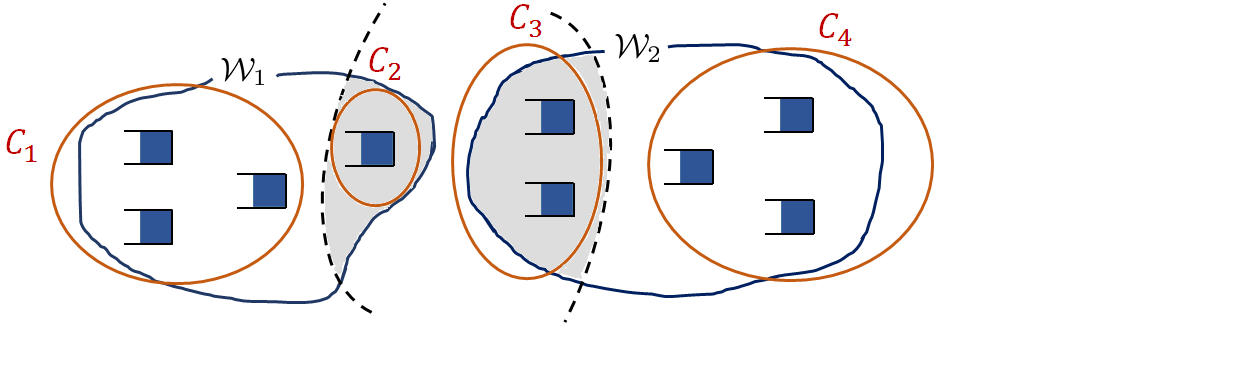}
\caption{An example of an LAoC. The communication links in the two regions \textbf{shaded gray} interfere with each other.
Since each of the two networks $\mathcal{W}_1$ and $\mathcal{W}_2$ is collocated, links within them interfere as well. The total network can hence, be decomposed into 4 cliques, viz. $\mathcal{C}_1,\cdots,\mathcal{C}_4.$ Fig.~\ref{Chapter05FigClustersOfCollocatedNetworksInterferenceGraph} shows the associated conflict graph.}
\label{Chapter05FigClusterOfCollocatedNetworks}
\end{figure}
	\item 
	We then present numerical results (Sec.~\ref{Chapter05SimulationResults}) showing the performance of our proposed policies, and 			comparisons with standard, high-overhead state-based policies such as the MaxWeight-$\alpha$ family \cite{shah-etal07heavy-traffic-optimal-scheduling-switched}. 
	

\end{itemize}

\section{The Scheduling Problem: Models and Notation}\label{secSystemModel}
\label{Chapter03SecSystemModel}
In Sec.~\ref{secTheGeneralQueueSchedulingModel}, we describe the general network model, and specify the optimal scheduling problem in Sec.~\ref{secPerformanceMetric}. Then, in Sec.~\ref{secSystemModelNWirelessLinksPathGraph} and Sec.~\ref{secSystemModelClusterOfCliques} we restrict the general model to the cases that we provide results for in the remainder of the paper.

There are several interfering links (\emph{transmitter-receiver pairs}), where each transmitting node has a stream of arriving packets. Time is slotted, and all links are synchronized to the time slots. In each slot, each scheduled link can transmit one packet. Packets that are not transmitted remain in the queues. Thus, we have a discrete time queue scheduling problem that belongs to the general class introduced in \cite{tassiulas92stability}. Note, from the preceding discussion, that activating a link in a time slot is the same as serving its associated queue. 
\subsection{The General Queue Scheduling Model}\label{secTheGeneralQueueSchedulingModel}
We consider a system comprising $N$ queues, where, as mentioned before, each queue models a radio link in a wireless network.  The {leading edges} of time slots are indexed $0,1,2,\cdots$. 
Exogenous arrivals to the queues are embedded \emph{at} slot boundaries, $t=0,1,2,\cdots,$ with the number of packets arriving to Queue $i$ at time $t$ being denoted by the random variable $A_i(t)$. 
$A_i(t)$ is assumed iid\footnote{\enquote{iid} stands for \emph{independent and identically distributed.}} across time and independent across queues and is modelled as a Bernoulli random variable with mean $\lambda_i$ i.e., $P\left(A_{i}(t)=1\right)=\lambda_{i},~\forall t\geq1.$ However, we will remove this restriction to include batch iid arrivals in Sec.~\ref{Chapter05SecTOSchedulingPolicies}. We use $\mathbf{Q}(t)=[Q_1(t),\dots,Q_N(t)]$ to denote the vector of all queue lengths at time $t$. The queue length process is embedded at the beginnings of time slots, so $Q_i(t),~\forall t\geq0$, is measured at $t+$, i.e., just \emph{after} the arrival. The duration of a slot is assumed to include packet transmission time, the receive-transmit turn around time at the receiver, the MAC layer acknowledgement (ACK) time\footnote{Most wireless systems require a MAC layer acknowledgement to combat high high packet error rates}, and any scheduling overhead. 
Packet transmissions are assumed to take exactly one time slot and succeed with probability\footnote{The effects of fading and channel errors are not considered here and are a subject of future research.} $1$. The random variable, $D_i(t)$, indicating the departure of a packet from Queue $i$ at time $t$, is such that $D_i(t)=1$ if and only if Queue $i$ is scheduled in slot $t$ \emph{and} $Q_i(t)>0$, else $D_i(t) = 0$; here, the departure is assumed to end just before the leading edge of slot $(t+1)$, i.e., at $(t+1)-$. 

The \emph{offered service} process to Queue $i$, $\left\lbrace S_i(t),t\geq0\right\rbrace$, is defined as follows: $S_i(t)=1$ whenever Queue $i$ is given access to the channel, so that 
$D_i(t)=S_i(t)\mathbb{I}_{\{Q_i(t)>0\}},~\forall t\geq0,~1\leq i\leq N$. Depending on the interference constraints, it may be possible to serve only a subset of queues in a given slot. For example, \eqref{eqnSystemModelPathGraphNetworksInterferenceConstraints} gives the constraints for path-graph interference networks and \eqref{Chapter05ActivationConstraintsForSoCModel} for Star-of-Cliques networks. The vector $\mathbf{S}(t):=[S_1(t),\dots,S_N(t)]$ satisfying the interference constraints is called an \emph{\color{blue}activation vector}.
Thus, for every queue $i,$ 
\begin{eqnarray}
Q_i(t+1)&=&Q_i(t)-D_i(t)+A_i(t+1)\nonumber\\
&=&(Q_i(t)-S_i(t))^++A_i(t+1),~\forall t\geq 0.\nonumber
\end{eqnarray}
 Denote by $\boldsymbol{\zeta}(t):=[\mathbb{I}_{\{Q_1(t)>0\}},\dots,\mathbb{I}_{\{Q_N(t)>0\}}]$ the system's \emph{\color{blue}occupancy vector} at time $t$, i.e., the empty-nonempty state of each of the $N$ queues.
%
 Let $\mathcal{V}\subset\{0,1\}^N$ be the set of all activation vectors. A scheduling policy 
 $\pi:=\{\mu_0,\mu_1,\dots\}$ decides which queues are allowed to transmit in each slot as a function of the available history $\mathcal{H}_t$, which comprises the past states and actions known to the controller, and the current (known) queue state. Specifically, $\mu_t: \mathcal{H}_t\rightarrow V$ is an $N\times1$ vector, and $S_i(t)=\mu_t(i)$. When the schedule depends only on state and not on time, the resulting policies are of the form $\pi=\{\mu,\mu,\dots\}$, and are said to be \emph{\color{blue}stationary.} We will focus on stationary policies in this article.
\subsubsection{Performance Metric}\label{secPerformanceMetric} 
By \emph{stability} of the process $\left\lbrace \mathbf{Q}(t),t\geq0\right\rbrace$ 
we will mean that
\begin{equation}
\limsup_{T\rightarrow\infty}\frac{1}{T}\sum_{t=0}^{T-1}\sum_{i=1}^N\mathbb{E}^{\left(\pi\right)}_{\mathbf{Q}(0)} Q_i(t)<\infty.
\label{eqnStrongStability}
\end{equation}
This condition is commonly known as \emph{strong stability} \cite{neely10stochastic-network-optimization-book}. A policy that ensures \eqref{eqnStrongStability} is said to be \emph{\color{blue}stabilizing,} and an arrival rate vector for which a stabilizing policy exists is said to be \emph{\color{blue}stabilizable}. The closure of the set of all stabilizable rate vectors is called the \emph{\color{blue}throughput capacity region} of the network \cite{tassiulas92stability}, and a policy that is stabilizing for every arrival rate vector in the interior of this region is called \emph{\color{blue}throughput-optimal} (T.O.). The set of arrival rates that are stabilizable under a given fixed policy is called \emph{\color{blue}stability region} of the policy. 


\subsection{Path Graph interference networks}\label{secSystemModelNWirelessLinksPathGraph}
\begin{figure}[tb]
\hspace{2.00cm}%
\centering
\includegraphics[height=5.00cm, width=10.10cm]{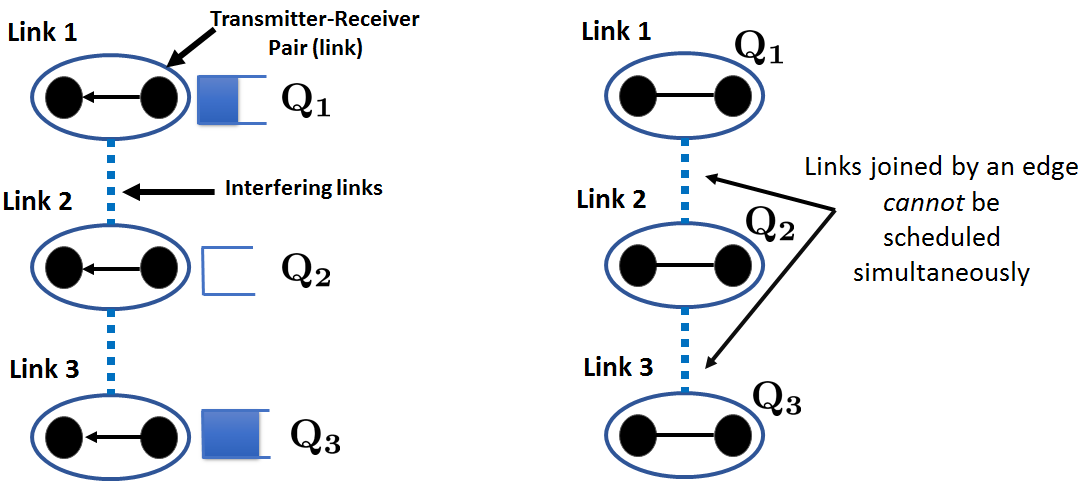}
\caption{The basic path-graph interference system with $N=3$ communication links along with the associated packet queues (left) and its conflict graph (right). 
The interference constraints are such that physically adjacent queues cannot be served simultaneously.}
\label{Chapter03FigBasicQueueingSytemCaricature}
\vspace{-0.5cm}
\end{figure}

The first system we will study in the subsequent sections is modelled by $N$ parallel queues (see Fig.~\ref{Chapter03FigBasicQueueingSytemCaricature}). 
The scheduling constraints are the same as the second model in Tassiulas and Ephremides 1994 \cite{tassiulas-ephremides94dynamic-scheduling-tandem-parallel}, namely that Queue $i$ and Queue $i+1$ cannot be served simultaneously 
for $1\leq i\leq N-1$. These interference constraints enforce the following rule on the offered service process $\mathbf{S}(t),~\forall t\geq0$
\begin{equation}
S_i(t)+S_{i+1}(t)\leq1,~\forall t\geq0,1\leq i\leq N-1.
\label{eqnSystemModelPathGraphNetworksInterferenceConstraints}
\end{equation}
The conflict graph associated with the system is a called \emph{path graph} \cite{gross-yellen05graph-theory-applications,diestel05graph-theory}.
%
%
%
%
 Standard analysis \cite{tassiulas92stability} show that the capacity region of this network is the set 
\begin{equation}
\Lambda_{N}:=\left\lbrace\boldsymbol{\lambda}\in\mathbb{R}^N_+\mid\lambda_i+\lambda_{i+1}\leq1,~\forall1\leq i\leq N-1\right\rbrace,
\label{eqnCapacityRegionTE93} 
\end{equation}
whose interior, $\Lambda^o_N$, is the set of all stabilizable rate vectors.


\subsection{The Cluster-of-Cliques (CoC) graph networks}\label{secSystemModelClusterOfCliques}
In the remainder of the paper, we will refer to the conflict graph associated with a collocated network, i.e., a fully connected graph or subgraph, as a \emph{clique.} The system under consideration comprises multiple cliques and the exact nature of the interference relations \emph{across} cliques are described in detail below. 
The number of packets arriving to Queue~$j$ in Clique~$i$ 
at time $t$ is denoted by the random variable $A_{i,j}(t)$. 
As before, $Q_{i,j}(t)$, the backlog of Queue~$j$ in Clique~$i$ is measured at $t+,~t\geq0$, i.e., just \emph{after} the arrival. 
Once again, as before, for every $(i,j),$ 
\begin{eqnarray}
Q_{i,j}(t+1)&=&Q_{i,j}(t)-D_{i,j}(t)+A_{i,j}(t+1)\nonumber\\
&=&(Q_{i,j}(t)-S_{i,j}(t))^++A_{i,j}(t+1),~\forall t\geq 0.\nonumber
\end{eqnarray}
\begin{figure*}[tbh]
\begin{subfigure}{.5\textwidth}
\hspace{1.0cm}
\includegraphics[height=3.0cm, width=7.0cm]{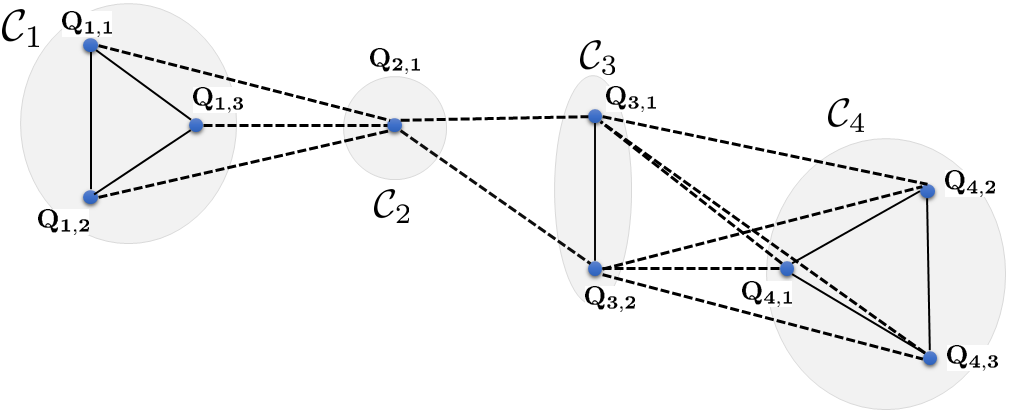}
\caption{The conflict graph associated with a Linear-Array-of-Cliques (LAoC) network. While this is clearly \emph{neither} fully connected nor a path-graph network, we will show how to extend ideas from the analysis of path-graph networks to construct scheduling protocols for such networks.}
\label{Chapter05FigClustersOfCollocatedNetworksInterferenceGraph}
\end{subfigure}
\qquad
\begin{subfigure}{.5\textwidth}
\centering
\includegraphics[height=3.0cm, width=6.0cm]{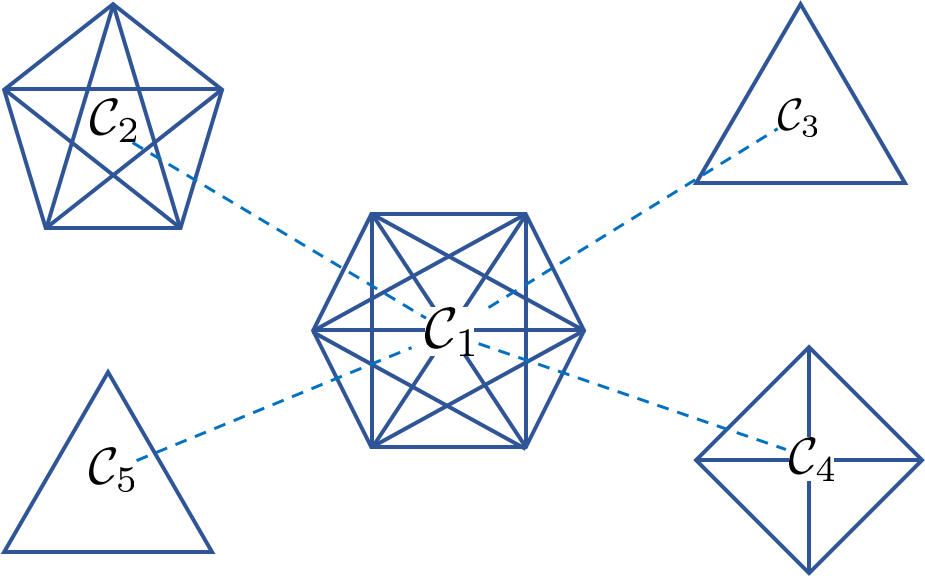}
\caption{The conflict graph associated with a Star-of-Cliques (SoC) network. A dotted line connecting cliques $\mathcal{C}_i$ and $\mathcal{C}_j$ means that transmissions in the two cliques cannot take place simultaneously.}
\label{Chapter05FigExtendingPi3PoliciesToOtherInterferenceGraphs}
\end{subfigure}
\caption{Cluster-of-Cliques networks.}
\vspace{-0.5cm}
\end{figure*}
Depending on the underlying conflict graph, the CoC networks studied in this paper are broadly classified into two classes

\textbf{Star-of-Cliques networks} (SoC): Consider an interference graph consisting of a central fully-connected subgraph (central clique) surrounded by $N-1$ cliques (see Fig.~\ref{Chapter05FigExtendingPi3PoliciesToOtherInterferenceGraphs}).
In other words, the network's conflict graph consists of $N$ cliques denoted $\mathcal{C}_1,\dots,\mathcal{C}_N$, and clique $\mathcal{C}_i$ consists of $\mathcal{N}_i$ vertices -- an \emph{arbitrary number} of cliques each having \emph{arbitrarily many} communication links (queues). 
Transmissions in $\mathcal{C}_1$ interfere with those in all other cliques while the transmissions in $\mathcal{C}_i,~i\geq2$ interfere with those in $\mathcal{C}_1$ only. 
Coming to the offered service processes, for any two queues $Q_{i,j}$ and $Q_{k,l}$ in the system, the interference constraints enforce the rule 
\begin{equation}
S_{i,j}(t)+S_{k,l}(t)\leq1,~\forall t\geq0,\text{ if } i=k,\text{ or }i=1,\text{ or }k=1.
\label{Chapter05ActivationConstraintsForSoCModel}
\end{equation}
Let $\mathcal{N}\equiv\sum_{i=1}^N\mathcal{N}_i$ denote the total number of queues in the system. The capacity region of this system is given by 
\begin{eqnarray}
\hspace{-0.50cm}
\Lambda^{(N)}_s &:=& \left\lbrace\boldsymbol{\lambda}\in\mathbb{R}^{\mathcal{N}}_+\bigg|\sum_{j=1}^{\mathcal{N}_1}\lambda_{1,j}\right.\nonumber\\
&& \left.+\sum_{k=1}^{\mathcal{N}_{i}}\lambda_{i,k}\leq1,~i\in \{2,\cdots,N\}\right\rbrace
\label{Chapter05EqnCapacityRegionOfSoCModel}
\end{eqnarray}
(the subscript $s$ highlights the fact that this is the Star-of-Cliques model).

\textbf{Linear-Array-of-Cliques} (LAoC): This system consists of N cliques $\mathcal{C}_1, \mathcal{C}_2,\cdots,\mathcal{C}_N$, but unlike the SoC model, all transmissions in $\mathcal{C}_{i-1}$ interfere with those in $\mathcal{C}_{i},~i\in\left\lbrace2,\cdots,N\right\rbrace$ and vice-versa (see Fig.~\ref{Chapter05FigClustersOfCollocatedNetworksInterferenceGraph}). As in the SoC model, Clique $\mathcal{C}_i$ comprises $\mathcal{N}_i$ queues and $\mathcal{N}\equiv\sum_{i=1}^N\mathcal{N}_i$ denotes the total number of queues in the system. 
Since transmissions in adjacent cliques interfere with each other, for the offered service processes of any two queues $Q_{i,j}$ and $Q_{k,l}$ in the system, we have 
\begin{equation}
S_{i,j}(t)+S_{k,l}(t)\leq1,~\forall t\geq0,\text{ if }k=i+1,~\forall 1\leq i\leq N-1.
\label{Chapter05ActivationConstraintsForLAoCModel}
\end{equation}
%
%
The capacity region of this system is given by 
\begin{eqnarray}
\Lambda^{(N)}_l &:=& \left\lbrace\boldsymbol{\lambda}\in\mathbb{R}^{\mathcal{N}}_+\bigg|\sum_{j=1}^{\mathcal{N}_i}\lambda_{i,j}\right.\nonumber\\
&& \left.+\sum_{k=1}^{\mathcal{N}_{i+1}}\lambda_{i+1,k}\leq1,~i\in \{1,\cdots,N-1\}\right\rbrace
\label{Chapter05EqnCapacityRegionOfLAoCModel}
\end{eqnarray}
(the subscript $l$ highlights the fact that this is the Linear-Array-of-Cliques model). 
As before, the vector $\mathbf{S}(t):=\left[S_1(t),\dots,S_\mathcal{N}(t)\right]\in\left\lbrace0,1\right\rbrace^\mathcal{N}$ is called an \emph{activation vector} if it satisfies the constraints in \eqref{Chapter05ActivationConstraintsForSoCModel} and \eqref{Chapter05ActivationConstraintsForLAoCModel} in the SoC and LAoC systems, respectively.
We now begin our study with path graph interference networks. 



\section{Maximum Size Matching (MSM) Policies}\label{secMSMPolicies}
We first define the subclass of scheduling policies to which we will restrict our attention and provide some motivation to do so in from the perspective of delay reduction. The latter will be stated and explained more formally in subsequent sections.
\begin{tcolorbox}
\begin{defn}
A policy $\pi$ is a \emph{Maximum Size Matching (MSM)} policy if in every slot the policy schedules the maximum number of nonempty queues subject to the inteference constraints.
\end{defn}
\end{tcolorbox}
For example, if $N=7$, and $\mathbf{Q}(t)=[1,2,0,0,4,3,3]$,  a policy that schedules queues $1,5\text{ and }7$ or $2,5\text{ and }7$ is MSM while a policy that schedules queues $1,7$ only, is not MSM. 
It might be expected that the policy must schedule as many queues as possible to maximise throughput and minimise delay. Indeed,  \cite{tassiulas-ephremides94dynamic-scheduling-tandem-parallel} shows that any policy defined on such path-graph networks can be improved into an MSM policy that will provide stochastically better delay. Interestingly, we show later that even non-MSM policies can be stabilising. 

\begin{figure}[tb]
\centering
\includegraphics[height=5.0cm, width=8cm]{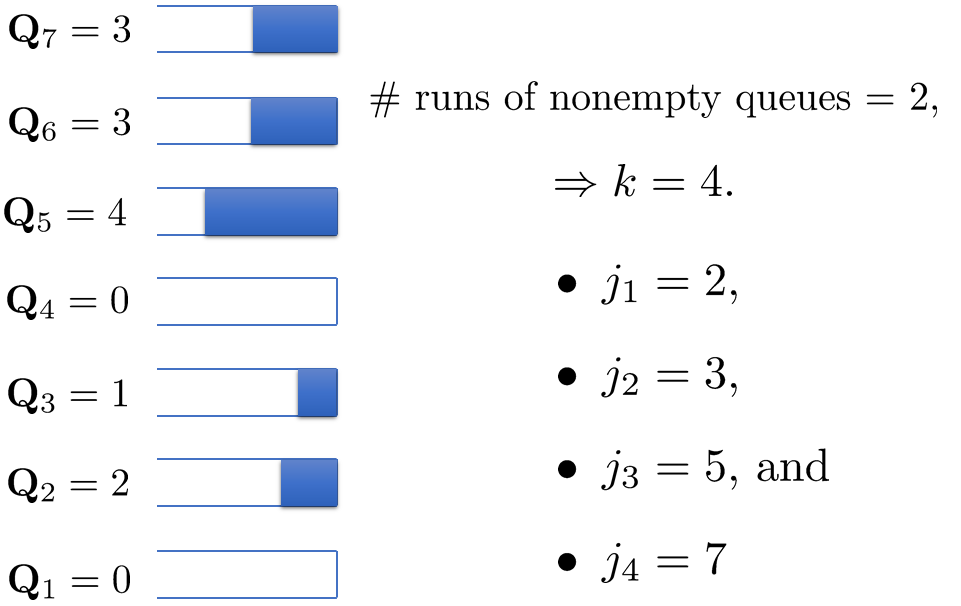}
\caption{Figure depicting how runs of non empty queues are numbered. Here, $N=7$ and since $\boldsymbol{\zeta}=[0,1,1,0,1,1,1]^T$, $S(\boldsymbol{\zeta})=\{[0,1,0,0,1,0,1]^T,[0,0,1,0,1,0,1]^T\}.$ There are two runs of nonempty queues, the first beginning at Queue~2 and ending at Queue~3, and the second beginning at Queue~5 and ending at Queue~7. Hence, $j_1=2,j_2=3,j_3=5\text{ and }j_5=7.$ Notice that odd subscripts indicate the \emph{beginning} of these runs, while even subscripts indicate their ends.}
\label{figSystemOccupancyShowingTheJs}
\end{figure}

Notice that in Fig.~\ref{Chapter03FigBasicQueueingSytemCaricature}, Queue~2 cannot be served in any slot in which either Queue~1 or Queue~3 is being served and similarly Queue~3 cannot be served in any slot in which either Queue~2 or Queue~4 is being served. In contrast, service to Queues 1 depends only on whether Queue~2 is being served, which makes it less constrained from the perspective of service. In this paper, we will refer to Queues~1 and $N$ in a path graph as the \enquote{outer} queues and the other $N-2$ queues as the \enquote{inner,} more constrained queues. Lemma~4.1 in \cite{tassiulas-ephremides94dynamic-scheduling-tandem-parallel}, which we state below (Lem.~\ref{Chapter03lemSufficientForActivationMSM}) since we will be invoking it often in the sequel, defines a class of policies that is more restrictive than MSM 
that can be described informally and succinctly as follows. 
\begin{enumerate}
\item the policy should be MSM (Conditions~\ref{Chapter03CondnOddNonEmptyRuns} \& \ref{Chapter03CondnEvenNonEmptyRuns} below), and
\item\label{Chapter03ConditionPrioritizeInnerQueuesPiTilde} the policy should prioritize \emph{inner} queues over \emph{outer} queues while breaking ties (Condition~\ref{Chapter03conditionMSMDelaySmall} below). 
\end{enumerate} 
Specifically, in \cite{tassiulas-ephremides94dynamic-scheduling-tandem-parallel}, the authors provide a sufficient but \emph{not necessary} condition for an activation vector to serve the largest number of nonempty queues in a slot. 
Define $S(\boldsymbol{\zeta}(t))\subset V$ as the set of all activation vectors that serve the largest number of queues in slot $t$ when the occupancy vector is $\boldsymbol{\zeta}(t)$. Given an occupancy vector $\boldsymbol{\zeta},$ let $k=k(\boldsymbol{\zeta})$ be the \emph{twice} the number of runs of nonempty queues, and $j_1=j_1(\boldsymbol{\zeta}),\dots,j_k= j_k(\boldsymbol{\zeta})$, the nonempty queues that mark the beginnings ($j_{\{odd~subscript\}}$) and ends ($j_{\{even~subscript\}}$) of the nonempty runs, or the two extreme queues (Queues 1 and $N$). Fig.~\ref{figSystemOccupancyShowingTheJs}  illustrates this numbering scheme. 
%
%
Clearly,
\begin{itemize}
\item If $j>j_k$ or $j<j_1$ Queue $j$ is empty,
\item If $j_{2m-1}\leq j\leq j_{2m},~m=1,2,\dots,\frac{k}{2},$ Queue $j$ is nonempty, and
\item If $j_{2m}\leq j\leq j_{2m+1},~m=1,2,\dots,\frac{k}{2}-1,$ Queue $j$ is empty.
\end{itemize}
\begin{lem}(Lem.~4.1 in \cite{tassiulas-ephremides94dynamic-scheduling-tandem-parallel})\label{Chapter03lemSufficientForActivationMSM}
$\mathbf{S}(t)\in S(\boldsymbol{\zeta}(t))$ if
\begin{enumerate}
\item\label{Chapter03CondnOddNonEmptyRuns}\textbf{Odd-length run condition:} If $j_{2m}-j_{2m-1}$ is even, then for all $j_{2m-1}\leq j\leq j_{2m},$ 
\[
\hspace{-1.0cm}   {S}_{j}(t)=
\begin{dcases}
    1~\text{if}~j-j_{2m-1}\text{ is even,}\\
    0~\text{otherwise},
\end{dcases}
\]
$m=1,\dots,k/2.$
\item\label{Chapter03CondnEvenNonEmptyRuns}\textbf{Even-length run condition:} If $j_{2m}-j_{2m-1}$ is odd, then any one of the following 3 conditions must be satisfied
\begin{enumerate}
\item\label{Chapter03CondnEvenNonEmptyRunsStartWithJ2mMinus1} For every $j_{2m-1}\leq j\leq j_{2m},$
\[
\hspace{-1.0cm}   {S}_{j}(t)=
\begin{dcases}
    1~\text{if}~j-j_{2m-1}\text{ is even,}\\
    0~\text{otherwise},
\end{dcases}
\]
\item For every $j_{2m-1}\leq j\leq j_{2m},$
\[
\hspace{-1.0cm}   {S}_{j}(t)=
\begin{dcases}
    1~\text{if}~j-j_{2m-1}\text{ is odd,}\\
    0~\text{otherwise},
\end{dcases}
\] 
or,
\item There exists an $l$ such that 
\[
\hspace{-1.0cm}   {S}_{j}(t)=
\begin{dcases}
    1~\text{if}~j-j_{2m-1}\text{ is even and }j_{2m-1}\leq j<l,\\
    ~~ \text{ or } j_{2m}-j\text{ is even and }j_{2m}\geq j>l+1,\\
    0~\text{otherwise},
\end{dcases}
\] 
\end{enumerate}
$m=1,\dots,k/2.$
\item\label{Chapter03conditionMSMDelaySmall}\textbf{Inner queues priority condition:} If $j_1=1,$ 
\[
\hspace{-1.0cm}   {S}_{j}(t)=
\begin{dcases}
    1~\text{if}~j_2-j\text{ is even, }j_1\leq j\leq j_2,\\
    0~\text{otherwise},
\end{dcases}
\] 
and similarly for the case with $j_2=N.$
\end{enumerate} 
\end{lem}
It is easily seen that condition~\ref{Chapter03conditionMSMDelaySmall} above is \emph{not necessary}, and any $\mathbf{S}$ that satisfies the first two will automatically exist in $S(\boldsymbol{\zeta})$. We will see, later that the third condition helps reduce delay by prioritizing \enquote{inner} queues. With this, one simply needs to ensure that in every slot, the policy chooses activation vectors only from $S(\boldsymbol{\zeta}(t)),$ to ensure that it is MSM. Note that there might exist several MSM activation vectors for a policy to choose from, in a given slot. The ultimate choice might depend on all of $\mathcal{H}_t$ (the history) and not only on $\boldsymbol{\zeta}(t)$. MaxWeight ($MW$) is the obvious example here, since it uses $\mathbf{Q}(t)$ for scheduling rather than just $\boldsymbol{\zeta}(t)$. So, for the same $\boldsymbol{\zeta}(t)$, MaxWeight could end up choosing different MSM vectors, depending on the actual queue lengths in those slots. But we will show that in several interference graphs, $\boldsymbol{\zeta}(t)$ is sufficient not only for stability but also for delay optimality. 

\textbf{Notation: }Classes of scheduling policies (see Fig.~\ref{Chapter03FigAllPi_sSetInclusions})
\begin{itemize}
\item $\Pi^{(N)}$: the class of all policies.
\item $\Gamma^{(N)}_M:$ the class of all MSM policies. 
\item $\Pi^{(N)}_{M}$: the class of all policies that take only the occupancy vector $\boldsymbol{\zeta}(t)$ as input and activate the largest number of non empty queues in every slot, .i.e., MSM policies that require only the empty or nonempty status of the queues in the network.
\item $\tilde{\Pi}^{(N)}$: the class of all MSM policies within $\Pi^{(N)}_{M}$ that additionally break ties in favour of inner queues (see condition~\ref{Chapter03ConditionPrioritizeInnerQueuesPiTilde} above).
\end{itemize} 
Note that $\Pi^{(N)}\supsetneq\Gamma^{(N)}_M\supsetneq\Pi^{(N)}_{M}\supsetneq\tilde{\Pi}^{(N)}$. 
Going back to our $7$-queue example, when $\boldsymbol{\zeta}(t)=[1,1,0,0,1,1,1],$ policies that choose $\mathbf{S}(t)=[{\color{blue}1},0,0,0,1,0,1]$ can be in $\Pi^{(7)}_M$ but not in $\tilde{\Pi}^{(7)}$, while those that choose $\mathbf{S}(t)=[0,{\color{blue}1},0,0,1,0,1]$ can be in $\tilde{\Pi}^{(7)}.$


\section{Queue Nonemptiness-Based (QNB) Scheduling}\label{secPrimerQLengthAgnostic}
While almost all well-known policies use full queue length-information ($\mathbf{Q}(t)$) to take scheduling decisions, e.g., MaxWeight \cite{tassiulas92stability}, a key objective of this paper is to show that for many classes of interference graphs, throughput-optimal policies can be designed that use much less information. By \enquote{queue nonemptiness-based policies,} we mean those that only require the knowledge of the occupancy 
vector, i.e., $\boldsymbol{\zeta}(t)$. Clearly, this contains much less information than the vector $\mathbf{Q}(t)$ that MaxWeight requires, and $\boldsymbol{\zeta}(t)$ can be transmitted across the network with just $1$ bit per queue per slot. 
The functions $\{\mu_t,t\geq0\}$ that constitute the policy are now maps of the form $\mu_t:\{0,1\}^N\rightarrow \mathcal{V} \subsetneq\{0,1\}^N$, the set of all activation vectors.

Although it is well-known that fully-connected interference graphs admit throughput-optimal, queue nonemptiness-based scheduling algorithms (e.g., schedule any nonempty queue), it is not immediately clear how to stabilize other interference graphs with reduced state policies. Moreover, the delay properties of such a reduced state scheduler are naturally suspect, since even MaxWeight, which uses complete knowledge of $\mathbf{Q}(t)$ in every slot, is only known to be asymptotically delay optimal in such networks \cite{maguluri-srikant12heavy-traffic-resource-cloud}. 

We now provide a sufficient condition that will later help construct strongly stable policies that use only $\{\boldsymbol{\zeta}(t),t\geq0,\}$, 
by proving a Lyapunov drift result that will be invoked often in the sequel.
\begin{tcolorbox}
\begin{lem}\label{lemPropertyPMeansTO}
Consider the class of systems described in Section~\ref{Chapter03SecSystemModel}, 
and define property $\mathcal{P}$ as 
\begin{equation}
D_i(t)+D_{i+1}(t)=0\iff Q_i(t)+Q_{i+1}(t)=0,
\tag{$\mathcal{P}$}
\label{eqnPropertyP}
\end{equation}
for all $t\geq0,$ and for $1\leq i\leq N-1.$ Any policy that satisfies property $\mathcal{P}$ in every slot $t$, is throughput-optimal.
\end{lem}
\end{tcolorbox}
\begin{rem}
Note that condition \eqref{eqnPropertyP} depends only on the reduced state $\boldsymbol{\zeta}(t)$. In words, \eqref{eqnPropertyP} reads: \enquote{for a pair of neighboring queues, there is no departure from either of these queues \emph{iff} both the queues are empty.} One direction is clear: when both queues are empty there can be no departures. For example, with $N=4$ and $\boldsymbol{\zeta}(t) = (1,1,1,1)$, $\mathbf{S}(t) = (1,0,1,0)$ satisfies condition \eqref{eqnPropertyP}, but $\mathbf{S}(t) = (1,0,0,1)$ does not. 
\end{rem}
\begin{pf}\label{AppendixProofOfPropertyPMeansTO}
We define a Lyapunov function $L(t):\mathbb{N}^N\rightarrow\R_+$ as 
\begin{equation}
L(\mathbf{Q}(t)):=\sum_{i=1}^{N-1}(Q_i(t)+Q_{i+1}(t))^2
\label{eqnLyapunovFunctionForPropertyP}
\end{equation}
With a slight abuse of notation, we denote $L(\mathbf{Q}(t))$ simply by $L(t)$. Using the Lyapunov drift argument and the telescoping sum method used in Sec.~\ref{AppendixTOFullyConnected}, 
we now show how property $\mathcal{P}$ ensures \emph{strong stability} of the system when the arrivals lie in $\Lambda^o$. To simplify notation, we denote $D_i(t)$ and $A_i(t+1)$ by $D_i$ and $A_i$ respectively, for every $i$ and let $\mathbf{Q}=[Q_1,\dots,Q_N]$.
\begin{eqnarray}
&&\mathbb{E}\left[L(t+1)-L(t)\mid\mathbf{Q}(t)=\mathbf{Q}\right]\nonumber\\
&&=\sum_{i=1}^{N-1}\mathbb{E}\bigg[(Q_i-D_i+A_i+Q_{i+1}-D_{i+1}+A_{i+1})^2\nonumber\\
&&-(Q_i+Q_{i+1})^2\mid\mathbf{Q}(t)=\mathbf{q}\bigg]\nonumber\\%
&&\stackrel{*}{\leq}\sum_{i=1}^{N-1}\bigg[(Q_i+Q_{i+1})^2+1+4-2(Q_i+Q_{i+1}) \nonumber\\
&&\left(\mathbb{E}\left[D_i+D_{i+1}\mid\mathbf{Q}\right]-\lambda_i-\lambda_{i+1}\right)-(Q_i+Q_{i+1})^2\bigg]\nonumber\\
&&=\sum_{i=1}^{N-1}\bigg[5-2(Q_i+Q_{i+1})\nonumber\\
&&\left(\mathbb{E}\left[D_i+D_{i+1}\mid\mathbf{Q}(t)=\mathbf{Q}\right]-\lambda_i-\lambda_{i+1}\right)\bigg],
\label{eqnPropertyPConditionalDriftSimplified}
\end{eqnarray}
where in inequality $*$, firstly, we have used the fact that for any 3 reals $x,y,z$, $(x-y+z)^2\leq x^2+y^2+z^2-2x(y-z)$ and set  $x=q_i+q_{i+1}$, $y=D_i+D_{i+1}$ and $z=A_i+A_{i+1}$. We then use the fact that $D_i+D_{i+1}\leq1$ in any time slot due to the scheduling constraints and since all arrivals are Bernoulli, $A_i+A_{i+1}\leq2,$ for all $1\leq i\leq N-1$. Taking expectation on both sides of Eqn.~\eqref{eqnPropertyPConditionalDriftSimplified}, and thus \emph{removing conditioning,} we get\\
\begin{eqnarray}
\mathbb{E}\left[L(t+1)-L(t)\right]
&\leq& \sum_{i=1}^{N-1}\bigg[5-2\mathbb{E}\left((Q_i+Q_{i+1})\right.\nonumber\\
&&\times\left.\mathbb{E}\left[D_i+D_{i+1}\mid\mathbf{Q}(t)\right]\right)\nonumber\\
&&+2(\lambda_i+\lambda_{i+1})\mathbb{E}\left(Q_i+Q_{i+1}\right)\bigg]
\label{eqnUnconditionedExpectedDrift}
\end{eqnarray}
We now use the fact that the policy satisfies property $\mathcal{P}$, to see that 
%
%
%
%
$\mathbb{E}\left[D_1+D_2\mid\mathbf{Q}(t)\right]=\mathbb{I}_{\{Q_i+Q_{i+1}>0\}},~w.p.1,$ and the fact that for any non negative random variable $Z,$ $\mathbb{E}\left(Z\mathbb{I}_{\{Z>0\}}\right)=\mathbb{E}Z,$ whereby,
\begin{eqnarray*}
\hspace{-1.5cm}
&&\mathbb{E}\left((Q_i+Q_{i+1})\mathbb{E}\left[D_i+D_{i+1}\mid\mathbf{Q}(t)\right]\right)=\mathbb{E}(Q_i+Q_{i+1}).\\
\end{eqnarray*}
This means that
\begin{eqnarray*}
\mathbb{E}\left((Q_i+Q_{i+1})\left(\mathbb{E}\left[D_i+D_{i+1}\mid\mathbf{Q}(t)\right]+(\lambda_i+\lambda_{i+1})\right)\right)\\
= \epsilon_i\mathbb{E}(Q_i+Q_{i+1}),
\end{eqnarray*}
 where $\epsilon_i=1-\lambda_i-\lambda_{i+1}$. Note that from the definition of $\Lambda^o$, $\epsilon_i>0,~\forall1\leq i\leq N-1$. 
 Substituting this in Eqn.~\eqref{eqnUnconditionedExpectedDrift}, we get
\begin{eqnarray}
&&\mathbb{E}\left[L(t+1)-L(t)\right]\nonumber\\
&&\leq 5(N-1)-2\sum_{i=1}^{N-1}\epsilon_i\mathbb{E}\left[Q_i(t)+Q_{i+1}(t)\right],\nonumber\\
&&\stackrel{\dagger}{\leq}5(N-1)-2\sum_{i=1}^{N-1}\epsilon\mathbb{E}\left[Q_i(t)+Q_{i+1}(t)\right],\nonumber\\
&&=5(N-1)-2\epsilon\mathbb{E}Q_1(t)-4\epsilon\sum_{i=2}^{N-2}\mathbb{E}Q_i(t)-2\epsilon\mathbb{E}Q_N(t),\nonumber
\end{eqnarray}
where in inequality $\dagger,$ $\epsilon:=\min_{1\leq i\leq N-1}\epsilon_i$. Since $\epsilon>0,$ $-4\epsilon\sum_{i=2}^{N-2}\mathbb{E}Q_i(t)<-2\epsilon\sum_{i=2}^{N-2}\mathbb{E}Q_i(t)$. Using this, we get
\begin{eqnarray}
\mathbb{E}\left[L(t+1)-L(t)\right]&\leq& 5(N-1)-2\epsilon\sum_{i=1}^{N}\mathbb{E}Q_i(t).
\label{eqnReadyForTelescopingPropertyP}
\end{eqnarray}
Using the telescoping sum technique (see \cite{neely10stochastic-network-optimization-book}) it can now be shown that the process $\left\lbrace \mathbf{Q}(t),~t\geq0\right\rbrace$, is strongly stable.
\end{pf}

\section{Path Graph Conflict Model with $N=3$:\\QNB Scheduling}\label{secTOSchedulingWith3Queues}
In this section, we first completely characterize $\Pi^{(3)}_{M}$ and the subclass $\tilde{\Pi}^{(3)}$, and explore stability and delay optimality for this
system. This study will provide some insights into the nature of MSM policies in general and, more importantly, in this process, the policies we propose here will act as \emph{building blocks} for policies for larger-$N$ systems. 
Before we embark on this analysis, we would like to make a few preliminary observations about $\Pi^{(3)}.$ 
Since this part of the thesis uses heavy notation, for the reader's convenience a glossary of notation is provided here: \ref{glossaryOfNotationForPart2}.

Note that with 3 queues, in any given slot $t$, a policy can choose either $\mathbf{S}(t)=[1,0,1]$ which serves Queues $1$ and $3$, or $[0,1,0]$ which serves Queue 2. So, a queue nonemptiness-based policy maps every state vector $\boldsymbol{\zeta}(t)$, of which there are 8 alternatives, to one of these two  activation vectors, giving us 
$2^8=256$ nonemptiness-based policies in all. Suppose $\mid A\mid$ denotes the cardinality of set $A.$ We prove that upon imposing the MSM condition, this number reduces to 4, i.e., $|\Pi^{(3)}_M|=4$ 
as follows. There is no choice to be made when $\boldsymbol{\zeta}(t)$ is either $\mathbf{0}$, a singleton, or $[1,0,1].$ The MSM condition also means that $\mathbf{S}(t)=[1,1,1]\mapsto\boldsymbol{\zeta}(t)=[1,0,1].$ This leaves only two states, viz., $[1,1,0]$ and $[0,1,1]$, each of which can be assigned either $[0,1,0]$ or $[1,0,1]$ and hence, $\mid\Pi^{(3)}_M\mid=4.$

\begin{figure}[t]
\centering
\includegraphics[height=4.5cm, width=6.5cm]{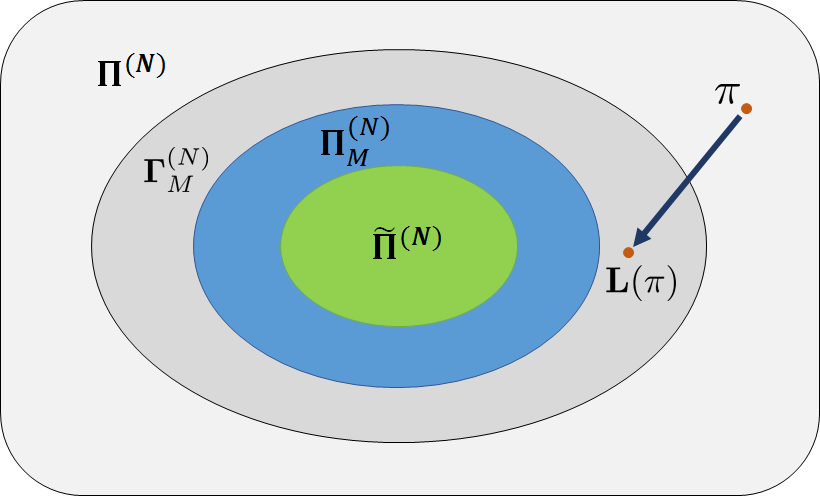}
\caption{Illustrating the four policy spaces and the action of the projection operator $L$. Recall that $\pi^{(3)}_{TD},\pi^{(3)}_{BU},\text{ and }\pi^{(3)}_{OQ}\in\Pi^{(3)}_{M}$, and that $\tilde{\pi}^{(3)}_{IQ}\in\tilde{\Pi}^{(3)}$. In Sec.~\ref{secNonMSMStablePolicy} we will encounter a policy, $\pi^{(3)}_{IQ}\in\Pi^{(3)}\setminus\Gamma^{(3)}_M.$}
\label{Chapter03FigAllPi_sSetInclusions}
\end{figure}

\textbf{Characterization of $\boldsymbol{\Pi}^{(3)}_M$}: We now show that this class contains throughput optimal, delay optimal, and also 
\emph{unstable} MSM policies. First, some additional notation is in order. Depending on the mapping from $\boldsymbol{\zeta}(t)$ to the activation vector, we denote the 4 MSM policies $\pi^{(3)}_{TD},\pi^{(3)}_{BU},\tilde{\pi}^{(3)}_{IQ},\pi^{(3)}_{OQ}$. We will follow the scheme below in the remainder of the thesis.
\begin{itemize}
\item The subscripts \enquote{TD} and \enquote{BU} stand for \enquote{Top-Down} and \enquote{Bottom-Up,} respectively and the reason for this nomenclature will become apparent shortly.
\item A \enquote{$\sim$} in the superscript always represents a policy in $\tilde{\Pi}^{N}$, regardless of any subscripts. It indicates that these policies always break ties in favor of inner queues. For example, $\tilde{\pi}^{(3)}_{IQ}\in\tilde{\Pi}^{3}$.
\end{itemize}
\begin{table}[htb]
\caption{Comparison of $\mathbf{S}(t)$ under $\pi^{(3)}_{TD}$, $\pi^{(3)}_{BU}$, $\tilde{\pi}^{(3)}_{IQ}$ and $\pi^{(3)}_{OQ}$} 
\centering 
\begin{tabular}{c c c c c} 
\hline\hline 
$\boldsymbol{\zeta}=\left[\zeta_1(t),\zeta_2(t),\zeta_3(t)\right]$ & $\pi^{(3)}_{TD}$ & $\pi^{(3)}_{BU}$ & $\tilde{\pi}^{(3)}_{IQ}$ & $\pi^{(3)}_{OQ}$ \\ [1ex] 
\hline 
\\
000 &    101 & 101 & 101 & 101 \\ 
001 &    101 & 101 & 101 & 101 \\
010 &    010 & 010 & 010 & 010 \\
\color{magenta}$\mathbf{011}$ &   \color{blue}010 & \color{red}101 & \color{ao(english)}010 & \color{darkorchid}101 \color{black} \\
\hline
\\
100 &    101 & 101 & 101 & 101 \\
101 &    101 & 101 & 101 & 101 \\
\color{magenta}$\mathbf{110}$ &   \color{blue}101 & \color{red}010 & \color{ao(english)}010 & \color{darkorchid}101  \color{black}\\ \color{black} 
111 &    101 & 101 & 101 & 101 \\ 
\hline 
\end{tabular}
\label{tableAll3QueueMSMPolicies} 
\end{table}
The complete descriptions of all these policies are given in Table.~\ref{tableAll3QueueMSMPolicies}, and the caption for  Fig.~\ref{Chapter03FigAllPi_sSetInclusions} specifies to which class each of these four policies belongs. 
In what follows we will describe and analyse each of these policies in detail. Notice from the entries corresponding to the rows $\boldsymbol{\zeta}=[011]$ and $\boldsymbol{\zeta}=[110]$ that $\pi^{(3)}_{TD}$\text{ and }$\pi^{(3)}_{BU}$ are complementary policies, and so are $\tilde{\pi}^{(3)}_{IQ}$ and $\pi^{(3)}_{OQ}$. Specifically, each of these four policies induces the following priority order, which will become clear when we consider each of them individually later:
\begin{itemize}
\item $\pi^{(3)}_{TD}$ gives \emph{decreasing} priority to Queues 1, 2 and 3 in that order,
\item $\pi^{(3)}_{BU}$ gives \emph{increasing} priority to Queues 1, 2 and 3 in that order,
\item $\tilde{\pi}^{(3)}_{IQ}$ gives maximum priority to Queue 2 the \emph{inner} queue (once again, check the rows in Table.~\ref{tableAll3QueueMSMPolicies} corresponding to $\boldsymbol{\zeta}=[011]$ and $\boldsymbol{\zeta}=[110]$), while not compromising the MSM property. This is, of course consistent with the fact that it lies in the $\tilde{\Pi}^{(3)}$ class where ties are always broken in favor of inner queues, and
\item $\pi^{(3)}_{OQ}$ prioritizes the two \emph{outer} queues. 
\end{itemize}

To begin with, we show that $\pi^{(3)}_{TD}$ and $\pi^{(3)}_{BU}$ are T.O. Both these policies will later be used as building blocks to construct T.O. policies for larger systems and are therefore very important to our study. 

\subsection{ Analysis of $\pi^{(3)}_{TD}$ and $\pi^{(3)}_{BU}$}\label{Chapter03SecAnalysisOfPi3_TDPi3_BU}

As the column corresponding to $\pi^{(3)}_{TD}$ in Table.~\ref{tableAll3QueueMSMPolicies} shows, this policy clearly gives absolute priority to Queue 1, i.e., serves Queue~1 whenever it is nonempty, and can be restated as follows.

\textsl{ At time $t$
\begin{enumerate}\label{enumerateDecisionTreeP3_1}
\item If $Q_1(t)>0$ choose $\mathbf{S}(t)=[1,0,1]$.
\item Else, if $Q_2(t)>0$, choose $[0,1,0]$.
\item Else choose $[1,0,1]$.
\end{enumerate}
}

In words, the policy $\pi^{(3)}_{TD}$ simply reads {\em \enquote{prioritize Queue 1 over Queue 2, and Queue 2 over Queue 3, while scheduling all possible non-interfering queues.}} Hence, the subscript \enquote{TD,} since this policy, in a sense, establishes a \enquote{Top-Down} priority.
\begin{thm}\label{thmPi3_1IsTO}
$\pi^{(3)}_{TD}$ and $\pi^{(3)}_{BU}$ are both throughput-optimal.
\end{thm}

The proof of Theorem \ref{thmPi3_1IsTO} uses the fact that under $\pi^{(3)}_{TD}$, Queues 1 and 2 form a priority queueing system and are stable. We then show that Queue 3 is served \enquote{sufficiently often} to ensure stability. $\pi^{(3)}_{BU}$ simply swaps the priorities of Queues 1 and 3 
and its proof proceeds mutatis mutandis. The complete proof is available in Sec.~\ref{AppendixProofOfPi3_1IsTO} in the Appendix. 

\subsection{Analysis of $\tilde{\pi}^{(3)}_{IQ}$}\label{Chapter03SecTildePi3_IQ}
This policy can be restated as follows. 

\textsl{At time $t$:\label{Chapter03DefinitionOfPolicyPi3DO}
\begin{enumerate}
\item If $Q_1(t)>0$ and $Q_3(t)>0$, choose $\mathbf{S}(t)=[1,0,1]$.
\item Else, if $Q_2(t)>0$, choose $[0,1,0]$.
\item Else choose $[1,0,1]$.
\end{enumerate}
}
In \cite{tassiulas-ephremides94dynamic-scheduling-tandem-parallel}, it has been asserted without formal proof that $\tilde{\pi}^{(3)}_{IQ}$ is delay optimal. We begin analysing the policy by proving that it is Throughput Optimal.
\begin{thm}\label{thmPi3_3IsTO}
$\tilde{\pi}^{(3)}_{IQ}$ is throughput-optimal.
\end{thm}

The proof of this result involves showing that $\tilde{\pi}^{(3)}_{IQ}$ satisfies property $\mathcal{P}$ in Lem.~\ref{lemPropertyPMeansTO} and is therefore T.O.
The proof is available in Sec.~\ref{AppendixProofOfPi3_3IsTO} in the Appendix. 

We next turn to the delay performance of the policy $\tilde{\pi}{(3)}_{IQ}$. Tassiulas and Ephremides \cite[Theorem~4.2]{tassiulas-ephremides94dynamic-scheduling-tandem-parallel} define a projection operator $L:\Pi^{(N)}\rightarrow\Gamma^{(N)}_M$ that takes any policy $\pi\in\Pi^{(N)}$ and produces an MSM policy, $L(\pi)$. They then show that the sum queue length with this MSM policy $L(\pi)$ is stochastically smaller than with $\pi$. Specifically, if $\mathbf{Q}^{\pi}(t)$ denotes the backlog induced by some policy $\pi$, then Theorem~4.2 in \cite{tassiulas-ephremides94dynamic-scheduling-tandem-parallel} shows that when the systems upon which $\pi$ and $L(\pi)$ act are started out in the same initial state and the arrivals have the same statistics, then
\begin{equation}
\sum_{i=1}^NQ^{L(\pi)}_i(t)\stackrel{st}{\leq}\sum_{i=1}^NQ^\pi_i(t),~\forall t\geq0,
\label{Chapter03eqnLReducesBacklogsGeneralN}
\end{equation}
where $st$ denotes stochastic ordering. Notice that in the above stochastic ordering relation is required to hold for any arrival rate vector in the system's capacity region. Extending this gives rise to the concept of a \emph{Uniformly Delay Optimal Policy:}
\begin{tcolorbox}
\begin{defn}\label{Chapter03DefnUniformlyDelayOptimal}
For a path graph interference network with $N$ queues, a policy $\pi^*\in\Pi^{(N)}$ is said to be \emph{Uniformly Delay Optimal} if, given \emph{any} policy $\pi\in\Pi^{(N)}$, when the systems upon which $\pi$ and $\pi^*$ act are started out in the same initial state and with the same arrivals statistics and for every arrival rate $\boldsymbol{\lambda}\in\Lambda_N,$
\begin{equation}
\sum_{i=1}^NQ^{\pi^*}_i(t)\stackrel{st}{\leq}\sum_{i=1}^NQ^\pi_i(t),~\forall t\geq0.
\label{Chapter03EqnDefnUniformlyDO}
\end{equation}
\end{defn}
\end{tcolorbox}
%
In \cite[Remark~2, pp.~353]{tassiulas-ephremides94dynamic-scheduling-tandem-parallel}, it is suggested that for $N=3$ there is exactly one MSM policy and that, as a result of Theorem~4.2 therein, is also sum queue length optimal (in the stochastic ordering sense). It is clear, however, that for $N=3$ there are 4 MSM policies. Indeed, the unique MSM policy that the authors refer to in \cite{tassiulas-ephremides94dynamic-scheduling-tandem-parallel} is $\tilde{\pi}^{(3)}_{IQ}$, which also prioritises inner queues. However, the projection operator $L(\cdot)$ does not ensure the inner queue prioritisation condition (3 in Lemma~4.1 therein). Thus, Theorem~4.2, as it stands, merely asserts that any one of the MSM policies could be delay optimal. It requires a further step in the proof to show that $\tilde{\pi}^{(3)}_{IQ}$ is, indeed, the unique uniformly delay optimal policy for $N=3$. We proceed to do so below.
\subsection{Improving Delay Performance via Projections}\label{Chapter03secProjectionInPolicySpace}
We use the operator $L:\Pi^{(N)}\rightarrow\Gamma^{(N)}_M$ (defined in \cite{tassiulas-ephremides94dynamic-scheduling-tandem-parallel}) that takes any policy $\pi\in\Pi^{(N)}$ and produces an MSM policy, $\pi_M\equiv L(\pi)$. 
In slot $t,$ suppose the occupancy vector is $\boldsymbol{\zeta}(t)$ and the system backlog is $\mathbf{Q}(t)$. Suppose also that $\pi$ chooses the activation vector $\mathbf{s'}$ in the slot. The operator $L$ produces another policy $\pi_M$ by constructing its activation vector $\mathbf{s}$, as follows. 
\begin{defn}\cite[Lem.~4.2]{tassiulas-ephremides94dynamic-scheduling-tandem-parallel} Given $\pi\in\Pi^{(N)}$ and occupancy vector $\boldsymbol{\zeta}$,
\begin{enumerate}
\item\label{Chapter03CondnProjectionL1} For all odd-length runs $\{j_{2m-1},\dots,j_{2m}\}$ of nonempty queues $\pi_M$ activates Queue $j_{2m-1}$ and every \emph{other} queue until and excluding $j_{2m}.$
\item When the run is of even length, $\pi_M$ follows $\pi$ in the following sense. 
\begin{enumerate}
\item If $\pi$ does not choose Queue $j_{2m-1}$, so does $\pi_M$, i.e., if $s'_{j_{2m-1}}=0$
\[
\hspace{-1.0cm}   {s}_{j}(t)=
\begin{dcases}
    1~\text{if}~j-j_{2m-1}\text{ is odd,}\\
    0~\text{otherwise},
\end{dcases}
\]
\item If $\pi$ chooses Queue $j_{2m-1}$, so does $\pi_M$, i.e., if $s'_{j_{2m-1}}=1$
\[
\hspace{-1.0cm}   {s}_{j}(t)=
\begin{dcases}
    1~\text{if}~j-j_{2m-1}\text{ is even,}\\
    0~\text{otherwise},
\end{dcases}
\]
\item Finally, if $s'_{j_{2m-1}}=1$ and $s'_{j_{2m}}=1$, there must exist\footnote{If not, one would begin with $s'_{j_{2m-1}}=1$ and set every \emph{other} $s'_j=1$ resulting in $s'_{j_{2m}}=0$, which results in a contradiction.}
two consecutive $0$ entries in $\mathbf{s}'$ between $j_{2m-1}$ and $j_{2m}$. Let $l$ be the smallest such number, i.e., $s'_{l}=s'_{l+1}=0.$ Then
\[
\hspace{-1.0cm}   {s}_{j}(t)=
\begin{dcases}
    1~\text{if}~j-j_{2m-1}\text{ is even, and }j_{2m-1}\leq j<l\\
    ~~or~j_{2m}-j\text{ is even, and }l+1<j\leq j_{2m}\\
    0~\text{otherwise},
\end{dcases}
\]
\end{enumerate}
\end{enumerate}
\end{defn}

Suppose we denote by $\mathbf{Q}^{\pi_M}(t)$ the backlog induced by $\pi_M$. Thm.~4.2 in \cite{tassiulas-ephremides94dynamic-scheduling-tandem-parallel} shows that when the systems upon which $\pi$ and $\pi_M$ act are started out in the same initial state and the arrivals have the same statistics, 
\begin{equation}
\sum_{i=1}^NQ^{\pi_M}_i(t)\stackrel{st}{\leq}\sum_{i=1}^NQ^\pi_i(t).
\end{equation}
It is easy to see that $L$ is a projection onto $\Gamma^{(N)}_M$, i.e.,
\begin{itemize}
\item $L^2:=L\circ L=L$, which means that $L(L(\pi))=L(\pi)$, $\forall\pi\in\Pi^{(N)},$ and 
\item On $\Gamma^{(N)}_M,$ $L$ is the identity map, i.e., for every $\pi_M\in\Gamma^{(N)}_M,$ $L(\pi_M)=\pi_M.$
\end{itemize}

\begin{rem}
 What is most important about this projection is that it does \emph{not} map directly to $\tilde{\Pi}^{(N)}$. Specifically, $L$ does not guarantee that the \emph{second} MSM condition (condition~\ref{Chapter03ConditionPrioritizeInnerQueuesPiTilde}) in Sec.~\ref{secMSMPolicies} is satisfied by $L(\pi)$ for any $\pi\in\Pi^{(N)}$. In other words, $L$ does not necessarily lead to an MSM policy that prioritises inner queues. Condition~\ref{Chapter03conditionMSMDelaySmall} in Lem.~\ref{Chapter03lemSufficientForActivationMSM} is necessarily satisfied by $L(\pi)$ for any $\pi\in\Pi^{(N)}$.
For example, $L\left(\pi^{(3)}_{TD}\right)=\pi^{(3)}_{TD}$ and $L\left(\pi^{(3)}_{BU}\right)=\pi^{(3)}_{BU}$, although they are both in $\Pi^{(3)}_M\setminus\tilde{\Pi}^{(3)}$.
\end{rem}
\begin{rem}\label{Chapter03remStabilityThroughProjections} 
Furthermore, it is easy to check that $\tilde{\pi}^{(3)}_{IQ}=L(\pi^{(3)}_{IQ})$, since they differ only when $\boldsymbol{\zeta}(t)=[1,1,1]$. In this case, $\tilde{\pi}^{(3)}_{IQ}$ schedules queues 1 and 3, while $\pi^{(3)}_{IQ}$ schedules Queue 2. But condition~\ref{Chapter03CondnProjectionL1} ensures that $L(\pi^{(3)}_{IQ})$ chooses queues 1 and 3, whereby we get that $\tilde{\pi}^{(3)}_{IQ}=L(\pi^{(3)}_{IQ})$.  Since $\pi^{(3)}_{IQ}$ is T.O., using Eqn.~\ref{Chapter03eqnLReducesBacklogsGeneralN}, we get, for all $t\geq0,$
\begin{eqnarray}
\sum_{i=1}^NQ^{\tilde{\pi}^{(3)}_{IQ}}_i(t)&\stackrel{st}{\leq}&\sum_{i=1}^NQ^{\pi^{(3)}_{IQ}}_i,\nonumber\\
\Rightarrow \sum_{i=1}^N\mathbb{E}Q^{\tilde{\pi}^{(3)}_{IQ}}_i(t)&\leq &\sum_{i=1}^N\mathbb{E}Q^{\pi^{(3)}_{IQ}}_i,\nonumber
\end{eqnarray}
which immediately gives us
\begin{eqnarray}
&&\limsup_{T\rightarrow\infty}\frac{1}{T}\sum_{t=0}^{T-1}\sum_{i=1}^N\mathbb{E}_{\tilde{\pi}^{(3)}_{IQ}}Q_i(t)\nonumber\\
&&<\limsup_{T\rightarrow\infty}\frac{1}{T}\sum_{t=0}^{T-1}\sum_{i=1}^N\mathbb{E}_{\pi^{(3)}_{IQ}}Q_i(t)\nonumber\\
&&<\infty,
\label{eqnStabilityThroughProjections}
\end{eqnarray}
and shows that $\tilde{\pi}^{(3)}_{IQ}$ is T.O. as well. This provides another way to check the stability of $\tilde{\pi}^{(3)}_{IQ}.$ This technique of producing a stable but \emph{non MSM} policy and projecting it onto $\Pi^{(N)}_M$ is very important and will be used repeatedly in this paper to prove the throughput-optimality of many of the scheduling policies that we propose. 
\end{rem}
\begin{rem}
Eqn.~\eqref{Chapter03eqnLReducesBacklogsGeneralN} shows that $L(MW)$ is always strongly stable, since MW is. 
The proof proceeds along the same lines as the argument in Remark.~\ref{Chapter03remStabilityThroughProjections}. In fact, the same proof can be easily generalized to show that 
\label{Chapter03ThmLPreservesTO}
If any policy $\pi\in\Pi^{(N)}$ is throughput-optimal, then so is $L(\pi).$
\end{rem}
Note that by our definition of the projection operator $L$ and that of (see glossary of notation: \ref{glossaryOfNotationForPart2}) the class $\Pi^{(N)}_M$, $L(MW)$ does \emph{not} reside in this class, since the same occupancy vector $\boldsymbol{\zeta}(t)$ can map to two different $\mathbf{s}(t)$'s depending on what MW chooses in that slot. However, 
$L(MW)$ is an important policy and we will use it for comparison in our simulation results later. 

Patently, all three policies we have proposed so far are MSM as they satisfy the 1$^{st}$ condition for a policy to be MSM (Sec.~\ref{secMSMPolicies}) 
but only $\tilde{\pi}^{(3)}_{IQ}$ satisfies the 2$^{nd}$ condition as well. 
%
The projection operator $L,$ as defined in \cite{tassiulas-ephremides94dynamic-scheduling-tandem-parallel}, only results in an MSM policy and \emph{not} an MSM policy that also prioritises inner queues. There  are 4 MSM policies, of which only prioritises inner queues. If $L$ was designed to map to MSM policies that also prioritise inner queues, then by Thm.~4.2 in \cite{tassiulas-ephremides94dynamic-scheduling-tandem-parallel} we could immediately conclude that in $\tilde{\pi}^{(3)}_{IQ}$, the unique MSM policy that also prioritises inner queues, is delay optimal in ${\Pi}^{(3)}$. However, $L$ does ensure that the delay with $L(\pi)$ is no worse than that with $\pi\in\Pi^{(3)}$, and hence, we need to look for a uniformly delay optimal (see Defn.~\ref{Chapter03DefnUniformlyDelayOptimal}) policy among the 4 MSM policies.
 
Now, since $\Pi^{(3)}_M$ also contains $\pi^{(3)}_{TD}$ and $\pi^{(3)}_{BU}$ and is \emph{not} a singleton, it becomes necessary to examine the delay performance of the proposed policies in greater detail. 
It should be noted that the ideas used in this proof will form the basis for analysing the delay performance of policies for larger systems later. 
\begin{thm}\label{Chapter03thmPi33isDelayOptimal}
For any policy $\pi\in\Pi^{(3)}$, let the system backlog vector at time $t$ be denoted by $\mathbf{Q}^\pi(t)$ and the backlog with $\tilde{\pi}^{(3)}_{IQ}$ be denoted by $\mathbf{Q}^{\tilde{\pi}^{(3)}_{IQ}}(t)$. Also let $\mathbf{Q}^\pi(0)=\mathbf{Q}^{\tilde{\pi}^{(3)}_{IQ}}(0).$
Then,
\begin{equation}
\sum_{i=1}^3Q^{\tilde{\pi}^{(3)}_{IQ}}_i(t)\stackrel{st}{\leq}\sum_{i=1}^3Q^\pi_i(t),~\forall t\geq0,
\label{eqnPi33isDelayOptimal}
\end{equation}
where \enquote{$st$} denotes stochastic ordering. 
\end{thm}

The proof technique is essentially the same as that of Theorem~4.2 in \cite{tassiulas-ephremides94dynamic-scheduling-tandem-parallel}, except that we make the observation that a key step in that proof has more general applicability. 
It involves constructing a sequence of policies each of which shows better delay than its predecessor and than a general policy $\pi.$ The limit of this sequence of policies is then shown to uniquely be $\tilde{\pi}^{(3)}_{IQ}$. The proof is deferred to Sec.~\ref{AppendixProofOfPi33isDelayOptimal} in the Appendix.

Directly analysing the stability and delay properties of the policies we propose in the sequel is very difficult. We therefore develop indirect methods to analyse them by first analysing non MSM policies whose behavior can be understood easily, but that do not show desirable delay properties and study the proposed policies as modifications (such as projection) of these simpler policies, with the modifications giving rise to better delay performance. 

\subsection{Analysis of $\pi^{(3)}_{OQ}$}\label{secAnalysisOfPi3Tilde}
This policy prioritizes the outer queues and can be restated as follows.

\textsl{At time $t$:
\begin{enumerate}
\item If either $Q_1(t)>0$ or $Q_3(t)>0$, choose $(1,0,1)$.
\item Else choose $(0,1,0)$.
\end{enumerate}
}
It turns out, analogous to the observation by McKeown et al \cite{mckeown-etal19achieving-100-percent-throughput-switch} that this MSM policy is, in fact, not throughput-optimal. 
\begin{prop}[MSM but not throughput-optimal] \label{thmPi3_4IsNotTO}
$\pi^{(3)}_{OQ}$ is \emph{not} throughput-optimal.
\end{prop}

The proof of this result involves constructing an arrival rate vector for which the offered service rate to one of the queues is strictly smaller than the arrival rate. It is available in Sec.~\ref{AppendixPi3_4IsNotTO} of the Appendix. Once again, this proof technique is important and we will repeatedly use it in the sequel.

This completes the characterization of $\Pi^{(3)}_M$. 

\subsection{Policies outside $\Pi^{(3)}_{M}$}\label{secNonMSMStablePolicy}
We now propose and analyse a policy that we denote $\pi^{(3)}_{IQ}$, and show the rather surprising result that it is T.O. despite not being MSM. This stability comes from the fact the policy prioritizes the inner queue. However, since it is not MSM, its delay performance is not very good (see simulation results in Sec.~\ref{Chapter05SimulationResults}).
This policy will become important shortly as a fundamental building block while constructing policies for larger systems using a novel \emph{Policy Splicing} technique.

\textsl{At time $t$\label{Chapter03DefinitionOfPi3IQNonMSMButTO}
\begin{enumerate}
\item If $Q_2(t)>0$ choose $\mathbf{S}(t)=[0,1,0],$
\item Else choose $\mathbf{s}(t)=[1,0,1]$.
\end{enumerate}
}
Since $\boldsymbol{\zeta}(t)=[1,1,1]\mapsto[0,1,0],$ this policy is not MSM. However, we have
\begin{prop}[A non-MSM but throughput-optimal policy]
\label{thmNonMSMPi3IsTO}
$\pi^{(3)}_{IQ}$ is throughput-optimal.
\end{prop}

\begin{pf}
The key tool behind the proof of this result is the throughput-optimality Lem.~\ref{lemPropertyPMeansTO}. It is easily checked that $\pi^{(3)}_{IQ}$ satisfies property $\mathcal{P}$ in every slot and thus, by Lemma~\ref{lemPropertyPMeansTO}, is throughput-optimal. 
\end{pf}


\section{A Randomized Policy: The Flow-in-the-Middle Problem}\label{Chapter03RandomizedPolicyFlowInTheMiddle}
%

The \enquote{Flow-in-the-middle} problem, or FIM for short, is a fundamental problem faced by all networks that employ CSMA at the MAC layer. This problem has been studied in detail both analytically and experimentally in asynchronous \emph{continuous-time} systems in the literature \cite{garetto-etal06modeling-per-flow-throughput-starvation-csma,chaudet-etal04impact-asymmetry-802-11-multihop-basic-case,nardelli-etal11experimental-evaluation-optimal-csma,warrier09diffq-practical-differential-backlog-congestion-control,maguluri11optimal-scheduling-ad-hoc-wireless,wang-kar05throughput-fairness-csma-ca-ad-hoc}. 
In this section, we aim to model such a scenario, albeit in slotted time, and understand whether such a phenomenon can occur in the network under study, which naturally leads to the central link (or flow) being starved for extended periods of time.
Recall that the occupancy vector is defined as $\boldsymbol{\zeta}(t):=\left[\mathbb{I}_{\lbrace Q_1(t)>0\rbrace},\mathbb{I}_{\lbrace Q_2(t)>0\rbrace},\mathbb{I}_{\lbrace Q_3(t)>0\rbrace}\right]$. 
Consider the policy $\rho^{(3)}_{\gamma}$ indexed\footnote{We use $\rho$ instead of $\pi$ to highlight the fact that this is a \emph{randomized} policy.} by a randomization parameter $\gamma\in[0,1]$ defined as follows.

\textsl{At time $t$:
\begin{itemize}
\item If $\boldsymbol{\zeta}(t)$ = $[1,1,1]$ or $[1,0,1]$, then $\mathbf{S}(t)$ = $[1,0,1]$.
\item Else, if $\boldsymbol{\zeta}(t)$ = $[1,1,0]$ or $[0,1,1]$, then 
\begin{enumerate}
\item $\mathbf{S}(t)=[1,0,1]$ w.p. $1-\gamma$ and
\item $\mathbf{S}(t)=[0,1,0]$ w.p. $\gamma$.
\end{enumerate}
\item Else, $\mathbf{S}(t)=\boldsymbol{\zeta}(t)$.
\end{itemize}
}
Clearly, this policy is a randomization between the two 3-queue MSM policies $\tilde{\pi}^{(3)}_{IQ}$ and $\pi^{(3)}_{OQ}$.
%
%
%
A comparison of the definitions of $\rho^{(3)}_{\gamma}$, $\tilde{\pi}^{(3)}_{IQ}$ and $\pi^{(3)}_{OQ}$ clearly shows that $\rho^{(3)}_{\gamma}$ essentially chooses $\tilde{\pi}^{(3)}_{IQ}$ w.p. $\gamma$ and $\pi^{(3)}_{OQ}$ w.p. $1-\gamma.$

\subsection{Analysis of $\rho^{(3)}_{\gamma}$}\label{Chapter03SecAnalysisOfPi3_Gamma}
\begin{prop}\label{propPie_GammaUnstableOver0To0Point5}
$\rho^{(3)}_{\gamma}$ is unstable for $\gamma\in[0,0.5).$
\end{prop}

\begin{pf}
The basic idea for proving instability is the same as the one we used to prove that $\pi^{(3)}_{OQ}$ is unstable (Prop.~\ref{thmPi3_4IsNotTO}). For every $\gamma\in[0, 0.5)$ we show the existence of an arrival rate vector sufficiently close to the boundary of the capacity region such that the policy is not able to stabilize it. 

Towards that end, consider the processes embedded at instant $t$ (we will drop the time index, i.e. $t$, for simplicity of notation)
\small
\begin{eqnarray*}
P\left\lbrace S_2=1\right\rbrace &=& P\left\lbrace S_2=1\bigg|Q_1+Q_3=0\right\rbrace P\left\lbrace Q_1+Q_3=0\right\rbrace\\
& +& P\left\lbrace S_2=1\bigg|Q_1>0,Q_3=0\right\rbrace P\left\lbrace Q_1>0,Q_3=0\right\rbrace \nonumber\\
&+& P\left\lbrace S_2=1\bigg|Q_1=0,Q_3>0\right\rbrace P\left\lbrace Q_1=0,Q_3>0\right\rbrace\\
& +& P\left\lbrace S_2=1\bigg|Q_1>0,Q_3>0\right\rbrace P\left\lbrace Q_1>0,Q_3>0\right\rbrace \nonumber\\
&=& 1\cdot P\left\lbrace Q_1+Q_3=0\right\rbrace + \gamma\cdot P\left\lbrace Q_1>0,Q_3=0\right\rbrace + \gamma\cdot P\left\lbrace Q_1=0,Q_3>0\right\rbrace\\
& + & 0\cdot P\left\lbrace Q_1>0,Q_3>0\right\rbrace \nonumber\\
&=& \gamma \left(1-P\left\lbrace Q_1>0,Q_3>0\right\rbrace\right) + (1-\gamma) P\left\lbrace Q_1=0,Q_3=0\right\rbrace. \nonumber
\end{eqnarray*}
\normalsize
Now, notice that $\lbrace A_1>0,A_3>0\rbrace\subset \lbrace Q_1>0,Q_3>0\rbrace$ which means that $P\lbrace A_1>0,A_3>0\rbrace=\lambda_1\lambda_3\leq P\lbrace Q_1>0,Q_3>0\rbrace$. Similarly, $\left(1-\lambda_1\right)\left(1-\lambda_3\right)\geq P\lbrace Q_1=0,Q_3=0\rbrace$. Using this in the above equation, we get
\begin{eqnarray}
P\left\lbrace S_2=1\right\rbrace &\leq& \gamma\cdot \left(1-\lambda_1\lambda_3\right)+(1-\gamma)\cdot \left(1-\lambda_1\right)\left(1-\lambda_3\right)\nonumber\\
&=& 1-(1-\gamma)(\lambda_1+\lambda_3)+(1-2\gamma)\lambda_1\lambda_3.\nonumber
\end{eqnarray}
Consider the arrival rate vector $\boldsymbol{\lambda}=\lambda\cdot[1,1,1]$, where, obviously, $\lambda<0.5$. For this vector,
\begin{eqnarray}
P\left\lbrace S_2=1\right\rbrace &\leq& (1-2\gamma)\lambda^2-2(1-\gamma)\lambda+1.\nonumber
\end{eqnarray}
Therefore, if we can show the existence of some $\lambda\in[0,0.5)$ such that $(1-2\gamma)\lambda^2-2(1-\gamma)\lambda+1<\lambda,$ we will be done. Consider the polynomial $p(\lambda):=(1-2\gamma)\lambda^2-(3-2\gamma)\lambda+1$. $p(0)=1>0,$ but
\begin{eqnarray}
p(0.5) &=& \frac{1-2\gamma}{4}-\frac{3-2\gamma}{2}+1\nonumber\\
&=& -\frac{5}{4}+\frac{\gamma}{2}+1\nonumber\\
&=& -\frac{1}{4}+\frac{\gamma}{2}\stackrel{*1}{<}0,
\end{eqnarray}
where, in $*1$ we have used the fact that $\gamma\in[0,0.5)$. By continuity, therefore, there exists some $\lambda\in[0,0.5)$ for every  such $\gamma$, such that Queue 2 cannot be stabilized.
\end{pf}

Now, consider the set of arrival rates
\small
\begin{eqnarray}
\Lambda^{(3)}_\gamma:=\left\lbrace\boldsymbol{\lambda}\in\mathbb{R}^3_+\bigg|\lambda_1+\lambda_2<\gamma,\lambda_2+\lambda_3<\gamma\right\rbrace
\label{Chapter03EqnCapacityRegionOfRho3Gamma}
\end{eqnarray}
\normalsize
\begin{figure}[tb]
\centering
\includegraphics[height=11.50cm, width=8.00cm]{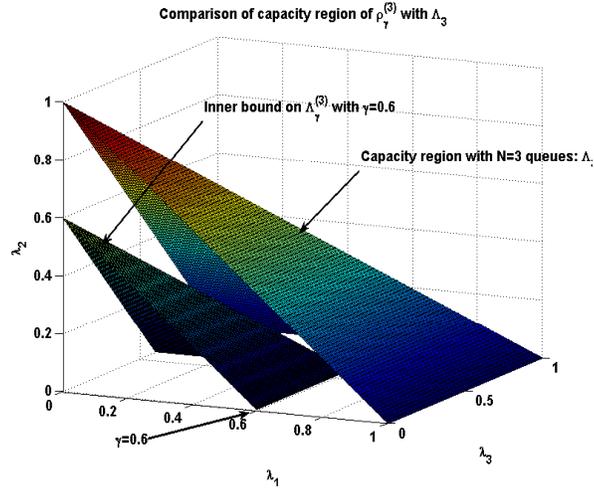}
\caption{Figure showing the inner bound on the stability region of $\rho^{(3)}_{\gamma}$ given in \eqref{Chapter03EqnCapacityRegionOfRho3Gamma}. The region below the outer surface, towards the origin, in the positive orthant is the capacity region $\Lambda_3$.} 
\label{Chapter03FigCapacityRegionOfRhoGamma}
\end{figure} 

\begin{prop}\label{Chapter03PropStabilityRegionOfRho3GammaExpandsToFullCapacityRegion}
For every $\gamma\in(0,1]$, $\rho^{(3)}_{\gamma}$ stabilizes all rate vectors in $\Lambda^{(3)}_\gamma$.
\end{prop}
\begin{rem}
The above result means that the stability region of $\rho^{(3)}_{\gamma}\nearrow\Lambda_3$ as $\gamma\uparrow1$, i.e., for all $0<\gamma_1\leq\gamma_2\leq1$, $\Lambda^{(3)}_{\gamma_1}\subseteq\Lambda^{(3)}_{\gamma_2}.$
This is consistent with the fact that $\rho^{(3)}_{\gamma}\bigg|_{\gamma=1}\equiv\tilde{\pi}^{(3)}_{IQ}$.
\end{rem}
\begin{pf}
Recall that $D_i(t)$ is the actual number of packets leaving Queue $i$ at the end of time slot $t$. We first focus our attention on the pair of queues 1 and 2. Now, from the definition of $\rho^{(3)}_{\gamma}$ we see that this pair receives service whenever 
\begin{itemize}
\item $\zeta_1(t)=1$, i.e., when $\boldsymbol{\zeta}(t)$ is $[1,0,0],[1,0,1],[1,1,0]\text{ or }[1,1,1]$, \emph{w.p.~1,}
\item $\boldsymbol{\zeta}(t)=[0,1,1]$, \emph{w.p.~$\gamma$,} and
\item $\boldsymbol{\zeta}(t)=[0,1,0]$, \emph{w.p.~1.}
\end{itemize}
Once again, we drop the time index for simplicity of notation. From the above discussion, we see that 
\small
\begin{eqnarray*}
\mathbb{E}\left[D_1+D_2\bigg|\mathbf{Q}\right] &=& 1\cdot \mathbb{I}_{\lbrace Q_1>0\rbrace} + \gamma\cdot \mathbb{I}_{\lbrace Q_1=0, Q_2>0, Q_3>0\rbrace} \nonumber\\
&& + 1\cdot \mathbb{I}_{\lbrace Q_1=0, Q_2>0, Q_3=0\rbrace} \nonumber\\
&=& (\gamma+1-\gamma)\cdot \left(\mathbb{I}_{\lbrace Q_1>0,Q_2>0\rbrace}+\mathbb{I}_{\lbrace Q_1>0,Q_2=0\rbrace}\right) \\
&& + \gamma\cdot \mathbb{I}_{\lbrace Q_1=0, Q_2>0, Q_3>0\rbrace} \\
&& + (\gamma+1-\gamma)\cdot \mathbb{I}_{\lbrace Q_1=0, Q_2>0, Q_3=0\rbrace} \nonumber\\
&=& \gamma\left[\mathbb{I}_{\lbrace Q_1>0, Q_2>0\rbrace} + \mathbb{I}_{\lbrace Q_1>0, Q_2=0\rbrace} + \mathbb{I}_{\lbrace Q_1=0, Q_2>0 \rbrace}\right]\\
&& +(1-\gamma)\left[\mathbb{I}_{\lbrace Q_1>0\rbrace}+\mathbb{I}_{\lbrace Q_1=0, Q_2>0, Q_3=0\rbrace}\right] \nonumber\\
&=& \gamma\cdot\mathbb{I}_{\lbrace Q_1+Q_2>0\rbrace}+(1-\gamma)\left[\mathbb{I}_{\lbrace Q_1>0\rbrace}+\mathbb{I}_{\lbrace Q_1=0, Q_2>0, Q_3=0\rbrace}\right] \nonumber\\
&\geq& \gamma\cdot\mathbb{I}_{\lbrace Q_1+Q_2>0\rbrace}.
\end{eqnarray*}
\normalsize
Using the same procedure as above, we also get $\mathbb{E}\left[D_2+D_3\bigg|\mathbf{Q}\right]\geq\gamma\cdot\mathbb{I}_{\lbrace Q_2+Q_3>0\rbrace}.$ Now, we are in a position to invoke a $\gamma$ randomized version of our Property $\mathcal{P}$, as follows.
Let $L(\mathbf{Q}(t)):=\left(Q_1(t)+Q_2(t)\right)^2+\left(Q_2(t)+Q_3(t)\right)^2$ be the Lyapunov function. We focus on the first term only; the analysis of the second term follows the same procedure.
\small
\begin{eqnarray*}
&&\left(Q_1(t+1)+Q_2(t+1)\right)^2-\left(Q_1(t)+Q_2(t)\right)^2 \\
&=& \left(Q_1(t)-D_1(t)+A_1(t+1)+Q_2(t)-D_2(t)\right.\\
&& +\left. A_2(t+1)\right)^2-\left(Q_1(t)+Q_2(t)\right)^2 \nonumber\\
&\stackrel{*2}{\leq}& 1+4-2(Q_1(t)+Q_2(t))\left(D_1(t)+D_2(t)\right.\\
&&\left. -A_1(t+1)-A_2(t+1)\right)
\end{eqnarray*}
\normalsize
In $*2,$ we have used several facts. For any 3 non negative reals, $x,y,z$, $(x-y+z)^2\leq x^2+y^2+z^2-2x(y-z).$ Furthermore, $D_i(t+1)\leq1,~\forall t$ and $(A_1(t+1)+A_2(t+1))^2\leq 4,~\forall t.$ So,
\small
\begin{eqnarray*}
&\mathbb{E}&\left[\left(Q_1(t+1)+Q_2(t+1)\right)^2-\left(Q_1(t)+Q_2(t)\right)^2\bigg|\mathbf{Q}(t)=\mathbf{Q}\right]\\ 
&{\leq}& 5-2(Q_1(t)+Q_2(t))\left(\mathbb{E}\left[D_1(t)+D_2(t)\bigg|\mathbf{Q}\right] - \left(\lambda_1+\lambda_2\right)\right) \nonumber\\
&\stackrel{*3}{\leq}& 5-2(Q_1(t)+Q_2(t))\left(\gamma\cdot\mathbb{I}_{\lbrace Q_1+Q_2>0\rbrace}- \left(\lambda_1+\lambda_2\right)\right). \nonumber 
\end{eqnarray*}
We now use the fact that for any non negative random variable $X$, $\mathbb{E}X\mathbb{I}_{\lbrace X>0\rbrace}=\mathbb{E}X.$
\small
\begin{eqnarray}
&&\mathbb{E}\left[\left(Q_1(t+1)+Q_2(t+1)\right)^2-\left(Q_1(t)+Q_2(t)\right)^2\right] \nonumber\\
&{\leq}& 5-2\mathbb{E}(Q_1(t)+Q_2(t))\left(\gamma-\left(\lambda_1+\lambda_2\right)\right). \nonumber\\
&\stackrel{*4}{\leq}& 5-2\epsilon_{1,2}\mathbb{E}(Q_1(t)+Q_2(t))\nonumber,\nonumber
\end{eqnarray}
\normalsize
where in $*4$ we have used the fact that $\boldsymbol{\lambda}\in\Lambda^{(3)}_{\gamma}$ to conclude that $\left(\gamma-\left(\lambda_1+\lambda_2\right)\right)=:\epsilon_{1,2}>0.$ The proof of strong stability follows by showing the existence of an $\epsilon_{2,3}$ for the queues 2 and 3 and putting together both these upper bounds to get an upper bound on the Lyapunov function and using the \emph{telescoping sum} argument.

\end{pf}


\section{Path Graph Conflict Models with $N > 3$:\\Policy Splicing for Throughput Optimal QNB Scheduling}\label{Chapter04SecQLenAgnosticTOSplicing}

 The previous section was devoted to introducing the reader to the idea of scheduling policies that rely only on occupancy information (empty-nonempty status of queues) and examining the behavior of such policies on a small network. We will use the knowledge gained therein to now propose such policies for larger systems while still confining ourselves to path-graph interference networks.

The path we shall follow uses a \enquote{policy splicing} technique to construct MSM policies for large systems by splicing together MSM policies for smaller systems. We first give a high-level overview of the technique. Recall the \enquote{Top-Down} and \enquote{Bottom-Up} policies, $\pi^{(3)}_{TD}$ and $\pi^{(3)}_{BU}$, discussed in Sec.~\ref{Chapter03SecAnalysisOfPi3_TDPi3_BU}. For a general path-graph network with $N$ queues (communication links), the \enquote{Top-Down} policy, $\pi^{(N)}_{TD}$, which maps an occupancy vector $\boldsymbol{\zeta}(t)$ to an activation vector $\mathbf{s}(t)$, is defined as follows. Before defining the policy, we assume the presence of two \emph{virtual queues}, Queue $0$ and Queue $N+1$, with $Q_0(t)=Q_{N+1}(t)=s_0(t)=s_{N+1}(t)=0,~\forall t\geq0.$ This is just to facilitate compact writing of the policy. These virtual queues do not play any actual role in the system. Recall that if $\mathbf{Q}(t)=[Q_1(t),\cdots,Q_N(t)]$ is the queue length vector at time $t$, then the \textbf{occupancy vector} at time $t$ is defined by $\boldsymbol{\zeta}(t)=[\mathbb{I}_{\left\lbrace  Q_1(t)>0\right\rbrace},\cdots,\mathbb{I}_{\left\lbrace  Q_N(t)>0\right\rbrace}]$.

\textsl{
At time $t$
\begin{enumerate}
\item For j=1:N\label{Chapter04SecQLenAgnosticTOSplicingGenericTDPolicyEnumeration}
\begin{enumerate}
\item\label{Chapter04TDPolicyCondnNonEmptyAndScheduled} If $\zeta_j(t)=1$ and $s_{j-1}(t)=0$, then $s_j(t)=1$ and $s_{j+1}(t)=0$.
\item\label{Chapter04TDPolicyCondnNonEmptyButNotScheduled} Else if $\zeta_j(t)=1$ and $s_{j-1}(t)=1$, then $s_j(t)=0$.
\item\label{Chapter04TDPolicyCondnEmpty} Else if $\zeta_j(t)=0$, then $s_j(t)=0$.
\end{enumerate}
\end{enumerate}
}
It is easy to see that this produces $\pi^{(3)}_{TD}$ for $N=3$, and $\pi^{(N)}_{BU}$ is defined similarly. The following important property follows immediately. 
\begin{prop}\label{Chapter04SecQLenAgnosticTOSplicingPropGenericTDPolicyIsTO}
$\pi^{(N)}_{TD}$ and $\pi^{(N)}_{BU}$ are MSM for all $N\in\mathbb{N}$.
\end{prop}

\begin{pf}
Lem.~\ref{Chapter03lemSufficientForActivationMSM} gives a sufficient condition for an activation vector to be of maximum size. We will now show that in every slot $t,$ the activation vector $\mathbf{s}(t)$ that $\pi^{(N)}_{TD}$ produces satisfies this condition, thereby establishing that the policy is MSM.

Recall that in Lem.~\ref{Chapter03lemSufficientForActivationMSM}, given an occupancy vector $\boldsymbol{\zeta},$ we defined $k=k(\boldsymbol{\zeta})$ to be the twice the number of runs of nonempty queues. Also, $j_1=j_1(\boldsymbol{\zeta}),\dots,j_k= j_k(\boldsymbol{\zeta})$, were defined to be the nonempty queues that mark the beginnings ($j_{\{odd~subscript\}}$) and ends ($j_{\{even~subscript\}}$) of the nonempty runs, or are the two outermost queues. We show that Conditions \ref{Chapter03CondnOddNonEmptyRuns} and \ref{Chapter03CondnEvenNonEmptyRuns} in the Lemma are both satisfied in every time slot, by the activation vector produced by $\pi^{(N)}_{TD}$.
\begin{enumerate}
\item When $j_{2m}-j_{2m-1}$ is even, i.e., we have an \emph{odd} length run of nonempty queues: By definition of the indices, this means that Queue~$\left(j_{2m-1}\right)-1$ is empty since a run of nonempty queues begins with $j_{2m-1}$, which from Condition~\ref{Chapter04TDPolicyCondnEmpty} in the definition of $\pi^{(N)}_{TD}$ ensures that $s_{j_{2m-1}-1}(t)=0$, which means that $s_{j_{2m-1}-1}(t)=1$ from Condition~\ref{Chapter04TDPolicyCondnNonEmptyAndScheduled}. Thereafter, since all queues between Queues $j_{2m-1}$ and $j_{2m}$ (including these two) are nonempty, the policy alternates between Conditions \ref{Chapter04TDPolicyCondnNonEmptyAndScheduled} and \ref{Chapter04TDPolicyCondnNonEmptyButNotScheduled}, scheduling every alternate queue and thus satisfying Condition~\ref{Chapter03CondnOddNonEmptyRuns} in Lem.~\ref{Chapter03lemSufficientForActivationMSM}.
\item When $j_{2m}-j_{2m-1}$ is odd, i.e., we have an \emph{even} length run of nonempty queues: Once again $\pi^{(N)}_{TD}$ schedules every alternate queue within this run starting with Queue~~$j_{2m-1}$, and in the process, satisfies Condition~\ref{Chapter03CondnEvenNonEmptyRunsStartWithJ2mMinus1} in Lem.~\ref{Chapter03lemSufficientForActivationMSM}.
\end{enumerate}
Since this holds true in every slot, the policy is MSM. 
\end{pf}

Before we venture into proving the throughput optimality of $\pi^{(N)}_{TD}$ and $\pi^{(N)}_{BU}$, we use these two policies to describe the policy splicing process. Consider a system of $2N-1$ queues, $N\geq1.$ In Algorithm~\ref{Chapter04SecQLenAgnosticTOSplicedPolicyEnumeration}, we splice the TD and BU policies and construct a scheduling policy\footnote{The subscript \enquote{SP} refers to the fact that this is a \emph{S}pliced \emph{P}olicy.} $\pi^{(2N-1)}_{SP}$ on this system. 
\textbf{Note:} We assume the presence of the two \emph{virtual queues} Queue~0 and Queue~$2N$ here as well.
%
\begin{algorithm}[bth]
 \KwData{Binary occupancy vector $\boldsymbol{\zeta}(t)$}
 \KwResult{Queue activation vector $\mathbf{S}(t)$}
 {\bf Initialize:} $j=1$, time$=t$, $\mathbf{S}(t)=\mathbf{0}$\\
 \eIf{$\zeta_N(t)=1$}{$S_N(t)=1$, $S_{N-1}(t)=0$ and $S_{N+1}(t)=0$}
 { 
	\For{\color{blue}$j=N-1:1$}
 	{
		\eIf{$\zeta_j(t)=1$ and $S_{j+1}(t)=0$}{$S_j(t)=1$}
		{
			\eIf{$\zeta_j(t)=1$ and $S_{j+1}(t)=1$}{$S_j(t)=0$}
			{
				\If{$\zeta_j(t)=0$}{$S_j(t)=0$}
			}
		}
	 }
	\For{\color{blue}$j=N+1:2N-1$}
 	{
		\eIf{$\zeta_j(t)=1$ and $S_{j-1}(t)=0$}{$S_j(t)=1$}
		{
			\eIf{$\zeta_j(t)=1$ and $S_{j-1}(t)=1$}{$S_j(t)=0$}
			{
				\If{$\zeta_j(t)=0$}{$S_j(t)=0$}
			}
		}
	 }
 }
 \caption{\footnotesize The spliced policy $\pi^{(2N-1)}_{SP}.$ The loop corresponding to {\color{blue}$j=N-1:1$} induces $\pi^{(N)}_{BU}$ on Queues~$N$ through $1$, while the latter loop induces $\pi^{(N)}_{TD}$ on Queues~$N$ through $2N,$ as depicted in Fig.~\ref{Chapter04FigTheGeneralSplicingProcessToGiveNonMSMPolicy}.}
\label{Chapter04SecQLenAgnosticTOSplicedPolicyEnumeration}
\end{algorithm}
Before proceeding to analyse this policy, we first need to make sure it really is a well-defined policy, i.e., it provides a valid activation vector for each of the $2^{2N-1}$ possible occupancy vectors.
\begin{lem}\label{Chapter04SecQLenAgnosticTOSplicingThmGenericSplicedPolicyIsAdmissible}
$\pi^{(2N-1)}_{SP},$ as defined above, is well-defined.
\end{lem}
\begin{pf}
See Sec.~\ref{Chapter04SecQLenAgnosticTOSplicingAppendixPfOfThmGenericSplicedPolicyIsAdmissible} in the Appendix. 
\end{pf}

A quick comparison with the definitions of $\pi^{(N)}_{TD}$ and $\pi^{(N)}_{BU}$ shows that $\pi^{(2N-1)}_{SP}$ induces the former two  policies on the subsets $\left\lbrace N,N+1,\cdots,2N-1\right\rbrace$ and $\left\lbrace 1,2,\cdots,N\right\rbrace$ respectively.
The following result follows from the definition of the splicing process. Recall from Sec.~\ref{secSystemModelNWirelessLinksPathGraph} that the capacity region of a path-graph interference network consisting of $N$ queues is defined by 
\begin{eqnarray}\label{Chapter04SecQLenAgnosticTOSplicingEqnCapacityPathGraphs}
\Lambda_N:=\left\lbrace \lambda\in\mathbb{R}_+^N \bigg| \lambda_i+\lambda_{i+1}<1,~1\leq i\leq N-1 \right\rbrace, N\in\mathbb{N}.
\end{eqnarray}

\begin{thm}\label{Chapter04SecQLenAgnosticTOSplicingThmGenericSplicedPolicyIsTO}
For every $N\in\mathbb{N}$, such that $\pi^{(N)}_{TD}$ and $\pi^{(N)}_{BU}$ are throughput optimal over $\Lambda_N$, $\pi^{(2N-1)}_{SP}$ is throughput optimal over $\Lambda_{2N-1}$.
\end{thm}

\begin{pf}
The policy $\pi^{(2N-1)}_{SP}$ is formed by splicing together $\pi^{(N)}_{TD}$ and $\pi^{(N)}_{BU}$. This means that $\pi^{(2N-1)}_{SP}$ restricted to Queues $1$ to $N$ is $\pi^{(N)}_{BU}$ and restricted to Queues $N$ to $2N-1$ is $\pi^{(N)}_{TD}$. The throughput optimality of $\pi^{(N)}_{TD}$ and $\pi^{(N)}_{BU}$ means that for every $\boldsymbol{\lambda}\in\Lambda_N$, 
\begin{eqnarray}
\limsup_{T\rightarrow\infty}\frac{1}{T}\sum_{t=0}^{t-1}\sum_{i=N}^{2N-1}\mathbb{E}_{\pi^{(N)}_{TD}} Q_i(t)&<&\infty,\text{ and }\label{eqnTDPartOfSplicedIsTO}
\\
\limsup_{T\rightarrow\infty}\frac{1}{T}\sum_{t=0}^{t-1}\sum_{i=1}^{N}\mathbb{E}_{\pi^{(N)}_{BU}} Q_i(t)&<&\infty.\label{eqnBUPartOfSplicedIsTO}
\end{eqnarray}
Notice in particular the indices of the inner summations in the above inequalities. Now, for every $\boldsymbol{\lambda}\in\Lambda_{2N-1}$ define $\boldsymbol{\lambda}_{1:N}=[\lambda_1,\cdots,\lambda_N]$ and $\boldsymbol{\lambda}_{N:2N-1}=[\lambda_N,\cdots,\lambda_{2N-1}]$ and notice that by the definition of ${\Lambda}_{N}$, $\boldsymbol{\lambda}_{1:N}\in\Lambda_{N}$ and $\boldsymbol{\lambda}_{N:2N-1}\in\Lambda_{N},$ which means that \eqref{eqnTDPartOfSplicedIsTO} and \eqref{eqnBUPartOfSplicedIsTO} are still separately true. The proof concludes when we observe that 
\begin{eqnarray*}
&&\limsup_{T\rightarrow\infty}\frac{1}{T}\sum_{t=0}^{t-1}\sum_{i=1}^{2N-1}\mathbb{E}_{\pi^{(2N-1)}_{SP}} Q_i(t)\\
&\leq& \limsup_{T\rightarrow\infty}\frac{1}{T}\sum_{t=0}^{t-1}\sum_{i=N}^{2N-1}\mathbb{E}_{\pi^{(N)}_{TD}} Q_i(t) \\
&&+ \limsup_{T\rightarrow\infty}\frac{1}{T}\sum_{t=0}^{t-1}\sum_{i=1}^{N}\mathbb{E}_{\pi^{(N)}_{BU}} Q_i(t)\\
&<&\infty
\end{eqnarray*}
\end{pf}

\begin{rem}
It is important to note that although $\pi^{(N)}_{TD}$ and $\pi^{(N)}_{BU}$ are MSM, $\pi^{(2N-1)}_{SP}$ is \emph{not.} For example, consider an occupancy vector such that $\zeta_{N-1}(t)=\zeta_{N}(t)=\zeta_{N+1}(t)=1$ and $\zeta_j(t)=0,~\forall j\in\{1,\cdots,N-2\}\cup\{N+2,\cdots,2N-1\}$, i.e., the central queue and both adjacent queues are nonempty and all other queues are empty. Any MSM policy would produce the activation vector with $s_{N-1}(t)=s_{N+1}(t)=1$ and $s_j(t)=0,~\forall j\notin \{N-1,N+1\}$, whereas $\pi^{(2N-1)}_{SP}$ produces the activation vector with $s_N(t)=1$ and $s_j(t)=0,~\forall j\neq N$ thereby scheduling one less queue for transmission.
\end{rem}

\begin{figure}[tb]
\centering
\includegraphics[height=6.05cm, width=10.0cm]{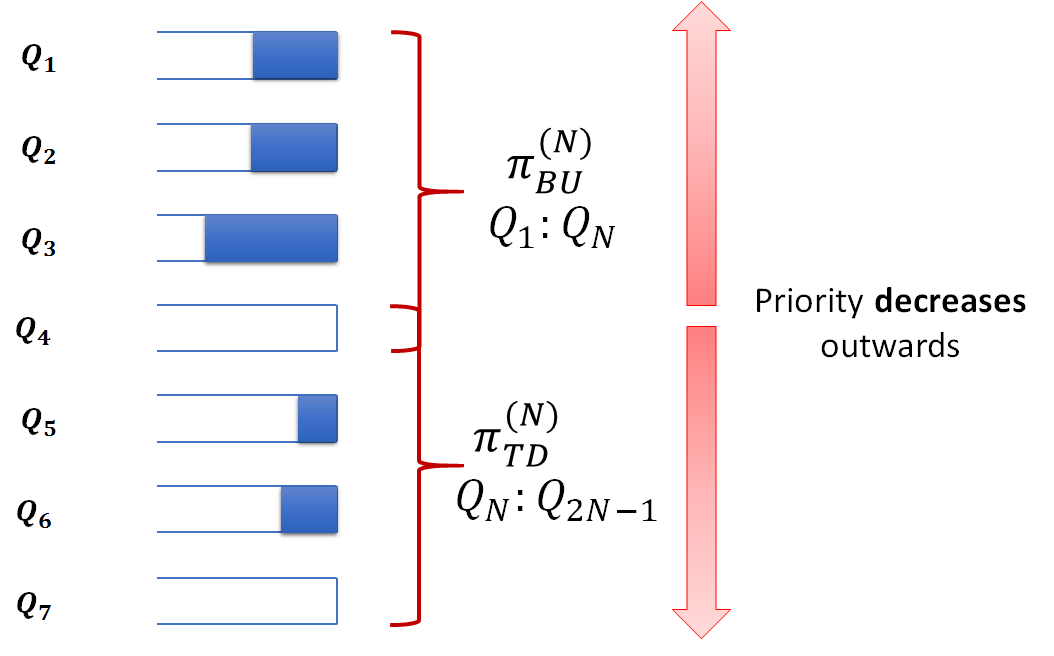}
\caption{Illustrating the manner in which the policies $\pi^{(N)}_{TD}$ and $\pi^{(N)}_{BU}$ are spliced together to form the \emph{non-MSM} policy $\pi^{(2N-1)}_{SP}$. Note that the splicing is consistent in that even though the two sub-policies schedule over overlapping sections of queues, their decisions do not contradict each other.}
\label{Chapter04FigTheGeneralSplicingProcessToGiveNonMSMPolicy}
\end{figure}

\begin{figure}[bt]
\centering
\includegraphics[height=4.15cm, width=6.50cm]{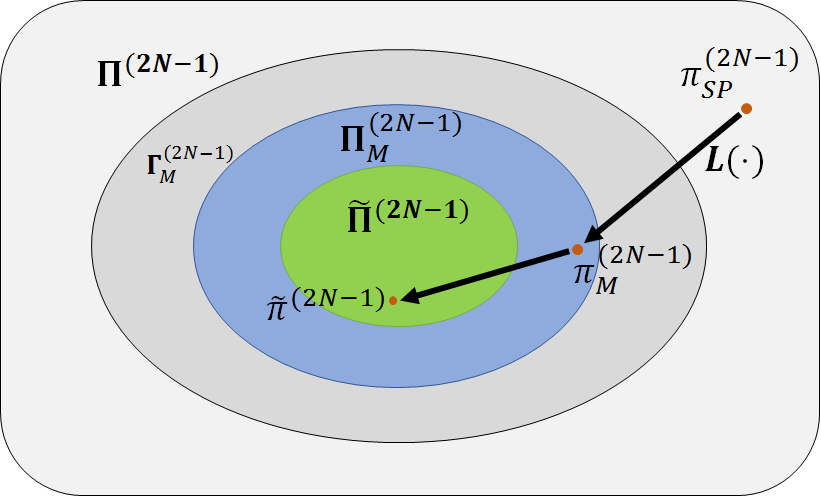}
\caption{Illustrating steps~\ref{Chapter04GeneralPSplicingProject} and \ref{Chapter04GeneralPSplicingPrioritizeInnerQueues} of the general policy splicing process. The queue nonemptiness-based, non-MSM policy $\pi^{(2N-1)}_{SP}$ is first projected into $\Pi^{(N)}_M$ to get the \emph{MSM} policy $\pi^{(2N-1)}_{M}\equiv L\left(\pi^{(2N-1)}_{SP}\right)$. Thereafter, $\pi^{(2N-1)}_{M}$ is modified to prioritize inner queues to get $\tilde{\pi}^{(2N-1)}\in\tilde{\Pi}^{(2N-1)}$.}
\label{Chapter04FigTheGeneralSplicingProcessStepsProjectionAndInnerQueuePriority}
\end{figure}

To reduce delay, one needs to extract an MSM policy from this spliced policy. Fortunately, the procedure to accomplish this has already been described in Sec.~\ref{Chapter03secProjectionInPolicySpace}. We simply project the policy onto the space $\Pi^{(N)}_M$ of MSM policies using the projection operator defined therein. Thereafter, we use some observations based on Condition~\ref{Chapter03conditionMSMDelaySmall} in Lem.~\ref{Chapter03lemSufficientForActivationMSM} to improve the delay performance of this projected MSM policy by finally obtaining a policy in $\tilde{\Pi}^{(N)}$. Figures~\ref{Chapter04FigTheGeneralSplicingProcessToGiveNonMSMPolicy} and \ref{Chapter04FigTheGeneralSplicingProcessStepsProjectionAndInnerQueuePriority} give a pictorial description of the entire process.

\begin{rem}
Another important observation is that $\pi^{(2N-1)}_{SP}$ is also a stabilizing policy for a system with $2N-k$ queues, with $1\leq k\leq 2N.$ All one needs is to begin with a system of $2N-k$ queues and append $k$ virtual queues that start out empty and receive no arrivals in any time slot and run $\pi^{(2N-1)}_{SP}$ on them. So, the focus of the remainder of this section is proving the throughput optimality of the Top-Down and Bottom-Up priority policies.
\end{rem}
 
To summarise, the general policy splicing process involves the following steps(see Figures~\ref{Chapter04FigTheGeneralSplicingProcessToGiveNonMSMPolicy} and  \ref{Chapter04FigTheGeneralSplicingProcessStepsProjectionAndInnerQueuePriority})
\paragraph{\bf General splicing procedure}\label{Chapter04GeneralPSplicingItemized}
\begin{enumerate}
\item\label{Chapter04GeneralPSplicingProposeTDAndBU} Proposing TD and BU policies for the $N$-queue system,
\item\label{Chapter04GeneralPSplicingSplice} Splicing them together to produce a non-MSM policy for the $(2N-1)$-queue system,
\item\label{Chapter04GeneralPSplicingProject} Projecting the spliced policy to get an MSM policy
\item\label{Chapter04GeneralPSplicingPrioritizeInnerQueues} Modifying the resulting policy to break all ties in favor of inner queues to get a policy with better delay performance than the MSM policy.
\end{enumerate}
We have already proposed and analysed two priority policies for 3-queue path graph networks that we named $\pi^{(3)}_{TD}$ and $\pi^{(3)}_{BU}$. As a quick illustration of the above procedure, we show how low delay policies for $2\times 3-1=5$-queue systems can be constructed from these two 3-queue policies.

\subsection{Low-delay Scheduling Policies for Systems with $N=5$ queues}\label{Chapter04sec5Queues}
Our goal is now to begin with the two 3-queue scheduling policies $\pi^{(3)}_{TD}$ and $\pi^{(3)}_{BU}$ and use the general policy splicing procedure \ref{Chapter04GeneralPSplicingItemized} to design a throughput optimal 5-queue scheduling policy $\tilde{\pi}^{(5)}$, in $\tilde{\Pi}^{(5)}$. The purpose of focusing on $\tilde{\Pi}^{(5)}$ is, as stated before, to ensure that the policy (if one such exists) will show desirable delay properties. 
We will begin with a non delay optimal policy, which we call ${\pi}^{(5)}_M$, since it is MSM. The stability of ${\pi}^{(5)}_M$ will be proved using a non MSM policy, $\pi^{(5)}_{SP}$, and finally, we will show that $\tilde{\pi}^{(5)}$ is throughput optimal by invoking certain properties of policies in $\tilde{\Pi}^{(5)}$. 
\subsubsection{Analysis of $\tilde{\pi}^{(5)}$}\label{Chapter04secPiTilde5}
The spliced policy $\pi_{SP}^{ (5)}$ , is defined as in the definition of the general spliced policy (Defn.~\ref{Chapter04SecQLenAgnosticTOSplicedPolicyEnumeration}) and is throughput optimal as shown in Thm.~\ref{Chapter04SecQLenAgnosticTOSplicingThmGenericSplicedPolicyIsTO}, since $\pi^{(3)}_{TD}$ and $\pi^{(3)}_{BU}$ have both been shown to be throughput optimal in Sec.~\ref{Chapter03SecAnalysisOfPi3_TDPi3_BU}. Since the projection operator $L$ preserves strong stability, the the MSM policy $\pi_{M}^{(5)}\equiv L(\pi_{SP}^{(5)})$, which lies in $\Pi_{M}^{(5)}$ , is also throughput optimal. The MSM policy $\pi_{M}^{(5)}$ is defined as follows

\textsl{At time $t:$
\begin{enumerate}
\item If $\boldsymbol{\zeta}(t)=[0,1,1,1,0],$ $\mathbf{s}(t)=[0,1,0,1,0]$
\item Else, if $Q_3(t)>0,$ $\mathbf{s}(t)=[1,0,1,0,1]$.
\item Else, if $Q_2(t)>0\text{ and }Q_4(t)>0,~\mathbf{s}(t)=[0,1,0,1,0]$.
\item Else, if $Q_2(t)>0\text{ or }Q_4(t)>0,$ 
\begin{enumerate}
\item $\mathbf{s}(t)=[0,1,0,0,1]$ if $Q_2(t)>0$ and
\item $\mathbf{s}(t)=[1,0,0,1,0]$ if $Q_4(t)>0$.
\end{enumerate}
\item Else, $\mathbf{s}(t)=[1,0,1,0,1]$.
\end{enumerate}
}
In an effort to improve the delay performance of the above policy, we modify ${\pi}^{(5)}_M$ to prioritise inner queues (here, Queues~2, 3 and 4). We accomplish this by using the conditions specified in Lem.~\ref{Chapter03lemSufficientForActivationMSM} that characterises MSM policies on path graph networks. The first two conditions by themselves form a necessary and sufficient condition for a policy to be MSM, and it is easy to show that ${\pi}^{(5)}_M$ satisfies both. To reduce delay, however, one can use Condition~\ref{Chapter03conditionMSMDelaySmall} therein that helps prioritise inner queues. Consider the definition of ${\pi}^{(5)}_M$ once more, and notice that under this policy,  
\begin{itemize}
\item $\boldsymbol{\zeta}(t)=[1,1,1,1,0]\mapsto\mathbf{s}(t)=[1,0,1,0,1]$, and 
\item $\boldsymbol{\zeta}(t)=[0,1,1,1,1]\mapsto\mathbf{s}(t)=[1,0,1,0,1]$.
\end{itemize}
Going back to the discussion on MSM policies, Condition~\ref{Chapter03conditionMSMDelaySmall} in Lem.~\ref{Chapter03lemSufficientForActivationMSM} states that when $j_1=1,$
\[
\hspace{-1.0cm}   {S}_{j}(t)=
\begin{cases}
    1~\text{if}~j_2-j\text{ is even, }j_1\leq j\leq j_2,\\
    0~\text{otherwise},
\end{cases}
\]
and similarly for the case with $j_2=N.$ This means when $\mathbf{s}(t)=[1,1,1,1,0]$, since $j_1=1,$ and $j_2=4,$ we really should have $\mathbf{s}(t)=[0,1,0,1,0]$. Similarly, when $\boldsymbol{\zeta}(t)=[0,1,1,1,1]$, $j_1=2$ and $j_2=5=N$, and using Condition~\ref{Chapter03conditionMSMDelaySmall} once again, we see that $\mathbf{s}(t)=[0,1,0,1,0]$. Including these two corrections in the definition of $\pi^{(5)}_M$ gives rise to the following policy, that we call $\tilde{\pi}^{(5)}$.

\textsl{At time $t:$
\begin{enumerate}
\item If $\boldsymbol{\zeta}(t)=[0,1,1,1,0],$ $\mathbf{s}(t)=[0,1,0,1,0]$.
\item If $\boldsymbol{\zeta}(t)=[1,1,1,1,0]\text{ or }[0,1,1,1,1],~\mathbf{s}(t)=[0,1,0,1,0]$. 
\item Else, if $Q_3(t)>0,$ $\mathbf{s}(t)=[1,0,1,0,1]$.
\item Else, if $Q_2(t)>0\text{ and }Q_4(t)>0,~\mathbf{s}(t)=[0,1,0,1,0]$.
\item Else, if $Q_2(t)>0\text{ or }Q_4(t)>0,$ 
\begin{enumerate}
\item $\mathbf{s}(t)=[0,1,0,0,1]$ if $Q_2(t)>0$ and
\item $\mathbf{s}(t)=[1,0,0,1,0]$ if $Q_4(t)>0$.
\end{enumerate}
\item Else, $\mathbf{s}(t)=[1,0,1,0,1]$.
\end{enumerate}
}
Clearly, $\tilde{\pi}^{(5)}$ satisfies all three requirements of Lem.~\ref{Chapter03lemSufficientForActivationMSM} and hence, resides in $\tilde{\Pi}^{(5)}$. As stated before. the objective is to show that $\tilde{\pi}^{(5)}\in\tilde{\Pi}^{(5)}$ gives rise to stochastically smaller sum queue lengths than $\pi^{(5)}_M\in\Pi^{(5)}_M.$ This, however, is quite obvious from the prior discussion in this section since  $\tilde{\pi}^{(5)}$ always satisfies Condition~\ref{Chapter03conditionMSMDelaySmall} in Lem.~\ref{Chapter03lemSufficientForActivationMSM}. Specifically, when $\boldsymbol{\zeta}(t)=[1,1,1,1,0]$ or $[0,1,1,1,1]$, $\pi^{(5)}_M$ chooses $\mathbf{s}(t)=[1,0,1,0,1]$, while $\tilde{\pi}^{(5)}$ chooses $\mathbf{s}(t)=[0,1,0,1,0]$. Using the same technique as in the proof of Prop.~\ref{propPoliciesInPi4TildeAreBetterThanPi4M}, we see that for any arrival rate $\boldsymbol{\lambda}\in\Lambda^o_5$,
\begin{eqnarray}
\sum_{i=1}^5Q^{\tilde{\pi}^{(5)}}_i(t)&\stackrel{st}{\leq}&\sum_{i=1}^5Q^{\pi^{(5)}_M}_i(t),~\forall t\geq0.\nonumber
\label{eqnStochasticOrderingPi5MAndPi5Tilde}
\end{eqnarray}
In fact, such a refinement of every policy in $\Pi^{(5)}_M\setminus\tilde{\Pi}^{(5)}$ can be obtained in a similar manner by enforcing Condition~\ref{Chapter03conditionMSMDelaySmall} in Lem.~\ref{Chapter03lemSufficientForActivationMSM} to reduce delay. With this, we see that. 
\begin{prop}\label{propPoliciesInPi5TildeAreBetterThanPi5M}
For every policy $\pi\in\Pi^{(5)}_M\setminus\tilde{\Pi}^{(5)},$ there exists a policy $\pi'\in\tilde{\Pi}^{(5)}$ such that
\begin{eqnarray}
\sum_{i=1}^5Q^{\pi'}_i(t)&\stackrel{st}{\leq}&\sum_{i=1}^5Q^{\pi}_i(t),~\forall t\geq0.
\label{eqnPoliciesInPi5TildeAreBetterThanPi5M}
\end{eqnarray}
\end{prop}
To summarize, in this section, we first proposed a general algorithm to generate MSM Top-Down and Bottom-Up priority policies for a system with any $N\in\mathbb{N}$ queues. We then showed how these policies can be combined to construct policies for larger systems and provided a sufficient condition for such a spliced policy to be throughput-optimal. We also provided an explicit example of the entire process using two of the policies already studied ($\pi^{(3)}_{TD}$ and $\pi^{(3)}_{BU}$). We now move on to proposing and analysing top-down and bottom-up priority policies for larger systems.
\section{Top-Down and Bottom-Up Policies for Systems with $N=4$ and $5$ queues}\label{Chapter04SecConstructAgnosticSplicingSubecTDandBUFor4Qs}
The Top-Down and Bottom-Up policies $\pi^{(4)}_{TD}$ and $\pi^{(4)}_{BU}$ will, as already discussed, be used to develop stabilizing policies for systems with $N=2\times4-1=7$ queues. An equivalent way to define the Top-Down policy $\pi^{(4)}_{TD}$ is as below. 

\textsl{At time $t$:
\begin{enumerate}
\item If $Q_1(t)>0$, 
\begin{enumerate}
\item If $Q_3(t)>0$, $\mathbf{s}(t) = (1,0,1,0)$.
\item Else\footnote{Strictly speaking, from the definition of $\pi^{(N)}_{TD}$ above, $s_4(t)$ should be set to $1$ iff $\zeta_4(t)=1$. But setting $s_4(t)=1$ when $\zeta_4(t) = 0$ doesn’t violate interference constraints since it means that Queue~4 is empty.}\label{footnoteLabel}, $\mathbf{s}(t) = (1,0,0,1)$.
\end{enumerate}
\item Else, if $Q_2(t) > 0$, $\mathbf{s}(t)=(0,1,0,1)$.
\item Else,
\begin{enumerate}
\item If $Q_3(t)>0$ $\mathbf{s}(t) = (1,0,1,0)$.
\item Else, $\mathbf{s}(t) = (1,0,0,1)$.
\end{enumerate}
\end{enumerate}
}

\begin{prop}\label{Chapter04SecAgnosticSplicingPropPi4_TDIsTO}
${\pi}^{(4)}_{TD}$ is throughput optimal.
\end{prop}
\begin{pf}
Notice that as far as the subsystem $\left[Q_1(t),Q_2(t),Q_3(t)\right]$ is concerned, this policy reduces to $\pi^{(3)}_{TD}$. That is, $\pi^{(4)}_{TD}$ restricted to the first three queues is $\pi^{(3)}_{TD}$. So, that subsystem is strongly stable. 
The remainder of the proof of the proposition can be found in Sec.~\ref{AppendixProofOfAgnosticSplicingPropPi4_TDIsTO} in the Appendix.
\end{pf}

It is easy to see that the Bottom-Up policy, $\pi^{(4)}_{BU}$, defined in a symmetric manner, giving highest priority to Queue 4 and lowest to Queue 1, is also throughput optimal using similar arguments. We then define $\pi^{(7)}_{SP}$ as in \ref{Chapter04SecQLenAgnosticTOSplicedPolicyEnumeration}.
%
Clearly, since $\pi^{(7)}_{SP}$ restricted to Queues 1, 2, 3, 4 is just $\pi^{(4)}_{BU}$ and restricted to Queues~4, 5, 6, 7 is $\pi^{(4)}_{TD},$ using the fact that both $\pi^{(4)}_{TD}$ and $\pi^{(4)}_{BU}$ are throughput optimal and  Thm.~\ref{Chapter04SecQLenAgnosticTOSplicingThmGenericSplicedPolicyIsTO}, we conclude that $\pi_{SP}^{(7)}$ is throughput optimal as well. However, since $\pi_{SP}^{(7)}$ is not MSM, some modifications are required to improve delay performance. This simply requires executing Steps~\ref{Chapter04GeneralPSplicingProject} and \ref{Chapter04GeneralPSplicingPrioritizeInnerQueues} in the general splicing procedure \ref{Chapter04GeneralPSplicingItemized}.


In a similar manner we will now show that the top-down and bottom-up policies for the 5 queue system ($\pi^{(5)}_{TD}$ and $\pi^{(5)}_{BU}$) are both throughput optimal, which will immediately yield a stabilizing policy ($\pi^{(9)}_{SP}$) for the 9-queue system. 

\begin{prop}\label{Chapter04SecAgnosticSplicingPropPi5_TDIsTO}
${\pi}^{(5)}_{TD}$ is throughput optimal.
\end{prop}
\begin{pf}
This analysis closely follows our analysis of $\pi^{(4)}_{TD}$. With $\pi^{(5)}_{TD}$ we only need to prove that Queue 5 receives \enquote{enough} service, since this policy restricted to the first 4 queues is just $\pi^{(4)}_{TD}$ which, as we have just shown, is throughput-optimal.
The remainder of the proof of the proposition can be found in Sec.~\ref{AppendixProofOfAgnosticSplicingPropPi5_TDIsTO} in the Appendix.
\end{pf}

The analysis of the bottom-up policy ($\pi^{(5)}_{BU}$) proceeds in a symmetric fashion. This means that the $9$-queue policy $\pi^{(9)}_{SP}$, that induces $\pi^{(5)}_{TD}$ on queues $1, 2, 3, 4,$ and $5$ and $\pi^{(5)}_{TD}$ on queues $5, 6, 7 , 8$ and $9$, is throughput-optimal.
%
%

\section{Path-Graph Conflict Models with $N>3$: Delay with QNB Policies}\label{chapter04SecDelayAnalysisAgnostic}
We now turn our attention to the vital aspect of delay. We have already proved, in Thm.~\ref{Chapter03thmPi33isDelayOptimal}, that for the system with $N=3$ queues, there exists a unique uniformly delay-optimal policy, that we named $\tilde{\pi}^{(3)}_{IQ}.$ The natural question to ask in this context is if one can find a delay optimal queue nonemptiness-based policy for larger systems as well. In this section, we will answer this question in the \emph{negative.} 

\textbf{General Flow of the Section:} We begin with a study of the class $\tilde{\Pi}^{(4)})$ of policies for systems with $N = 4$ queues. This class contains policies that are both MSM and prioritize inner queues (here, Queues~2 and 3), i.e, they satisfy all three conditions in Lem.~\ref{Chapter03lemSufficientForActivationMSM}. The number of policies in this class, $\mid\tilde{\Pi}^{(4)}\mid$, is 4 and we show that all four are throughput optimal. Thereafter, we show that for every policy in $\Pi^{(4)}_M$, the class of MSM queue nonemptiness-based policies, there exists a policy in $\tilde{\Pi}^{(4)}$ that shows better delay performance. However, we finally show that different policies in $\tilde{\Pi}^{(4)}$ perform better than the others (in terms of delay) over different portions of the capacity region $\Lambda_4$, which means that there is no \emph{uniformly delay optimal} (see glossary \ref{glossaryOfNotationForPart2}) policy for 4-queue systems. The general proof then follows from a contradiction argument.

\subsection{Characterizing the Class $\tilde{\Pi}^{(4)}$}\label{Chapter04SecDelayAnalysisPoliciesWith4Queues}
In Sec.~\ref{secTOSchedulingWith3Queues} we saw how the MSM property helped whittle down the number of queue nonemptiness-based policies in $\Pi^{(3)}$, making the final analysis of the class $\Pi^{(3)}_M$ tractable. Continuing along the same lines, with 4 queues, we have three activation vectors to choose from in any given slot, viz., $[1,0,1,0],$ $[1,0,0,1]$ and $[0,1,0,1],$ and $\boldsymbol{\zeta}(t)$ can take one of $2^4=16$ values. The number of queue noneptiness-based policies within $\Pi^{(4)}$ is, therefore, $3^{2^4}>43\times10^6.$ However, we must keep in mind that we will be dealing with MSM policies that serve the largest number of noninterfering queues in every slot. 

Observe Columns~1 and 2 in Table~\ref{Chapter04TablePoliciesInPi4TildeAreBetterThanPi4M}. The second gives all possible activation vectors an MSM policy can choose from for every occupancy vector $\boldsymbol{\zeta}(t)$ (the choice with $\zeta(t) = 0$ is inconsequential since it refers to when the system is empty). Most have a single activation vector associated with them meaning MSM policies don’t have any choice for such occupancies, but there are 5 vectors each with 2 valid activation vectors and 1 with 3 valid activation vectors, giving rise to $25 \times 3 = 96$ policies. So, the number of policies gets whittled down to $96$ policies once the MSM condition is imposed. But since this number is also inordinately large, we will restrict our study to $\tilde{\Pi}^{(4)}$, which contains 4 policies as shown in Column~3 in Table~\ref{Chapter04TablePoliciesInPi4TildeAreBetterThanPi4M}. We denote them by $\{\tilde{\pi}^{(4)}_i,1\leq i\leq4\}.$ 

In what follows, we provide a complete characterization of $\tilde{\Pi}^{(4)}.$ We first show that the policies $\tilde{\pi}^{(4)}_1$ and $\tilde{\pi}^{(4)}_2$ are stable using the policy splicing technique on some of the 3-queue policies whose stability we have already
established. Then, we prove the stability of the other two policies using an extension
of the stochastic ordering technique we used to show the delay optimality of $\tilde{\pi}^{(3)}_{IQ}$ in Thm.~\ref{Chapter03thmPi33isDelayOptimal}.

\subsubsection{Analysis of $\tilde{\pi}^{(4)}_1$ and $\tilde{\pi}^{(4)}_2$}\label{secPi4_1andPi4_2}
We define the policy $\tilde{\pi}^{(4)}_1$ as follows. 

\textsl{At time $t,$
\begin{enumerate}
\item If $\boldsymbol{\zeta}(t)=[1,1,1,0]$, $\mathbf{s}(t)=[1,0,1,0]$.
\item Else, if $Q_2(t)>0,$ $\mathbf{s}(t)=[0,1,0,1]$.
\item Else, if $Q_3(t)>0,~\mathbf{s}(t)=[1,0,1,0]$.
\item\label{condnPi41DefaultCondition} Else, $\mathbf{s}(t)=[1,0,0,1]$.
\end{enumerate}
}
The policy $\tilde{\pi}^{(4)}_2$ simply swaps the priorities of Queues $2$ and $3$. For clarity and completeness, we define $\tilde{\pi}^{(4)}_2$ explicitly below.

\textsl{At time $t,$
\begin{enumerate}
\item If $\boldsymbol{\zeta}(t)=[0,1,1,1]$, $\mathbf{s}(t)=[0,1,0,1]$.
\item Else, if $Q_3(t)>0,$ $\mathbf{s}(t)=[1,0,1,0]$.
\item Else, if $Q_2(t)>0,~\mathbf{s}(t)=[0,1,0,1]$.
\item\label{condnPi42DefaultCondition} Else, $\mathbf{s}(t)=[1,0,0,1]$.
\end{enumerate}
}

Table.~\ref{Chapter04tableAll4QueuePolicies} provides the complete enumeration of $\tilde{\pi}^{(4)}_1$, i.e., for every occupancy vector $\boldsymbol{\zeta}(t)$.
\begin{rem}
Firstly, the presence of a default, i.e., condition~\ref{condnPi41DefaultCondition}, ensures that the above definitions are exhaustive, in the sense that they cover all $2^4=16$ states of $\boldsymbol{\zeta}(t)$. This means, using the terminology in Sec.~\ref{Chapter04SecConstructAgnosticSplicingSubecTDandBUFor4Qs}, that both these policies are \emph{admissible.}
This will become important when we restate the same policies in a different manner while proving their stability.
\end{rem}
\begin{rem} We next note that while these policies have been proposed in \cite{ji-srikant10optimal-scheduling-small-switches}, only an informal argument regarding their stability properties has been provided therein, followed by a study of their performance in the Halfin Whitt regime. 
The informal argument therein asserts that the fraction of time for which Queue $2$ is nonempty equals its arrival rate $\lambda_2$, and this
claim is crucial to their stability argument. 

By Little’s Theorem applied to the HOL position, this assertion holds only if the mean waiting time in the HOL position of Queue 2 is exactly 1 slot. The actual fraction of time for which Queue $2$ is nonempty converges to $\lambda_2\cdot\mathbb{E}B_2$,
where $\mathbb{E}B_2$ is the mean service time of packets in Queue $2$. Since Queue $2$ is not always served whenever it is nonempty, $\mathbb{E}B_2>1$ (\emph{strict} inequality), so the fraction of time left to offer service to Queue $1$ is strictly smaller than $1-\lambda_2$. Moreover, as can be seen from the definition of $\tilde{\pi}^{(4)}_1$, $\mathbf{s}(t)$ is not necessarily $[1, 0, 1, 0]$ whenever Queues $1$ and $3$ are nonempty (take $\boldsymbol{\zeta}(t) = [1, 1, 1, 1]$, for instance). It is, therefore, unclear whether Queue $1$ is offered service often enough to stabilize it. We here provide a formal proof of the throughput optimality of $\tilde{\pi}^{(4)}_1$ and $\tilde{\pi}^{(4)}_2$.
\end{rem}
Hence, the fact that this policy is stabilizing requires an actual argument. We here provide a formal proof of the throughput optimality of $\tilde{\pi}^{(4)}_1\text{ and }\tilde{\pi}^{(4)}_2$. In the process, we derive another throughput optimal policy which is \emph{not} MSM and complete the proof using a stochastic dominance argument.

\begin{prop}
$\tilde{\pi}^{(4)}_1$ is throughput optimal.
\end{prop}

\begin{pf}
Before proceeding with the proof, notice that the definition of $\tilde{\pi}^{(4)}_1$ can be restated as follows. 

\textsl{At time $t:$
\begin{enumerate}
\item\label{condn4QueuesThatMakesThisMSM} If $\boldsymbol{\zeta}(t)=[1,1,1,0]$, $\mathbf{s}(t)=[1,0,1,0]$.
\item Else, 
\begin{enumerate}
\item\label{condn4QueuesOnlyFirst3} Check $[\zeta_1(t), \zeta_2(t), \zeta_3(t)]$
\begin{enumerate}
\item If $\zeta_2(t)=1,~\mathbf{s}(t)=[0,1,0,1].$
\item Else if $\zeta_3(t)=1,~\mathbf{s}(t)=[1,0,1,0].$
\item Else ${s}(t)=[1,0,0,1].$
\end{enumerate}
\item\label{condn4QueuesOnlyLast3} Check $[\zeta_2(t), \zeta_3(t), \zeta_4(t)]$
\begin{enumerate}
\item If $\zeta_2(t)=1,~\mathbf{s}(t)=[0,1,0,1].$
\item Else if $\zeta_3(t)=1,~\mathbf{s}(t)=[1,0,1,0].$
\item Else ${s}(t)=[1,0,0,1].$
\end{enumerate}
\end{enumerate}
\end{enumerate}
}
Clearly, for a given $\boldsymbol{\zeta}(t)$ the actions recommended in \ref{condn4QueuesOnlyFirst3} and \ref{condn4QueuesOnlyLast3} above never contradict each other despite depending on overlapping subsets of queues. From the above definition, if condition \ref{condn4QueuesThatMakesThisMSM}, i.e., the check for $\boldsymbol{\zeta}(t)=[1,1,1,0]$ is removed, the resulting policy, which we call $\pi^{(4)}_{TI}$, is no longer MSM. Table.~\ref{Chapter04tableAll4QueuePolicies} shows the difference between the two policies. 
It should be noted that the policy $\pi^{(4)}_{TI}$ is also constructed by splicing together two 3-queue policies. However, since the subscript \enquote{SP} has been reserved for policies constructed by only splicing $\pi^{(N)}_{TD}$ and $\pi^{(N)}_{BU}$, and since the constituent policies of $\pi^{(4)}_{TI}$ are \emph{not} the Top-Down Bottom-Up pair, we have used a different subscript. We now state the following result.

\begin{lem}\label{lemSplicedPolicy4QueuesIsTO}
$\pi^{(4)}_{TI}$ is Throughput Optimal.
\end{lem}
\begin{pf}
See Sec.~\ref{AppendixTOofSplicedPolicy4Queues} in the Appendix. 
\end{pf}

Recall the projection operator $L:\Pi^{(N)}\rightarrow\Gamma^{(N)}_M$ defined in Sec.~\ref{Chapter03secProjectionInPolicySpace}. 
It can now be observed that $\tilde{\pi}^{(4)}_1=L(\pi^{(4)}_{TI})$, since the only occupancy vector for which $\pi^{(4)}_{TI}$ does not serve the maximum number of queues is $[1,1,1,0].$ When this shortcoming is rectified, we get $\tilde{\pi}^{(4)}_1$. Now, we use the discussion in Sec.~\ref{secTOSchedulingWith3Queues}, specifically, Eqn.~\eqref{Chapter03eqnLReducesBacklogsGeneralN}, to see that
\begin{equation}
\sum_{i=1}^4Q^{\tilde{\pi}^{(4)}_1}_i(t)\stackrel{st}{\leq}\sum_{i=1}^4Q^{\pi^{(4)}_{TI}}_i(t),~\forall t\geq0.
\end{equation}
Finally, since $\pi^{(4)}_{TI}$ is strongly stable, so is $\tilde{\pi}^{(4)}_1$ as shown below
\begin{eqnarray}
\sum_{i=1}^4Q^{\tilde{\pi}^{(4)}_1}_i(t)&\stackrel{st}{\leq}&\sum_{i=1}^4Q^{\pi^{(4)}_{TI}}_i,\nonumber\\
\Rightarrow \sum_{i=1}^4\mathbb{E}Q^{\tilde{\pi}^{(4)}_1}_i(t)&\leq &\sum_{i=1}^4\mathbb{E}Q^{\pi^{(4)}_{TI}}_i,\nonumber
\end{eqnarray}
which immediately gives us
\begin{eqnarray}
&&\limsup_{T\rightarrow\infty}\frac{1}{T}\sum_{t=0}^{T-1}\sum_{i=1}^4\mathbb{E}_{\tilde{\pi}^{(4)}_1}Q_i(t)\nonumber\\
&&<\limsup_{T\rightarrow\infty}\frac{1}{T}\sum_{t=0}^{T-1}\sum_{i=1}^4\mathbb{E}_{\pi^{(4)}_{TI}}Q_i(t)\nonumber\\
&&<\infty,
\label{Chapter04eqnStabilityThroughProjections}
\end{eqnarray}
\end{pf}

As mentioned before, $\tilde{\pi}^{(4)}_2$ is defined by simply replacing checking the status of Queue 3 with that of Queue 2 and vice versa. 
%
A similar proof of throughput optimality holds for this policy as well. 
In fact, it is easily seen that when $N$ is even, every policy $\pi\in\tilde{\Pi}^{(N)}$ has a dual that breaks ties by prioritizing an alternate set of queues than $\pi.$ With $N=4$, we obtained two policies that prioritized Queue 2 and Queue 3 respectively.

\begin{rem}
In the above proof, the manner in which $\pi^{(3)}_{TD}$ and $\pi^{(3)}_{IQ}$ were spliced to form $\pi^{(4)}_{TI}$ is of particular note. While $\pi^{(3)}_{TD}$ is a policy that prioritizes \enquote{outer} queues (Queue 1), $\pi^{(3)}_{IQ}$ prioritizes \enquote{inner} queues, i.e., Queue 2. Since, in the four queue system, Queue 2 becomes the outer and inner queue for these two policies respectively, both subsystems are simultaneously stabilized. 
\end{rem}

\subsubsection{Analysis of $\tilde{\pi}^{(4)}_3$ and $\tilde{\pi}^{(4)}_4$}\label{secPi4_3andPi4_4}
Table.~\ref{Chapter04tableAll4QueuePolicies} compares the activation vectors of policies $\pi^{(4)}_{TI}$, $\tilde{\pi}^{(4)}_1$ and $\tilde{\pi}^{(4)}_3$. Note that in column 4, only the places where $\tilde{\pi}^{(4)}_3$ differs from $\tilde{\pi}^{(4)}_1$ are shown in blue. The unspecified entries, therefore, follow those in the corresponding rows of the third column. Also note that the first coordinate of the first column is $i_4(t)$ and not $i_1(t).$
\begin{table}[tbh]
\caption{Comparison of $\mathbf{S}(t)$ under $\pi^{(4)}_{TI}$, $\tilde{\pi}^{(4)}_1$ and $\tilde{\pi}^{(4)}_3$} 
\centering 
\begin{tabular}{c c c c} 
\hline\hline 
$[i_4(t),i_3(t),i_2(t),i_1(t)]$ & $\pi^{(4)}_{TI}$ & $\tilde{\pi}^{(4)}_1$ & $\tilde{\pi}^{(4)}_3$ \\ [0.5ex] 
\hline 
0000 &    1001 & 1001 &  \\ 
0001 &    1001 & 1001 &  \\
0010 &    1010 & 1010 &  \\
0011 &    1010 & 1010 &  \\

0100 &    0101 & 0101 & \\
0101 &    0101 & 0101 & \\
${\color{black}0110}$ & 1010 & 1010 & 1010 \\
${\color{blue}\mathbf{0111}}$ &   ${\color{blue}\mathbf{1010}}$ &  0101 & 0101 \\

1000 &    1001 & 1001 & \\
1001 &    1001 & 1001 & \\
1010 &    1010 & 1010 & \\
1011 &    1010 & 1010 & \\
 
1100 &    0101 & 0101 & \\
1101 &    0101 & 0101 & \\
1110 &    1010 & 1010 & \\
${\color{blue}\mathbf{1111}}$ & 1010 & 1010 & ${\color{blue}\mathbf{0101}}$ \\ [1ex] \\
\hline 
\end{tabular}
\label{Chapter04tableAll4QueuePolicies} 
\end{table}

\begin{prop}\label{Chapter04propPiTilde4Other2PoliciesAlsoStable}

$\tilde{\pi}^{(4)}_3$ and $\tilde{\pi}^{(4)}_4$ are Throughput Optimal.
\end{prop}

\begin{pf}
The proof essentially proceeds by using a modification of the technique used to prove the delay optimality of $\tilde{\pi}^{(3)}_{IQ}$. This is eventually used to show that the total system backlog with $\tilde{\pi}^{(4)}_3$ and $\tilde{\pi}^{(4)}_1$ are identically distributed, and since $\tilde{\pi}^{(4)}_1$ is throughput-optimal, so is $\tilde{\pi}^{(4)}_3$. The proof is available in Sec.~\ref{AppendixProofOfPropPiTilde4Other2PoliciesAlsoStable} in the Appendix.
\end{pf}

\subsection{Analysis of Delay in $\Pi^{(4)}$}\label{secDelayAnalysis4and5Queues}
We now show that ${\Pi}^{(4)}$ \textbf{does not contain} any queue length agnostic policy that is \emph{uniformly} delay optimal over the entire set $\Lambda^o_{4}$. This is unlike the case with $N=3,$ where $\tilde{\pi}^{(3)}_{IQ}$ produced the lowest possible delay regardless of the arrival rate. We first prove in Prop.~\ref{Chapter04Prop2LambdasForNonDOinPi4} that $\tilde{\Pi}^{(4)}$ does not contain any uniformly delay optimal policy. 

\begin{prop}\label{Chapter04Prop2LambdasForNonDOinPi4}
There exist arrival rate vectors $\boldsymbol{\lambda}_1$ and $\boldsymbol{\lambda}_2$ within $\Lambda^{0}_{4}$, such that, under $\boldsymbol{\lambda}_1$,
\begin{eqnarray}
\sum_{i=1}^4Q^{\tilde{\pi}^{(4)}_1}_i(t)&\stackrel{st}{\leq}&\sum_{i=1}^4Q^{\tilde{\pi}^{(4)}_2}_i(t),~\forall t\geq0,\text{ and, }\nonumber\\
\sum_{i=1}^4Q^{\tilde{\pi}^{(4)}_3}_i(t)&\stackrel{st}{\leq}&\sum_{i=1}^4Q^{\tilde{\pi}^{(4)}_4}_i(t),~\forall t\geq0,
\label{eqnLamForWhich1and3BetterThan2and4}
\end{eqnarray}
and under $\boldsymbol{\lambda}_2$,
\begin{eqnarray}
\sum_{i=1}^4Q^{\tilde{\pi}^{(4)}_2}_i(t)&\stackrel{st}{\leq}&\sum_{i=1}^4Q^{\tilde{\pi}^{(4)}_1}_i(t),~\forall t\geq0,\text{ and, }\nonumber\\
\sum_{i=1}^4Q^{\tilde{\pi}^{(4)}_4}_i(t)&\stackrel{st}{\leq}&\sum_{i=1}^4Q^{\tilde{\pi}^{(4)}_3}_i(t),~\forall t\geq0.
\label{eqnLamForWhich2and4BetterThan1and3}
\end{eqnarray}
\end{prop}

\begin{pf}
The proof can be found in Sec.~\ref{AppendixProofOf2LambdasForNonDOinPi4} of the Appendix.
\end{pf}

Next, we show that policies in $\tilde{\Pi}^{(4)}$ show better delay performance than those in ${\Pi}^{(4)}_M.$

\begin{prop}\label{propPoliciesInPi4TildeAreBetterThanPi4M}
Given any policy $\pi\in\Pi^{(4)}_M\setminus\tilde{\Pi}^{(4)},$ there exists a policy $\pi'\in\tilde{\Pi}^{(4)}$ such that
\begin{eqnarray}
\sum_{i=1}^4Q^{\pi'}_i(t)&\stackrel{st}{\leq}&\sum_{i=1}^4Q^\pi_i(t),~\forall t\geq0.
\label{eqnPoliciesInPi4TildeAreBetterThanPi4M}
\end{eqnarray}
\end{prop}
\begin{pf}
Proof available in Sec.~\ref{AppendixProofOfPoliciesInPi4TildeAreBetterThanPi4M} of the Appendix.
\end{pf}

 We already know, from Eqn.\eqref{Chapter03eqnLReducesBacklogsGeneralN}, that the delay of any policy in $\Pi^{(N)}$ can be improved by projecting it onto $\Pi^{(N)}_M.$ Now, Prop.~\ref{propPoliciesInPi4TildeAreBetterThanPi4M}, along with Eqn.\eqref{Chapter03eqnLReducesBacklogsGeneralN}, shows that delay optimal policies, when they exist, must necessarily lie in $\tilde{\Pi}^{(N)}$. This observation, along with the nonexistence of delay optimal policies in $\tilde{\Pi}^{(4)}$ (Prop.~\ref{Chapter04Prop2LambdasForNonDOinPi4}), has one far reaching consequence.
\begin{tcolorbox}
\begin{thm}\label{Chapter04ThmNonexistenceOfDOPoliciesForNGreaterThan4}
For all $N\geq4,$ there does not exist any policy in $\Pi^{(N)}_M$ that is uniformly delay optimal over all of $\Lambda^o_N.$
\end{thm}
\end{tcolorbox}
\begin{pf}
Suppose $\pi\in\Pi^{(N)}_M$ was a delay optimal policy for the $N$ queue system for some $N,$ then, by setting $\lambda_i=0,$ for all $i\geq5$ we would get a delay optimal policy for the 4 queue system, contradicting Prop.~\ref{Chapter04Prop2LambdasForNonDOinPi4}. Hence, such a policy cannot exist. 
\end{pf}

\begin{rem} A few important remarks are in order.
\begin{enumerate}
\item It is possible that policies which take decisions based on $\mathbf{Q}(t)$ rather than just $\boldsymbol{\zeta}(t)$ might perform better. MaxWeight ($MW$) and its projection into $\Gamma^{(N)}_M$, viz $L(MW)$, use all of $\mathbf{Q}(t)$ and hence, perform dynamic randomization between $s_1$ and $s_2$ when $\boldsymbol{\zeta}(t)=[0,1,1,0]$, based on which queue is larger. Their performance with respect to the policies developed in this section is further explored through simulations in Sec.~\ref{Chapter05SimulationResults}. 
\item\label{remDelayAnalysis4and5queuesPriorityToInnerQueues} It is to be noted that, as Column~3 in Table.~\ref{Chapter04TablePoliciesInPi4TildeAreBetterThanPi4M} shows, $\{\tilde{\pi}^{(4)}_i,1\leq i\leq 4\}$ satisfy Condition~\ref{Chapter03conditionMSMDelaySmall} in Lem.~\ref{Chapter03lemSufficientForActivationMSM}. 
It appears that prioritizing \enquote{inner,} and hence more constrained, queues improves not only throughput but also delay. 
\end{enumerate}
\end{rem}
So in essence, Thm.~\ref{Chapter04ThmNonexistenceOfDOPoliciesForNGreaterThan4} shows that while \emph{throughput} optimality in these interference graphs only requires knowledge of queue occupancy (i.e., $\boldsymbol{\zeta}(t)$), \emph{delay} optimality potentially requires more information from the filtration, 
$\mathcal{H}_t:=\sigma\left(\left\lbrace\mathbf{Q}(t),\cdots,\mathbf{Q}(0),\mathbf{s}(t-1)\cdots\mathbf{s}(0)\right\rbrace\right)$. We explore this in greater detail in Sec.~\ref{Chapter05SimulationResults} where we compare the performance of these policies with those of $MW$ and $L(MW)$.


\section{Cluster-of-Cliques Interference Networks: Throughput Optimal Scheduling}\label{Chapter05SecTOSchedulingPolicies}
We will now show that some of the scheduling policies developed for path-interference graph networks extend in a natural manner to policies for the SoC and the LAoC networks. 

\textbf{Notation:} We denote policies designed for Star-of-Cliques networks by \enquote{$\phi$} and include an \enquote{$(S)$} in the superscript to emphasize this. On the other hand, \enquote{$\theta$} with and \enquote{$(L)$} in the superscript specifies an LAoC network policy. We will begin with centralized scheduling in SoC networks. 

\subsection{Scheduling in the Star-of-Cliques Network}\label{Chapter05SecSchedulingInSoC}
Consider the following policy that we denote $\tilde{\phi}^{(S)}_{IC}$, which is motivated by the 3-node path graph policy $\tilde{\pi}^{(3)}_{IQ}$ which we discussed in Sec.~\ref{Chapter03SecTildePi3_IQ}. Recall that we defined $\mathcal{N}$ to be the total number of queues in the network. In keeping with the objective of developing queue nonemptiness-based policies, in every slot, $\tilde{\phi}^{(S)}_{IC}$ maps the occupancy vector $\boldsymbol{\zeta}(t)\in\left\lbrace0,1\right\rbrace^\mathcal{N}$ to an activation vector $\mathbf{s}(t)\in \left\lbrace0,1\right\rbrace^\mathcal{N}$. We define $\tilde{\phi}^{(S)}_{IC}$ as follows.

\textsl{At each time $t:$
\begin{enumerate}\label{Chapter05DefinitionOfPhi3Tilde}
\item If $\prod_{m=2}^N\left(\sum_{l\in\mathcal{C}_m}\zeta_l(t)\right)$ $>$ $0$ serve any nonempty queue in every clique $\left\lbrace\mathcal{C}_m,~m\geq2\right\rbrace$ having nonempty queues.
\item Else, if $\sum_{l\in\mathcal{C}_1}\zeta_l(t)$ $>$ $0,$ serve any nonempty queue in $\mathcal{C}_1$.
\item Else, serve one nonempty queue (if it exists) in each of $\left\lbrace\mathcal{C}_m,~m\geq2\right\rbrace.$
\end{enumerate}
}
In words, the above policy states that if, at time $t$, 
\begin{itemize}
\item every peripheral clique has at least one non empty queue, then serve one non empty queue in each of these cliques,
\item else, if the inner clique has a non empty queue, serve one non empty queue in that clique,
\item else, serve one non empty queue in every peripheral clique that has a non empty queue.
\end{itemize}

\begin{prop}[]
\label{Chapter05PropPhiNTildeIsTO}
$\tilde{\phi}^{(S)}_{IC}$ is throughput-optimal.
\end{prop}
\begin{pf}
The main idea behind the proof of this proposition is to prove a more general version of Property $\mathcal{P}$ (which we defined in Lem.~\ref{lemPropertyPMeansTO} for path-graph networks) and use and use the total per-clique backlog as inputs to a new Lyapunov function to prove strong stability. See \ref{AppendixProofOfPhiNTildeIsTO} for details.
\end{pf}

Towards the end of our discussion on queue nonemptiness-based scheduling for path-graph networks with $N=3$ queues (see Sec.~\ref{Chapter03DefinitionOfPi3IQNonMSMButTO}), we defined a \emph{non-MSM} policy $\pi^{(3)}_{IQ}$. Extending this to the SoC network model gives us a second queue nonemptiness-based policy $\phi^{(S)}_{IC}$, which we define as follows.

\textsl{ At time $t,$
\begin{enumerate}\label{Chapter05DefinitionOfPhi3_5}
\item If $\sum_{l\in\mathcal{C}_1}\zeta_l(t)$ $>$ $0,$ serve any nonempty queue in $\mathcal{C}_1$.
\item Else, serve one nonempty queue (if it exists) in each of $\left\lbrace\mathcal{C}_m,~m\geq2\right\rbrace.$
\end{enumerate}
}
\begin{prop}\label{Chapter05PropPhi3_5IsTO}
$\phi^{(S)}_{IC}$ is throughput-optimal.
\end{prop}
\begin{pf}
Once again, the proof of this result rests on  proving the new version of Property $\mathcal{P}$ for this policy, followed by Lyapunov analysis. 
The proof is available in Sec.~\ref{AppendixProofOfPhi3_5IsTO} in the Appendix. 
\end{pf}

We end this section with some remarks about implementation and delay performance. From the point of view of implementation, the latter, $\phi^{(S)}_{IC}$ is actually easier to implement than $\tilde{\phi}^{(S)}_{IC}$. We discuss this in detail in Sec.~\ref{Chapter05SecRemarksDecenralized} which is completely dedicated to implementation issues. 
However, $\tilde{\phi}^{(S)}_{IC}$ has its own advantages.
With respect to the packet delay, recall that we had used a stochastic ordering argument to prove the delay optimality of Policy~$\tilde{\pi}^{(3)}_{IQ}$ and later used a similar technique to show the absence of uniformly delay-optimal queue nonemptiness-based policies for path-graph networks. Along similar lines, we compare the delays induced by $\tilde{\phi}^{(S)}_{IC}$ and $\phi^{(S)}_{IC}$ below. 

\subsubsection{Comparison of delay with $\tilde{\phi}^{(S)}_{IC}$ and $\phi^{(S)}_{IC}$} 
\begin{prop}\label{Chapter05PropPhi3_TildeIsBetterThanPhi3_1}
Let the system backlog at time $t\geq0$ with $\phi^{(S)}_{IC}$ and $\tilde{\phi}^{(S)}_{IC}$ be denoted by $\mathbf{Q}^{\phi^{(S)}_{IC}}(t)$, and $\mathbf{Q}^{\tilde{\phi}^{(S)}_{IC}}(t)$ respectively. 
Then, with $\mathbf{Q}^{\tilde{\phi}^{(S)}_{IC}}(0)\stackrel{s}{=}\mathbf{Q}^{\phi^{(S)}_{IC}}(0)$, and arrivals to corresponding queues having the same statistics in both systems,
\begin{equation}
\sum_{m=1}^N\sum_{j=1}^{\mathcal{N}_m}Q^{\tilde{\phi}^{(S)}_{IC}}_{m,j}(t)\stackrel{st}{\leq}
\sum_{m=1}^N\sum_{j=1}^{\mathcal{N}_m}Q^{\phi^{(S)}_{IC}}_{m,j}(t),~\forall t\geq0.
\label{eqnPhi3_TildeisBetterThanPhi3_1}
\end{equation}
\end{prop}
\begin{pf}This proof proceeds along the same lines as the proof of delay optimality of Policy~$\tilde{\pi}^{(3)}$ that we presented in Sec.~\ref{AppendixProofOfPi33isDelayOptimal}. It can be found in Sec.~\ref{AppendixProofOfPropPhi3_TildeIsBetterThanPhi3_1} in the Appendix.  
\end{pf}

We now begin our study of scheduling in LAoC networks. However, as mentioned before, we will return to these policies once again when we shift our focus to decentralized implementation. 

\subsection{Scheduling in Linear-Arrays-of-Cliques}\label{Chapter05SecSchedulingInLAoC}
The technique we use to propose scheduling policies for LAoC networks is the policy splicing technique we developed in Sec.~\ref{Chapter04SecQLenAgnosticTOSplicing}. The proofs therein cannot be directly used to assess the stability of policies designed for LAoC networks since the proofs are designed for Bernoulli arrival processes to queues and require some more work to be extended to handle scheduling over cliques. However, one could argue that a clique can, in essence, be treated as a queue with an arrival process which is simply the sum of the arrivals to the constituent queues. For example, Clique $\mathcal{C}_1$ in Fig.~\ref{Chapter05FigClustersOfCollocatedNetworksInterferenceGraph} can be treated as a single queue with an arrival process that is the sum of the processes to Queues $Q_{1,1},Q_{1,2}\text{ and }Q_{1,3}$ therein. The resulting arrival process to the queue would then be a batch arrival process with \emph{arbitrary batch size} (there can be any number of queues in a clique), and simple extensions of the proofs supplied hitherto can be shown to suffice. 

As before, we begin with Top-Down and Bottom-Up policies for the $3$-clique LAoC and splice them to construct policies for the LAoC's with 4 and 5 cliques. Note once again, that we place no restrictions on the number of queues within any clique. 
\subsubsection{Scheduling Policies for Systems with $N=3$ Cliques}\label{secCliquePolicies3CliqueSystems}

The policy $\theta^{(3L)}_{TD}$ is described as follows. 

\textsl{At time $t:$
\begin{itemize}
\item If $\sum_{j=1}^{\mathcal{N}_1}\zeta_{1,j}(t)$ $>$ $0$ schedule any non-empty queue in $\mathcal{C}_1$.
\begin{itemize}
\item If $\sum_{j=1}^{\mathcal{N}_3}\zeta_{3,j}(t)$ $>$ $0$ schedule any non-empty queue in $\mathcal{C}_3$.
\end{itemize}
\item Else, if $\sum_{j=1}^{\mathcal{N}_2}\zeta_{2,j}(t)$ $>$ $0$ schedule any non-empty queue in $\mathcal{C}_2$.
\item Else schedule any non-empty queue in $\mathcal{C}_3$.
\end{itemize}
}
In other words, if in slot $t$ 
\begin{itemize}
\item there is a non-empty queue in $\mathcal{C}_1,$ then $\theta^{(3L)}_{TD}$ serves one non-empty queue in $\mathcal{C}_1$ and $\mathcal{C}_3$.
\item if $\mathcal{C}_1$ is empty but $\mathcal{C}_2$ has a non-empty queue in it, then $\theta^{(3L)}_{TD}$ serves that queue.
\item if $\mathcal{C}_1$ and $\mathcal{C}_1$ are both empty, then $\theta^{(3L)}_{TD}$ serves any non-empty queue in $\mathcal{C}_3.$
\end{itemize}

\begin{prop}\label{Chapter05PropPhi3TDIsTO}
$\theta^{(3L)}_{TD}$ is throughput-optimal.
\end{prop}

\begin{pf}
This proof uses the ideas involved in proving the throughput optimality of $\pi^{(3)}_{TD}$ and simply extends them to incorporate batch arrivals. The proof is available in Sec.~\ref{AppendixProofOfPropPhi3TDIsTO} in the Appendix.
\end{pf}

A similar proof shows that $\theta^{(3L)}_{BU}$ is also throughput-optimal. 
\subsubsection{Scheduling Policies for Systems with $N=4$ and $5$ Cliques}\label{secCliquePolicies4and5QueueSystems}

Now, by splicing together $\theta^{(3L)}_{TD}$ and $\theta^{(3L)}_{BU}$, one can construct stable policies for the system with 5 cliques and hence, systems with 4 cliques. The spliced policy, $\theta^{(5L)}_{SP}$ is defined as 

\textsl{At time $t:$
\begin{enumerate}
\item If $\sum_{j=1}^{\mathcal{N}_3}\zeta_{3,j}(t)$ $>$ $0$ then schedule a nonempty queue in $\mathcal{C}_3.$
\begin{enumerate}
\item If $\sum_{j=1}^{\mathcal{N}_1}\zeta_{1,j}(t)$+$\sum_{j=1}^{\mathcal{N}_5}\zeta_{5,j}(t)$ $>$ $0$ then schedule any nonempty queue each in $\mathcal{C}_1$ and $\mathcal{C}_5.$
\end{enumerate}
\item Else if $\sum_{j=1}^{\mathcal{N}_2}\zeta_{2,j}(t)$ $\times$ $\sum_{j=1}^{\mathcal{N}_4}\zeta_{4,j}(t)$ $>$ $0$ then schedule a nonempty queue each in $\mathcal{C}_2$ and $\mathcal{C}_4.$
\item Else if $\sum_{j=1}^{\mathcal{N}_2}\zeta_{2,j}(t)$ $\times$ $\sum_{j=1}^{\mathcal{N}_5}\zeta_{5,j}(t)$ $>$ $0$ then schedule a nonempty queue each in $\mathcal{C}_2$ and $\mathcal{C}_5.$
\item Else if $\sum_{j=1}^{\mathcal{N}_4}\zeta_{4,j}(t)$ $\times$ $\sum_{j=1}^{\mathcal{N}_1}\zeta_{1,j}(t)$ $>$ $0$ then schedule a nonempty queue each in $\mathcal{C}_1$ and $\mathcal{C}_4.$
\item Else schedule any nonempty queue each in $\mathcal{C}_1$ and $\mathcal{C}_5.$
\end{enumerate}
}

\begin{prop}\label{Chapter05PropPhi5_SPIsTO}
The policy $\theta^{(5L)}_{SP}$ is throughput optimal.
\end{prop}

\begin{pf}
The policy $\theta^{(5L)}_{SP}$ is formed by splicing together $\theta^{(3L)}_{TD}$ and $\theta^{(3L)}_{BU}$. This means that $\theta^{(5L)}_{SP}$ restricted to Queues $1$ to $3$ is $\theta^{(3L)}_{BU}$ and restricted to Queues $3$ to $5$ is $\theta^{(3L)}_{TD}$. The throughput optimality of $\theta^{(3L)}_{TD}$ and $\theta^{(3L)}_{BU}$ means that for every $\boldsymbol{\lambda}\in\Lambda^{(5)}_l$ (defined in \eqref{Chapter05EqnCapacityRegionOfLAoCModel}), 
\begin{eqnarray}
\limsup_{T\rightarrow\infty}\frac{1}{T}\sum_{t=0}^{t-1}\sum_{i=3}^{5}\sum_{j=1}^{\mathcal{N}_i}\mathbb{E}_{\theta^{(3L)}_{TD}} Q_{i,j}(t)&<&\infty,\text{ and }\label{Chapter05EqnTDPartOfSplicedIsTO}
\\
\limsup_{T\rightarrow\infty}\frac{1}{T}\sum_{t=0}^{t-1}\sum_{i=1}^{3}\sum_{j=1}^{\mathcal{N}_i}\mathbb{E}_{\theta^{(3L)}_{BU}} Q_{i,j}(t)&<&\infty.\label{Chapter05EqnBUPartOfSplicedIsTO}
\end{eqnarray}
Now, as in Thm.~\ref{Chapter04SecQLenAgnosticTOSplicingThmGenericSplicedPolicyIsTO}, for every $\boldsymbol{\lambda}\in\Lambda^{(5)}_l$ we define two new arrival rate vectors $\boldsymbol{\lambda}_{1:3}=[\lambda_1,\lambda_2,\lambda_3]$ and $\boldsymbol{\lambda}_{3:5}=[\lambda_3,\lambda_4,\lambda_{5}]$ and note that by the definition of the set $\Lambda^{(N)}_l$, $\boldsymbol{\lambda}_{1:3}$ and $\boldsymbol{\lambda}_{3:5}$ are both in $\Lambda^{(3)}_l$, which means that \eqref{Chapter05EqnTDPartOfSplicedIsTO} and \eqref{Chapter05EqnBUPartOfSplicedIsTO} are still separately true. We conclude the proof by observing that 
\begin{eqnarray*}
&&\limsup_{T\rightarrow\infty}\frac{1}{T}\sum_{t=0}^{t-1}\sum_{i=1}^{5}\sum_{j=1}^{\mathcal{N}_i}\mathbb{E}_{\phi^{(5)}_{SP}} Q_{i,j}(t)\\
&\leq& \limsup_{T\rightarrow\infty}\frac{1}{T}\sum_{t=0}^{t-1}\sum_{i=3}^{5}\sum_{j=1}^{\mathcal{N}_i}\mathbb{E}_{\theta^{(3L)}_{TD}} Q_{i,j}(t) \\
&&+ \limsup_{T\rightarrow\infty}\frac{1}{T}\sum_{t=0}^{t-1}\sum_{i=1}^{3}\sum_{j=1}^{\mathcal{N}_i}\mathbb{E}_{\theta^{(3L)}_{BU}} Q_{i,j}(t)\\
&<&\infty
\end{eqnarray*}
\end{pf}

To summarize, in this section, we studied scheduling in the Star-of-Cliques and Linear-Array-of-Cliques models that occur frequently in IoT-type sensor network applications. Having characterized the capacity region of such networks, we proposed an analysed multiple scheduling policies. However, as mentioned before, these policies depend on being able to find a nonempty queue in every slot in which the system is not empty. While disseminating this occupancy information across the network is certainly not as expensive as sharing queue length information (required by the MaxWeight family of scheduling algorithms) it would be beneficial to sensor network designers if we could produce scheduling policies that worked with even less information. In the following section, we attempt to do precisely that.







\subsection{Preamble to the rest of the Paper}\label{secPreambleToRestOfPaper}
Since we will be introducing several scheduling policies in the sequel, we will now present a short preamble to the rest of the paper for the convenience of the reader. Note that all the necessary notation is also present in the glossary of notation \ref{glossaryOfNotation} that the reader can refer to for clarification.
\begin{itemize}
	\item
	We first propose queue nonemptiness-based policies for SoC and LAoC networks. We begin with SoC networks and propose two policies $\phi^{(S)}_{IC}$ and $\tilde{\phi}^{(S)}_{IC}$ that are based on the policies $\pi^{(3)}_{IQ}$ and $\tilde{\pi}^{(3)}_{IQ}$ respectively, that were proposed and analysed in Sections~\ref{Chapter03SecTildePi3_IQ} and \ref{secNonMSMStablePolicy}. The \enquote{$S$} in the superscript stands for SoC network, and \enquote{$IC$} in the subscript shows that they prioritize the \emph{inner clique,} i.e., $\mathcal{C}_1.$
	We analyse their throughput and delay properties.
	\item 
	We then move on to LAoC networks for which we, once again, propose three policies. We begin with 3-clique networks and propose extensions of the Top-Down and Bottom-Up policies we developed in Sec~\ref{Chapter03SecAnalysisOfPi3_TDPi3_BU}. We call these policies $\theta^{(3L)}_{TD}$ and $\theta^{(3L)}_{BU}$, where the \enquote{$L$} in the superscript stands for LAoC network.
	\item
	In the sequel, $\phi$ will always represent a policy for SoC networks and $\theta$ for LAoC network (recall that $\pi$ was used to denote policies on path-graph networks).
	\item
	We then move on to decentralized implementation, introduce the concept of \enquote{minislots} show how they can be used to implement $\phi^{(S)}_{IC}$ in a distributed fashion. We then propose a class of policies that we denote $\phi^{(S)}_{IC}(T)$ for SoC networks, that require knowledge of the occupancy vector $\boldsymbol{\zeta}=[\mathbb{I}_{\{Q_1(t)>0\}},\dots,\mathbb{I}_{\{Q_\mathcal{N}(t)>0\}}]$ only every $T$ time slots and show that the class is throughput optimal. Following this, we propose a policy $\phi^{(S)}_{CS}$ that\footnote{The \enquote{CS} in the subscript stands for channel sensing.}, like the QZMAC protocol in \cite{mohan-etal16hybrid-macsMASSversion}, takes scheduling decisions based solely on the information gathered by sensing the channel for activity.
\end{itemize}


\section{Some Remarks on Decentralized Implementation}\label{Chapter05SecRemarksDecenralized}
\begin{figure}[tb]
\centering
\includegraphics[height=4.5cm, width=8.250cm]{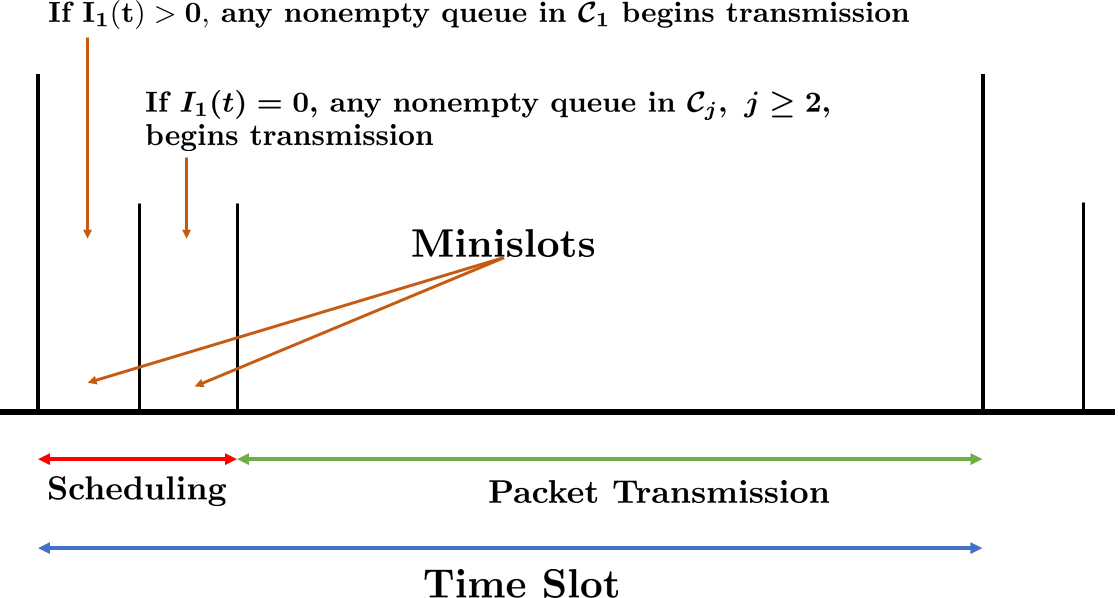}
\caption{Implementing $\phi^{(S)}_{IC}$ using the slot and minislot structures. It is important to note that the number of minislots is O(1), i.e., \emph{does not scale} with the number of communication links the network. This results in constant scheduling overhead, which is patently desirable.}
\label{Chapter05FigMinislotStructureForPhi_1}
\end{figure}

In this section, we discuss several ways in which the policies developed and analysed hitherto can be made amenable to decentralized implementation. 
To accomplish our stated objective of constructing a distributed scheduling protocol, we take the help of what are known as \emph{minislots} \cite{rhee08zmac,mohan-etal16hybrid-macsMASSversion} which we describe in detail, below.\\ 
\textbf{Transmission Sensing:}\label{Chapter05ParagraphTransmissionSensing}
We assume that all nodes transmit at the same fixed power, and the maximum internode distance is such that every node in clique $\mathcal{C}_j,j\geq2$ can sense the power from a transmitting node in clique $\mathcal{C}_1$ and vice versa, as dictated by the interference constraints\footnote{Obviously, for all $2\leq j,k\leq N,$ nodes in $\mathcal{C}_j$ \underline{cannot} sense transmissions in $\mathcal{C}_k$ and vice versa.}. Suppose a node has been scheduled to transmit in a slot. Then whether or not the node actually transmits can be determined by the other nodes by averaging the received power over a small interval (akin to the \enquote{Clear Channel Assessment} or CCA mechanism 
\cite{kinney01cca-method-wpan-working-group}). For reliable assessment, the interval will need to be of a certain length, and the distance between the nodes will need to be limited. As before, we refer to \emph{this} activity-sensing interval a \textbf{minislot} (see \cite{mohan-etal16hybrid-macsMASSversion} and Fig.~\ref{Chapter05FigMinislotStructureForPhi_1}). 

Decentralized methods of implementing both $\phi^{(S)}_{IC}$ and $\tilde{\phi}^{(S)}_{IC}$ immediately follow from the minislot structure. Let $I_j(t):=\sum_{k\in\mathcal{C}_j}i_{j,k}(t),~j\geq1$, indicate if clique $j$ has any nonempty nodes at the beginning of time slot $t$. We will first discuss implementing $\phi^{(S)}_{IC}$ in detail, and then $\tilde{\phi}^{(S)}_{IC}$ in Sec.~\ref{Chapter05SecProtocolTildePhiMinislots}.

\subsection{Decentralized Implementation of $\phi^{(S)}_{IC}$}\label{Chapter05SecProtocolPhi_1Minislots}
\textsl{At time $t$,
\begin{enumerate}
\item\label{step1Phi1Clique1} If $I_1(t)>0$, then one nonempty node from clique $\mathcal{C}_1$ is allowed to transmit (see Fig.~\ref{Chapter05FigMinislotStructureForPhi_1}). Nodes in the other cliques sense this transmission in the first minislot and refrain from transmitting during that slot.
\item\label{step2Phi1OtherCliques} If no power is sensed in the first minislot, it means $I_1(t)=0,$ and each of the other cliques choose one nonempty queue (if any) for transmission during that slot.
\end{enumerate}
}
This, of course, assumes that one is somehow able to identify a nonempty queue, if one exists, in each clique. 
So this implementation is, by itself, centralized \emph{within} a clique and \emph{decentralized} across cliques.
We now propose methods to determine which (nonempty) queue \emph{within} a clique actually gets to transmit in either of the two steps above. 

One method is for the nodes in a clique to periodically share occupancy information which could be accomplished by having a sink node in every clique. The sink node of each clique periodically aggregates occupancy information from its nodes and uses it to schedule nonempty queues in some order. We discuss this in later in this section. In Sec.~\ref{Chapter05SecTowardsFullyDecentralizedTheGammaProtocol} we discuss a version of $\phi^{(S)}_{IC}$ that requires no explicit information exchange between queues. 
\subsubsection{$\phi^{(S)}_{IC}$ with Periodic Occupancy Information: The $\phi^{(S)}_{IC}(T)$ Policy}
\label{Chapter05SecPhiWithPeriodicOccupancyInformation}
Suppose the nodes share 
their occupancy status 
at slots $\left\lbrace kT,~T\geq1,k\in\mathbb{N}\cup\{0\} \right\rbrace$. 
By setting $T$ large enough, the cost of exchanging this information can be amortized over this interval. We group slots $kT,kT+1,\dots,kT+T-1$ into the $k^{th}$ \emph{frame} ($k\geq0$). In each frame $k$, only the packets that have arrived \emph{until} time $kT$ are served and any new packets arriving in the frame are queued. 
With this, we get a new \emph{class} of protocols parameterized by the information sharing interval $T$, denoted $\left\lbrace\phi^{(S)}_{IC}(T),~T\geq1\right\rbrace$, defined as follows

\textsl{Over $kT,\dots,kT+T-1$, 
\begin{enumerate}
\item If $I_1(kT)>0$, serve nonempty queues in $\mathcal{C}_1$ (in any arbitrary order) until either the next frame begins or all the packets in $\mathcal{C}_1$ have left the system. The other cliques detect this when they sense no power in the first minislot of such a slot. 
\item Any slots left after $\mathcal{C}_1$ is served, is used by the other cliques for transmission until either frame $k+1$ begins or they run out of packets. 
\end{enumerate}
}
We prove that $\phi^{(S)}_{IC}(T)$ is \textbf{throughput optimal} in Sec.~\ref{Chapter05AppendixProofOfTOOfPhiT_1} in the Appendix and therein also show some desirable delay properties of the protocol. While $\phi^{(S)}_{IC}(T)$ requires state information dissemination only every $T$ slots, we would like to explore the possibility of implementing $\phi^{(S)}_{IC}$ without \emph{any} explicit information dissemination.
\subsection{$\phi^{(S)}_{IC}$ without Occupancy Information: Towards Fully Decentralized Policies}\label{Chapter05SecWSNProtocolsPartialInformation}
\label{Chapter05SecTowardsFullyDecentralizedTheGammaProtocol}
First consider a clique, say $\mathcal{C}_i$, in isolation. This is, by itself, a fully connected interference graph. 
Suppose the nodes in $\mathcal{C}_i$ could determine the backlog of a node in $\mathcal{C}_i$ each time it transmitted a packet\footnote{The backlog  information could be quantized and contained in the packet header, for example.}. Then, at the beginning of slot $t$, the information common to all nodes in $\mathcal{C}_i$ would consist of the number of slots $V_i(t)$ since node $i$ last transmitted\footnote{If the node were empty at this instant, it wouldn't have actually transmitted anything. The others can infer its \enquote{emptiness} by sensing no power in a minislot.} and its backlog $Q_i(t-V_i(t))$ at that instant. With this partial information structure, we have already shown, in \cite{mohan-etal16hybrid-macsMASSversion}, that exhaustively serving a nonempty queue minimizes mean delay.
\begin{figure}[tb]
\centering
\includegraphics[height=5.5cm, width=8.2cm]{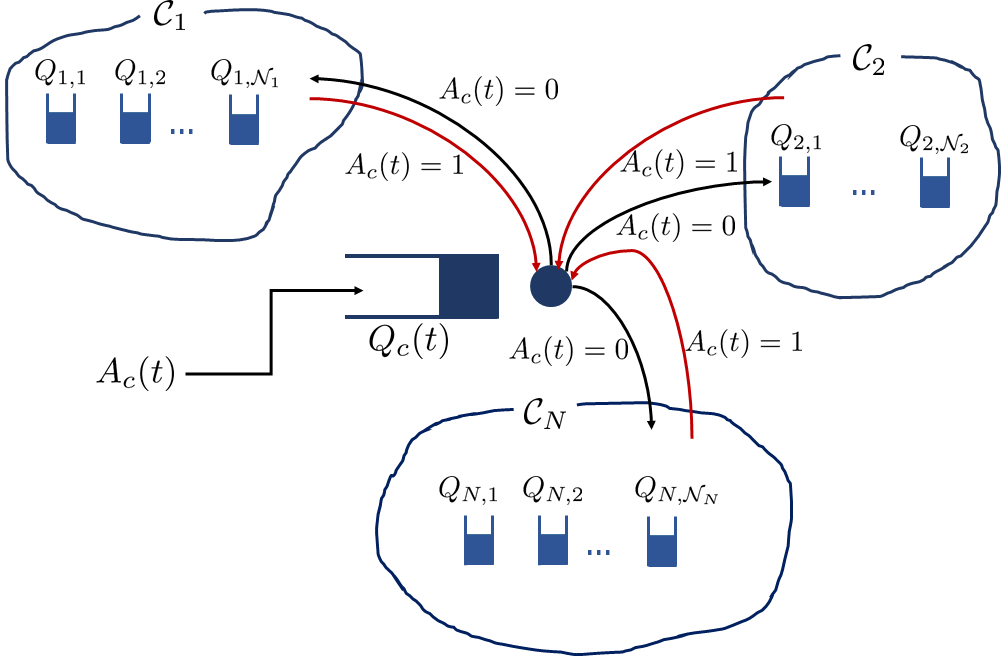}
\caption{The system upon which the policy $\phi^{(S)}_{CS}$ is analysed. $Q_c(t)$ is the central queue which has highest priority. This means that the queue is served whenever it has packets, i.e., whenever the arrival $A_c(t)=1.$ This is indicated by the \textbf{red arrows } that show access to the channel being granted by $\phi^{(S)}_{CS}$ to $Q_c$. However, when $Q_c(t)=0,~\phi^{(S)}_{CS}$ enters Step~\ref{Chapter05ProtocolGammaCentralQueueEmptyGoToPeriphery} and serves the appropriate queue in the peripheral cliques, as indicated by the \textbf{black arrows}.}
\label{Chapter05FigQueueingSystemForGamma}
\end{figure}
With exhaustive service, $Q_i(t-V_i(t))$ is always 0, which obviates the need to transmit queue lengths. When the queue under service, called the \textbf{incumbent} in the sequel, becomes empty  
we have already shown that 
scheduling node $\argmax{j\in\mathcal{C}_i}V_j(t)$ is throughput-optimal, and under certain conditions, also mean delay optimal.
Motivated by this, we define another partial information version $\phi^{(S)}_{CS}$, of $\phi^{(S)}_{IC}$ below \textbf{under the assumption} that the inner clique $\mathcal{C}_1$ has exactly \emph{one} node, i.e., $|\mathcal{C}_1|=1.$ We refer to this queue as Queue~$c$ (see Fig.~\ref{Chapter05FigQueueingSystemForGamma}).

\textsl{At time $t:$
\begin{enumerate}
\item If $I_1(t)$ $>$ $0,$ then the queue in $\mathcal{C}_1$ transmits its packet. Nodes in the other cliques sense this transmission in \underline{minislot 1} (see Fig.~\ref{Chapter05FigMinislotStructureForPhi_1}) and refrain from attempting any transmissions.
\item\label{Chapter05ProtocolGammaCentralQueueEmptyGoToPeriphery} If no power is sensed in minislot 1, every clique $\mathcal{C}_i,~i$ $\geq$ $2$ does the following
\begin{enumerate}
\item\label{Chapter05ProtocolGammaStepSenseAndScheduleProcedure} The incumbent begins to transmit at the end of minislot 1 if it is nonempty. The other nodes in $\mathcal{C}_i$ sense this in \\ \underline{minislot 2} and refrain from attempting transmissions.
\item\label{Chapter05ProtocolGammaNoPowerInMinislot2MeansArgMaxVi} If no power is sensed in minislot 2, then the incumbent is empty and 
 $\argmax{j\in\mathcal{C}_i}V_j(t)$ is now allowed to transmit. 
\end{enumerate}
\end{enumerate}
}
Adding more minislots reduces chances of slot wastage, but also reduces system throughput since it increases time wasted in not actually transmitting a packet. Hence, this parameter represents a tradeoff between throughput and delay. It is not clear if this policy is throughput-optimal and we now provide a formal argument.

\begin{thm}\label{Chapter05ThmGammaForQZSoCIsTO}
The policy $\phi^{(S)}_{CS}$ is throughput-optimal.
\end{thm}

\begin{pf}
The proof uses a Lyapunov drift argument and invokes the Foster-Lyapunov theorem to prove that the system backlog process $\mathbf{Q}(t):=[Q_1(t),\cdots,Q_{\mathcal{N}}(t)],~t\geq0$ is positive recurrent. The details of the proof can be found in the Appendix in Sec.~\ref{AppendixProofOfThmGammaForQZSoCIsTO}.
\end{pf}

We now move on to $\tilde{\phi}^{(S)}_{CS}$, the low-delay counterpart of $\phi^{(S)}_{CS}$. We now show how to use the minislot structure to implement $\tilde{\phi}^{(S)}_{CS}$ in a decentralized manner. 
\subsection{Decentralized Implementation of $\tilde{\phi}^{(S)}_{IC}$}\label{Chapter05SecProtocolTildePhiMinislots}
While $\phi^{(S)}_{CS}$ works even without a sink node in each clique to determine which node transmits in its clique in each slot, this implementation of $\tilde{\phi}^{(S)}_{IC}$ requires a sink. This protocol, while throughput-optimal, is tailored for \textbf{heavy-traffic} applications, when the queues in the system are rarely empty. The protocol requires \underline{three minislots} and proceeds as follows.

\textsl{At time $t$
\begin{enumerate}
\item Every \emph{empty} queue in clique $\mathcal{C}_j,~j$ $\geq$ $2,$ makes a small transmission during \underline{minislot 1}. The sink uses this to determine if its clique is empty, i.e., if $I_j(t)=0$.
\item In \underline{minislot 2}, every peripheral sink with $I_j(t)$ $=$ $0$, $j$ $\geq$ $2$ makes a small transmission which is sensed by the sink of Clique~1. 
\item We now have three possibilities:
\begin{itemize}
\item Suppose one of the outer cliques is empty and $I_1(t)$ $>$ $0$. Then $\mathcal{C}_1$ senses power in minislot 2 infers that at least one outer clique is empty and begins transmission in minislot 3.
\item If none of the outer cliques is empty then the outer cliques do not sense any power in \underline{minislot 3} and those cliques that have nonempty queues begin transmission.  
\item Finally, if one of the outer cliques is empty but $I_1(t)=0$ then, once again, the outer cliques do not sense any power in \underline{minislot 3} and begin transmission.
\end{itemize}
\end{enumerate}
}
\begin{rem}
Prop.~\ref{Chapter05PropPhi3_TildeIsBetterThanPhi3_1} shows that policy $\tilde{\phi}^{(S)}_{IC}$ induces stochastically smaller delays than $\phi^{(S)}_{IC}$.
However, implementing the former consumes more energy than implementing $\phi^{(S)}_{IC}$ and its partial knowledge versions, since even empty nodes need to be ON. This is why the present protocol is more suited to heavy-traffic applications, when the arrival rates to the system are heavy enough that queues do not remain empty for long.
\end{rem}
\subsection{Policies with Periodic State Information for Linear-Arrays-of-Cliques}\label{Chapter05SecLAoCPoliciesWithPeriodicStateInformation}
We now move on to proposing a protocol similar to $\phi^{(S)}_{IC}(T)$ which we discussed in Sec.~\ref{Chapter05SecPhiWithPeriodicOccupancyInformation}, that does not require the empty-nonempty status of queues in the system in every time slot, for LAoC networks. Once again, we assume the presence of a sink node in each clique, that periodically aggregates and disseminates this information from the nodes in its own clique to take scheduling decisions. Suppose the system receives the occupancy information of cliques only every $T$ time slots, i.e., information about whether there exists a non empty queue in clique $i$ ($1\leq i\leq 3$) is disseminated only at $kT, k\in\lbrace0,1,2,\cdots\rbrace$. Then we would like to know if $\theta^{(3L)}_{TD}$ and $\theta^{(3L)}_{BU}$ (both defined in Sec.~\ref{Chapter05SecSchedulingInLAoC}) can be suitably modified to ensure stability. Thereafter, our splicing technique will immediately provide us with policies for systems with $N=4$ and $N=5$ cliques. 

Towards that end, first define for every slot $t$ the \emph{clique occupancy vector} $\boldsymbol{\zeta}(t):=\left[i_1(t),i_2(t),i_3(t)\right]$, where $\zeta_j(t):=\mathbb{I}_{\left\lbrace \sum_{k=1}^{\mathcal{N}_j}i_{j,k}(t)>0\right\rbrace}$ indicates whether Clique~$j$ is empty or not. We call the slots $kT, kT+1, \cdots, kT+(T-1)$ together, the $k^{th}$ frame. Similar to $\phi^{(S)}_{IC}(T)$ consider the class of policies  $\theta^{(3L)}_{TD}(T)$ indexed by the frequency of information dissemination, $T$, and defined as follows. Note that these policies only require knowledge of $\boldsymbol{\zeta}(kT),k=0,1,\cdots.$

\textsl{During Frame $k$, i.e., over slots $\left\lbrace t:kT\leq t\leq kT+(T-1)\right\rbrace$
\begin{itemize}
\item If $\zeta_1(kT)=1$ serve queues in $\mathcal{C}_1$ until either $T$ slots elapse or all queues in $\mathcal{C}_1$ become empty, whichever occurs first.
\begin{itemize}
\item Simultaneously serve any non empty queues in $\mathcal{C}_3$.
\end{itemize}
\item Else, serve queues in $\mathcal{C}_2$, 
\begin{itemize}
\item if $\zeta_1(kT)=1$ but $\mathcal{C}_1$ became empty at some $t<kT+(T-1)$ and $\zeta_2(kT)=1$, or
\item if $\zeta_1(kT)=0$ and $\zeta_2(kT)=1$.
\end{itemize}
In both of the above cases, $\mathcal{C}_2$ is served either the frame ends or all queues in $\mathcal{C}_2$ become empty, whichever occurs first.
\item Else, serve non empty queues in $\mathcal{C}_3$ (if any).
\end{itemize}
}
$\theta^{(3L)}_{BU}(T)$ is defined as above, replacing $\mathcal{C}_3$ with $\mathcal{C}_1$ and vice versa.
\begin{rem}\label{Chapter05PropPsi3_1TIsTO}
$\theta^{(3L)}_{TD}(T)$ and $\theta^{(3L)}_{BU}(T)$ are both throughput-optimal. 
\end{rem}

\begin{pf}
The proof is available in Sec.~\ref{AppendixProofOfPropPsi3_1TIsTO} in the Appendix.
\end{pf}

Clearly, splicing these two policies yields a periodic state information-based policy for $5$-clique and hence, 4-clique, LAoCs. This policy is obviously throughput optimal.
%

\section{Simulation Results}\label{Chapter05SimulationResults}
\begin{figure*}[bth]
\begin{subfigure}{.475\textwidth}
\hspace{-2.0cm}
\centering
\includegraphics[height=6.50cm, width=10.50cm]{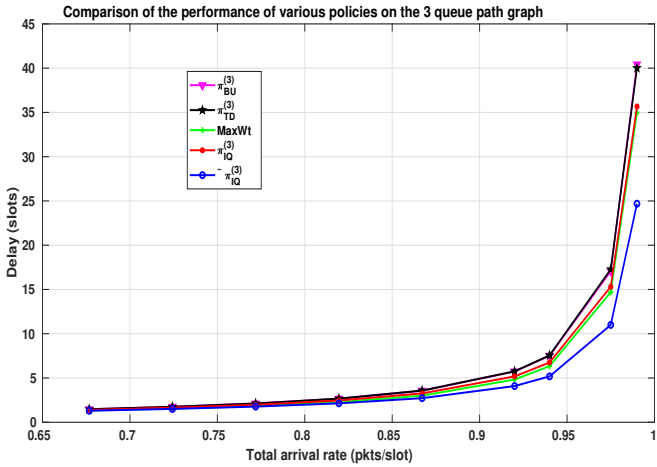}
\caption{Delay performance of the policies $\tilde{\pi}^{(3)}_{IQ},~\pi^{(3)}_{IQ}$, MaxWeight, $\pi^{(3)}_{TD}$ and $~\pi^{(3)}_{BU}$ 
along the trajectory $\boldsymbol{\lambda}(s)=s\times[0.25, 0.74, 0.25],~s\in[0,1],$ in the capacity region $\Lambda_3$.} 
\label{Chapter03FigBernoulliArrivalsComparingAllPoliciesMiddleQueueMostLoaded}
\end{subfigure}
\qquad
\begin{subfigure}{.475\textwidth}
\hspace{0.50cm}
\centering
\includegraphics[height=6.50cm, width=10.50cm]{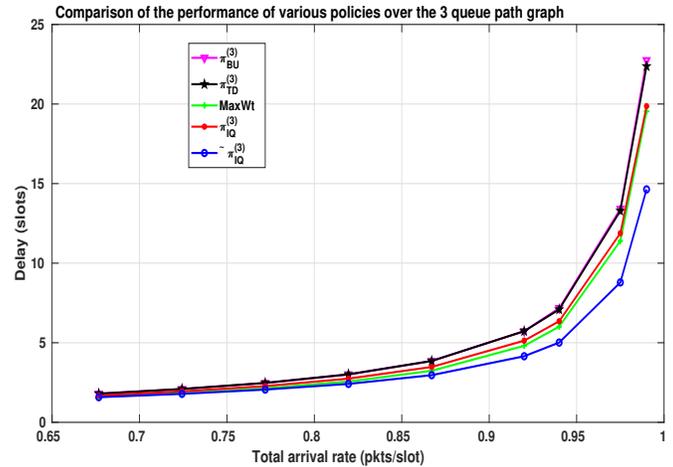}
\caption{Delay performance of the policies $\tilde{\pi}^{(3)}_{IQ},~\pi^{(3)}_{IQ}$, MaxWeight, $~\pi^{(3)}_{TD}$ and $~\pi^{(3)}_{BU}$ 
along the trajectory $\boldsymbol{\lambda}(s)=s\times[0.74, 0.25,0.74],~s\in[0,1]$, in the capacity region $\Lambda_3$.} 
\label{Chapter03FigBernoulliArrivalsComparingAllPoliciesOuterQueuesMostLoaded}
\end{subfigure}
\qquad
\caption{Simulation results for the path-graph network with $N=3$ for \emph{\color{blue}Bernoulli} packet arrival processes. The mean delay performances of all deterministic policies discussed in Sec.~\ref{secTOSchedulingWith3Queues} are shown in Figures (a) and (b), and compared with the MaxWeight scheduling policy \cite{tassiulas92stability}.}
\end{figure*}

In this section we numerically compare the performance of the various policies we have proposed and analysed in the preceding sections. To begin with, we simulate the mean delay performances of the policies for the path graph network with $N=3$, discussed in Sec.~\ref{secTOSchedulingWith3Queues} and compare them against the MaxWeight scheduling policy. To recapitulate, $\pi^{(3)}_{TD},\text{ and }\pi^{(3)}_{BU}$ are the Top-Down and Bottom-Up policies respectively, $\tilde{\pi}^{(3)}_{IQ}$ is the delay optimal policy defined in \ref{Chapter03DefinitionOfPolicyPi3DO} and $\pi^{(3)}_{IQ}$ is the throughput-optimal \emph{non-MSM} policy defined in \ref{Chapter03DefinitionOfPi3IQNonMSMButTO}. In every slot $t\geq0,$ MaxWeight simply serves Queues~1 and 3 if $Q_1(t)+Q_3(t)>Q_2(t)$ and Queue~2 otherwise. Obviously, this policy requires more state-information than any of the others. 
We simulate these policies when the arrival processes to the three queues are independent Bernoulli processes of rates $s\times[0.25,0.74,0.25],~s\in[0,1)$, i.e., the inner queue has a high arrival rate, and $s \times [0.74,0.25,0.74]$, i.e., the outer two queues have the high arrival rates.
The results are shown in
Figures~\ref{Chapter03FigBernoulliArrivalsComparingAllPoliciesMiddleQueueMostLoaded} and  \ref{Chapter03FigBernoulliArrivalsComparingAllPoliciesOuterQueuesMostLoaded}. As claimed in Thm.~\ref{Chapter03thmPi33isDelayOptimal}, $\tilde{\pi}^{(3)}_{IQ}$ performs best, showing in mean delay of up to 30\% less than MaxWeight near $s=1$ (in fact, the reduction in delay becomes more pronounced as $s$ approaches 1) and 38\% less than $\pi^{(3)}_{TD}$. Notice that in both plots MaxWeight does not perform as well as $\tilde{\pi}^{(3)}_{IQ}$ showing that it does not prioritize the middle queue frequently \enquote{enough.}

Moving on, although we have proved our stability results with Bernoulli arrival processes, we now provide a simulation study  which suggests that these results seem to hold for more general arrival processes. Also note that the stochastic ordering proof of Thm.~\ref{Chapter03thmPi33isDelayOptimal}, i.e., the delay optimality of policy $\tilde{\pi}^{(3)}_{IQ}$, being a sample path optimality argument, does not take into account the fact that the arrival processes to the queues are Bernoulli. It is, hence, equally valid for other types of arrival processes as well. Our simulations bear out this fact. Once again, we simulate our policies on  the 3-queue path-interference graph with Markovian arrival processes as described below.
\begin{figure}[tbh]
\centering
\includegraphics[height=3.750cm, width=9.50cm]{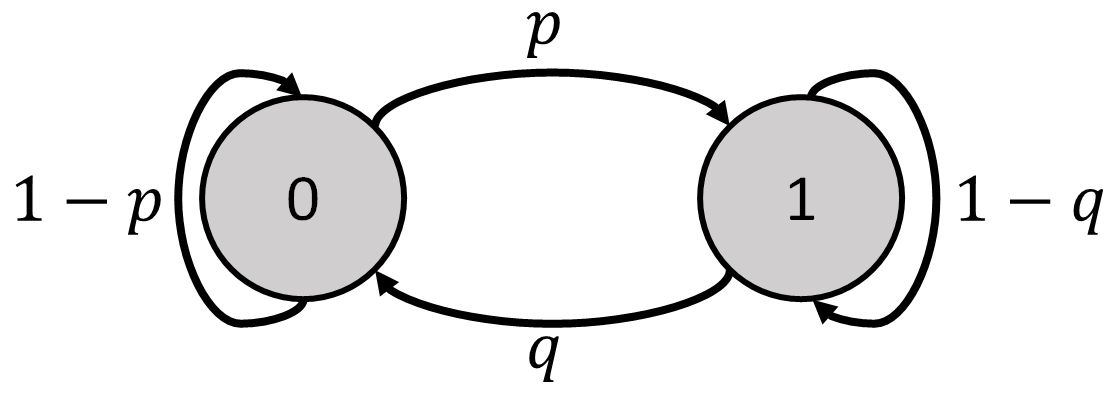}
\caption{The transition probability diagram of the Markovian arrival process. If, in slot $t$, the arrival process was in State~$0$, i.e., $A(t)=0,$ then in slot $t+1$, $A(t+1)=1$ with probability $p$ and $A(t+1)=0$ with probability $1-p$.} 
\label{Chapter03FigDTMCArrivalProcessTPDiagram}
\end{figure}
The arrivals to every queue form a two-state stationary discrete-time Markov chain (DTMC), i.e., $\left\lbrace A_i(t),~t\geq1\right\rbrace$ forms a DTMC. As before, $A_i(t)=1$ refers to the arrival of a packet into Queue~$i$ and $A_i(t)=0$ refers to no arrivals. Fig.~\ref{Chapter03FigDTMCArrivalProcessTPDiagram} shows the transition probability diagram of a generic two-state DTMC. For Queue~$j,$ the stationary probability of the arrival being in State $i,~i\in\{0,1\}$ is given by $\xi_{i,j};$ obviously, $\xi_{0,j}+\xi_{1,j}=1,~\forall j\in\{1,2,3\}.$ Suppose the transition probabilities of the process for Queue~$j$ are given by $P\{A_j(t+1)=1|A_j(t)=0\}=p_j$, and $P\{A_j(t+1)=0|A_j(t)=1\}=q_j.$ From basic Markov chain theory, we know that $\xi_{1,j}=\frac{p_j}{p_j+q_j},~\xi_{0,j}=\frac{q_j}{p_j+q_j}$ and for every $t\geq1$, $\mathbb{E}A_j(t)=\xi_{1,j}=\lambda_j.$ 
\begin{figure*}[tbh]
\begin{subfigure}{.475\textwidth}
\hspace{-1.9cm}
\includegraphics[height=6.50cm, width=10.50cm]{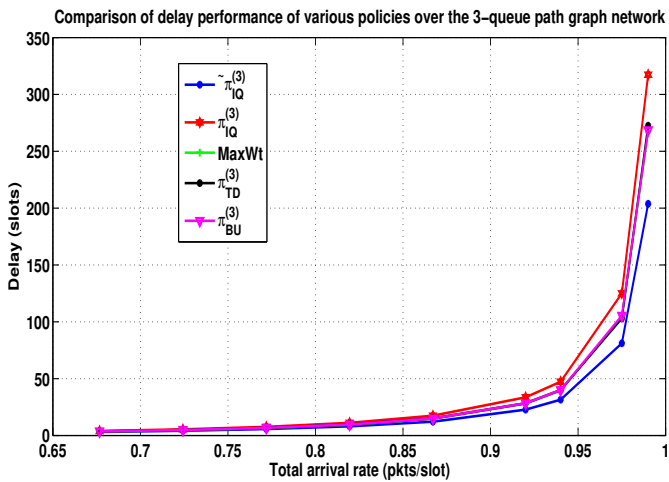}
\caption{Delay performance of the policies $\tilde{\pi}^{(3)}_{IQ},~\pi^{(3)}_{IQ}$, MaxWeight, $\pi^{(3)}_{TD}$ and $~\pi^{(3)}_{BU}$ 
along the trajectory $\boldsymbol{\lambda}(s)=s\times[0.25, 0.74, 0.25],~s\in[0,1],$ in the capacity region $\Lambda_3$.} 
\label{Chapter03FigDTMCArrivalsComparingAllPoliciesMiddleQueueMostLoaded}
\end{subfigure}
\qquad
\begin{subfigure}{.475\textwidth}
\hspace{0.50cm}
\includegraphics[height=6.50cm, width=10.50cm]{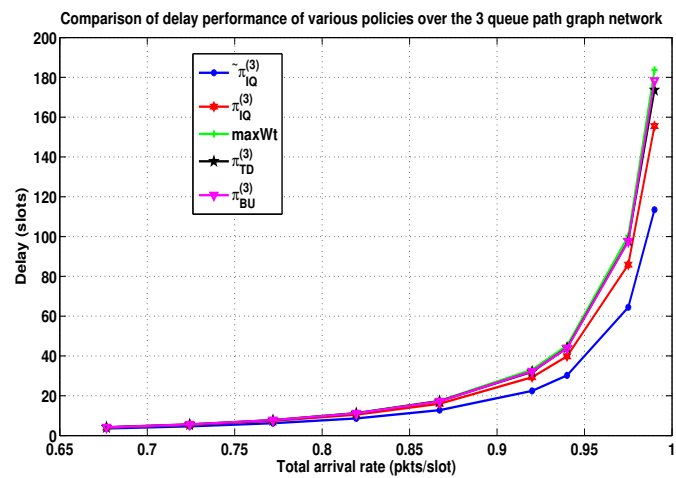}
\caption{Delay performance of the policies $\tilde{\pi}^{(3)}_{IQ},~\pi^{(3)}_{IQ}$, MaxWeight, $~\pi^{(3)}_{TD}$ and $~\pi^{(3)}_{BU}$ 
along the trajectory $\boldsymbol{\lambda}(s)=s\times[0.74, 0.25,0.74],~s\in[0,1]$, in the capacity region $\Lambda_3$.} 
\label{Chapter03FigDTMCArrivalsComparingAllPoliciesOuterQueuesMostLoaded}
\end{subfigure}
\qquad
\caption{Simulation results for the path-graph network with $N=3$ for \emph{\color{blue}Markovian} packet arrival processes. For all plots, and every $i\in\{1,2,3\}$ the transition probabilities of the arrival process (see Fig.~\ref{Chapter03FigDTMCArrivalProcessTPDiagram}) are chosen as follows $p_i=0.10$, and $q_i=(\frac{1}{\lambda_i}-1)p_i.$ 
}
\end{figure*}

\begin{figure}
\centering
\includegraphics[height=6.50cm, width=10.50cm]{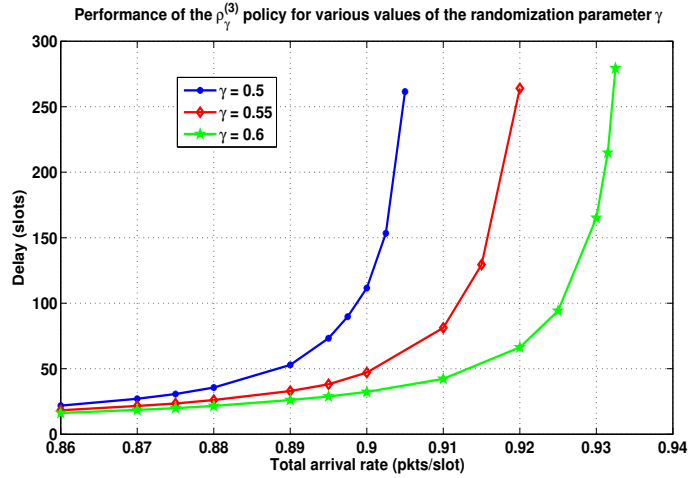}
\caption{Illustrating the loss of stability due to the "Flow-in-the-Middle" problem discussed in Sec.~\ref{Chapter03RandomizedPolicyFlowInTheMiddle}. We compare the delay performance of the policy $\rho^{(3)}_{\gamma}$ along the trajectory $\boldsymbol{\lambda}(s)=s\times[0.74, 0.25,0.74],~s\in[0,1]$, in the capacity region $\Lambda_3$. For every $s\leq1,$ the arrival rate vector lies within the interior of $\Lambda_3$ and is, hence, stabilizable. The policies can be seen to render the system unstable much before the system load parameter $s$ hits $1$.} 
\label{Chapter03FigDTMCArrivalsInstabilityOfRho3GammaBeyondHalf}
\end{figure}

Both plots (Figures~\ref{Chapter03FigDTMCArrivalsComparingAllPoliciesMiddleQueueMostLoaded} and \ref{Chapter03FigDTMCArrivalsComparingAllPoliciesOuterQueuesMostLoaded}) bear out the fact that even with non-Bernoulli arrivals, $\tilde{\pi}^{(3)}_{IQ}$ shows the best delay performance beating the closest competitor by at least $34\%.$  Again, in both plots we see that MaxWeight performs about just as  well as the Top-Down and Bottom-Up policies, suggesting that it does not prioritize the inner queue \enquote{enough.}
We now move on to the performance of the randomized policy $\rho^{(3)}_{\gamma},$ indexed by the randomization parameter $\gamma\in[0,1].$ In Sec.~\ref{Chapter03RandomizedPolicyFlowInTheMiddle} we derived an inner bound on the set of arrival rates that the policy can stabilize for a given $\gamma$, its stability region\footnote{See Sec.~\ref{secTheGeneralQueueSchedulingModel} for details.}, and showed that $\Lambda^{(3)}_{\gamma}\nearrow\Lambda_3$ as $\gamma\uparrow1.$ The plot in Fig.~\ref{Chapter03FigDTMCArrivalsInstabilityOfRho3GammaBeyondHalf}, simulated with $\gamma=0.5,~0.55,\text{ and }0.6$ help corroborate our analysis. However, the plots also suggest that the inner bound $\Lambda^{(3)}_{\gamma}$ is actually \emph{not very tight.} Further study is required to establish better bounds on this region. 

Moving on to larger path graphs, recall that in Sec.~\ref{Chapter04GeneralPSplicingItemized} we proposed a policy-splicing procedure to derive low delay QNB-MSM scheduling policies for path graphs with arbitrary number of queues. We demonstrate the performance of these policies in Table~\ref{table5QueuesSimulationsPiTilde5MWAndLOfMWAlpha}, where we compare our proposed policies with the benchmark MaxWeight ($MW$) and a third policy that is based on a popular scheduler called the \enquote{MaxWeight-$\alpha$} scheduler. This last policy, that we denote by $L(MW\alpha)$, is an MSM policy, obtained by using the operator $L$ (see Sec.~\ref{Chapter03SecTildePi3_IQ}) to project a modification of MaxWeight ($MW$) called $MW\alpha$ onto $\Gamma^{(N)}_M.$ The $MW\alpha$ policy, studied in \cite{stolyar04maxweight-generalized-switch-collapse} and \cite{shah-etal07heavy-traffic-optimal-scheduling-switched}, is essentially $MW$  with all queue lengths raised to their $\alpha$\textsuperscript{th} powers, with $\alpha>0$. This policy has been observed to show smaller sum queue lengths (than $MW$) with smaller $\alpha$ \cite{keslassy-mackoen01analysis-scheduling-full-throughput-switches}.

\begin{figure}[tb]
\centering
\includegraphics[height=2.5cm, width=6.75cm]{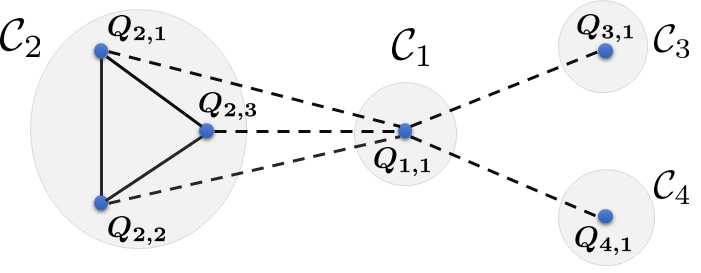}
\caption{The Star-of-Cliques (SoC) network used to study the performance of $\tilde{\phi} $. Simulation results are reported in Table~\ref{tableNonPathGraphSimulationsPhi3TildePhi3_1MWAndLOfMW}.}
\label{Chapter05FigNonPathNetworkForSimulations}
\end{figure}

The table shows that our proposed policies do outperform MaxWeight in all cases\footnote{The construction of the QNB-MSM policy in the first row which, by our nomenclature, is denoted $\tilde{\pi}^{(5)}$, is discussed in depth in \cite{mohan-etal18reduced-state-optimal-decentralized-TECHREPORT} and serves as a good example to illustrate the general splicing process discussed in \ref{Chapter04GeneralPSplicingItemized}.}. Note that the arrival rate vectors have not been shown in Table~\ref{table5QueuesSimulationsPiTilde5MWAndLOfMWAlpha} due to space constraints. We have reported the vectors in Sec.~\ref{AppendixSimulationDetails} of the Appendix. 
Recall that the analysis of throughput optimality was limited to $N=9$ queue systems. In Row~3, we perform the splicing procedure (Sec.~\ref{Chapter04GeneralPSplicingItemized}) to produce a QNB-MSM policy for a system with $N=15$ queues and show that it outperforms both the benchmark policies. Finally, Row~1 of the table shows an arrival rate vector for which our proposed queue length-agnostic policy does worse and $L(MW\alpha)$ shows the smallest sum queue length. In light of Thm.~\ref{Chapter04ThmNonexistenceOfDOPoliciesForNGreaterThan4}, this should not be entirely surprising. Moreover, the loss in performance is small.\\
%
\begin{table}[tb]
\centering 
\footnotesize
\begin{tabular}{c | c | c | c} 
\hline\hline 
& \multicolumn{3}{c}{Mean sum queue length (packets)} \\ 
& \multicolumn{3}{c}{Bernoulli arrivals} \\ \cline{2-4}
Number of  & QNB-MSM & MaxWeight & $L(MW\alpha)$\\ 
queues ($N$) & $\left(\tilde{\pi}^{(N)}\right)$ & &\\[0.5ex]
\hline 
\\
4  & 45.963 & 57.302 & \textbf{\color{blue}43.508} \\
5  & \textbf{\color{blue}61.537} & 88.243 & 75.642 \\ 
15 & \textbf{\color{blue}76.72} & 107.88 & 92.100\\  \\
\hline 

\end{tabular}
\caption{\footnotesize Path graph interference models with $N=4, 5$ and $15$. Comparison of sum queue length with Bernoulli arrivals, under the proposed QNB MSM policies with MaxWeight and $L(MW\alpha)$ with $\alpha=0.01$. Details about the arrival rate vectors can be found in Sec.~\ref{AppendixSimulationDetails}  of the Appendix.}
\label{table5QueuesSimulationsPiTilde5MWAndLOfMWAlpha} 
\end{table}
We move on to simulations of the policies proposed for the second class of conflict graphs discussed in this article, namely, Cluster-of-cliques graphs. The first, shown in Fig.~\ref{Chapter05FigNonPathNetworkForSimulations}, is a \emph{Star-of-Cliques} (SoC) networks comprising 4 cliques and a total of 6 queues. The second network is the LAoC network shown in Fig.~\ref{Chapter05FigClustersOfCollocatedNetworksInterferenceGraph}. It consists of 4 cliques and a total of 9 queues. 
Table~\ref{tableNonPathGraphSimulationsPhi3TildePhi3_1MWAndLOfMW} shows the result of simulating $\tilde{\phi}$, $\tilde{\theta}^{(5L)}$ (the projected version of $\theta^{(5L)}_{SP}$, defined in Sec.~\ref{Chapter05SecSchedulingInLAoC}) and $MW$ on these networks. We see that the proposed policies consistently perform better than the benchmarks. 

The result for the SoC networks is, in particular, quite interesting, since one expects that situations may arise wherein only two of the three peripheral cliques and $\mathcal{C}_1$ are nonempty. In such a case, $\tilde{\phi}$ would serve $\mathcal{C}_1$, giving up the chance to serve both the peripheral nonempty cliques simultaneously and remove 2 packets from the system in a single slot, which is what $MW$ might have attempted, if the queues therein were large enough. 
If, for example, in some slot $t$, $\mathcal{C}_2$ is empty, while $Q_{1,1}(t)=1$, $Q_{3,1}(t)=5$ and $Q_{4,1}(t)=2$, $\tilde{\phi} $ still serves only $Q_{1,1}$ (1 packet transmitted) while MaxWeight serves both $Q_{3,1}$ and $Q_{4,1}$ (2 packets transmitted). Why $\tilde{\phi}$ still performs better requires more investigation and will be a focus of our future work. 

\begin{table}[tb]
\footnotesize
\centering 
\begin{tabular}{c | c | c} 
\hline\hline 
Cluster of Cliques & \multicolumn{2}{c}{Mean sum queue length (packets)} \\ \cline{2-3}
Network & QNB-MSM & MaxWeight \\ [0.5ex]  
\hline 
\\
Star of Cliques (Fig.~\ref{Chapter05FigNonPathNetworkForSimulations}) & \textbf{\color{blue}45.535} & 57.861\\ 
Linear Array of Cliques (Fig.~\ref{Chapter05FigClustersOfCollocatedNetworksInterferenceGraph}) & \textbf{\color{blue}245.038} & 309.450\\ \\
\hline 

\end{tabular}
\caption{Comparison of sum queue length under the proposed Cluster of Cliques policies, and MaxWeight acting on the networks in Fig.~\ref{Chapter05FigClustersOfCollocatedNetworksInterferenceGraph} and Fig.~\ref{Chapter05FigNonPathNetworkForSimulations}. Details about the arrival rate vectors can be found in Sec.~\ref{AppendixSimulationDetails} of the Appendix.} 
\label{tableNonPathGraphSimulationsPhi3TildePhi3_1MWAndLOfMW} 
\end{table}


\section{Conclusion and Future work}

In this paper, we began by studying the scheduling of transmissions over a class of non collocated interference networks that we called \enquote{path-graph interference networks.} We provided sufficient conditions for queue nonemptiness based (QNB) policies to be throughput-optimal over these networks. We then provided a complete characterization of the class of MSM-QNB policies on path-graphs with 3 queues and showed that it contains stable, delay-optimal and even unstable policies. 

We then saw how priority policies for smaller path-graphs can be combined to construct QNB policies for larger networks. Next, we showed that policies so constructed are not MSM, but can be made MSM using a projection operator. 
We also showed how the delay properties of these MSM policies can be further improved by using certain observations of the nature of scheduling policies in $\tilde{\Pi}^{(N)}$.
We then showed that there cannot exist QNB policies that are uniformly delay optimal over the entire capacity region, for any path graph network with $N\geq4$ links. 

Motivated by wireless networks commonly used for IoT-type applications, we introduced a new class of interference networks, called the \enquote{Cluster-of-Cliques} networks and studied two subclasses, namely, the \emph{Star-of-Cliques} and the \emph{Linear-Arrays-of-Cliques} networks. We then constructed QNB scheduling policies for both these classes and studied their stability and delay properties. We showed how the minislot structure can be used to implement these policies in a decentralized manner, and also developed a protocol that requires no explicit exchange of even occupancy information and proved that it is throughput-optimal. 
Our simulation results showed that the QNB policies we have developed, in fact, perform better than existing scheduling policies that require complete knowledge of the system backlog in every slot.

In short, MaxWeight and policies based on it (such as MaxWeight-$\alpha$) have been known to suffer from two major implementation issues, namely (i) disseminating queue length information across the network (or reporting it to some centralized scheduling entity), and (ii) finding the maximum weight independent set (MWIS), which for general conflict graphs, is famously an NP-hard problem. However, in the context of the current article, the latter problem is simplified. In fact, there exist dynamic programming approaches to solve the MWIS problem in linear time for path graphs. The outstanding issue in computing schedules, therefore, is one of information dissemination. Our work provides rigorous theoretical evidence that suggests that once the MWIS problem is simplified, detailed queue length information is (almost) irrelevant. 

Future work will include extending these throughput-optimality results to non-Bernoulli arrival processes, obtaining better bounds on the stability region of the policy $\rho^{(3)}_{\gamma}$ and proving the throughput-optimality of the Top-Down and Bottom-Up policies for general $N$-queue path graph networks. We would like to find techniques to encode occupancy information $\boldsymbol{\zeta}(t)$ using fewer than $\mathcal{N}$ bits, getting as close as possible to $\log_2(\mathcal{N})$ bits, albeit in a decentralized manner. Finally, we would also like to explore such reduced state information based scheduling policies for more general conflict graphs and the existence of graphs that do not permit stable QNB scheduling policies.



\section{Appendix}\label{Chapter05SecAppendix}

\subsection{Glossary of Acronyms}\label{glossaryOfAbbreviations}
\begin{itemize}
	\item BU: Bottom-up
	\item CoC: Cluster-of-Cliques
	\item CSMA: Carrier Sense Multiple Access
	\item D.O.: Uniformly Delay Optimal (see Defn.~\ref{Chapter03DefnUniformlyDelayOptimal}).	
	\item IID: Independent and Identically Distributed
	\item IoT: Internet of Things
	\item LAOC: Linear-Array-of-Cliques
	\item MAC: Medium Access Control
	\item MSM: Maximum Size Matching
	\item $MW:$ The MaxWeight algorithm defined in \cite{tassiulas92stability}.
	\item $MW\alpha:$ The MaxWeight-$\alpha$ algorithm defined in \cite{stolyar04maxweight-generalized-switch-collapse}	
	\item MWIS: Maximum Weight Independent Set
	\item QNB: Queue Nonemptiness Based (policy)
	\item SoC: Star-of-Cliques
	\item T.O.: Throughput Optimal
	\item TD: Top-down.
\end{itemize}

\subsection{Glossary of Notation}\label{glossaryOfNotation}
\label{glossaryOfNotationForPart2}
\begin{enumerate}
	\item $\mathbb{I}_{\left\lbrace\text{\textsc{condition}}\right\rbrace}$: the indicator function, which evaluates to $1$ whenever \textsc{condition} is true, and $0$ otherwise.
	\item $\boldsymbol{\zeta}(t):$ the occupancy vector or nonemptiness vector at time $t$, defined as $\boldsymbol{\zeta}(t)=\left[\mathbb{I}_{\left\lbrace Q_1(t)>0\right\rbrace},\cdots,\mathbb{I}_{\left\lbrace Q_N(t)>0\right\rbrace}\right]$.
	\item $\mathcal{V}:$ is the set of all activation vectors. Clearly, in a system of $N$ queues, $\mathcal{V}\subsetneq\{0,1\}^N$ due to interference constraints.
	\item $\Lambda_N:$ The capacity region of a path graph interference network with $N$ queues (communication links).
	\item $\sigma\left(X\right):$ The sigma algebra generated by the random variable $X.$
	\item $\Pi^{(N)}$: the class of all scheduling policies defined on path graphs.
	\item $\Gamma^{(N)}_M:$ the class of all Maximum Size Matching (MSM) policies. 
	\item $\Pi^{(N)}_{M}$: the class of all policies that take only the occupancy vector $\boldsymbol{\zeta}(t)$ as input and activate the largest number of non empty queues in every slot, .i.e., MSM policies that require only the empty or nonempty status of the queues in the network.
	\item $\tilde{\Pi}^{(N)}$: the class of all MSM policies within $\Pi^{(N)}_{M}$ that additionally break ties in favour of inner queues (see condition~\ref{Chapter03ConditionPrioritizeInnerQueuesPiTilde}).
	\item $\mid A \mid:$ represents the cardinality of set $A.$
	\item $Geo(\lambda)/Geo(\mu)/1$ queue: A queue with a Bernoulli arrival process whose interarrival periods are geometrically distributed with mean $\lambda,$ and whose service times are IID and geometrically distributed with mean $\mu.$
	\item $\pi^{(N)}_{TD}\text{ and }\pi^{(N)}_{BU}:$ Top-Down and Bottom-Up policies for path graph networks with $N$ queues.
	\item $\pi^{(2N-1)}_{SP}:$ The policy obtained by \emph{splicing} $\pi^{(N)}_{TD}\text{ and }\pi^{(N)}_{BU}.$ This policy is not MSM.
	\item $\{\tilde{\pi}^{(4)}_i,1\leq i\leq4\}:$ These are the four policies within the class $\tilde{\Pi}^{(4)}$ that are both MSM and break ties in favor of the inner queues, i.e., Queues~2 and 3.
	\item $\pi^{(4)}_{TI}:$ The policy on 4-queue path graphs obtained by splicing $\pi^{(3)}_{TD}$ and $\pi^{(3)}_{IQ}$, the Top-Down and Bottom-Up policies on 3-queue path graphs. Since it is not obtained by splicing Top-Down and Bottom-Up policies, we do not give it the subscript \enquote{$SP$.}
	
	\item $\mathcal{N}:$ The total number of queues in an Linear Array of Cliques (LAoC) or Star-of-Cliques (SoC) network.
	\item $\phi^{(S)}_{IC}:$ The policy defined on Star-of-Cliques (SoC) networks that prioritizes the inner clique over all the peripheral cliques. It is defined in Sec.~\ref{Chapter05DefinitionOfPhi3_5}.
	\item $\tilde{\phi}^{(S)}_{IC}$ The \enquote{$S$} in the superscript stands for SoC network, and \enquote{$IC$} in the subscript shows that they prioritize the \emph{inner clique,} i.e., $\mathcal{C}_1.$  It is defined in Sec.~\ref{Chapter05DefinitionOfPhi3Tilde}.
	\item $\theta^{(3L)}_{TD}$ and $\theta^{(3L)}_{BU}:$ Top-Down and Bottom-Up policies defined over LAoC networks with 3 cliques. Here, the \enquote{$L$} in the superscript stands for LAoC network.
	\begin{rem} 
		Throughout Sec.~\ref{Chapter05SecTOSchedulingPolicies}, $\phi$ will always represent a policy for SoC networks and $\theta$ for LAoC network.
	\end{rem}
	\item $\phi^{(S)}_{IC}(T):$ The version of $\phi^{(S)}_{IC}$ defined, once again for SoC networks, that requires knowledge of the vector $\boldsymbol{\zeta}$ only every $T$ time slots.
	\item $\phi^{(S)}_{CS}:$ that, like the QZMAC protocol in \cite{mohan-etal16hybrid-macsMASSversion}, takes scheduling decisions based solely on the information gathered by sensing the channel for activity. Obviously, the \enquote{CS} in the subscript stands for channel sensing.
	
\end{enumerate}

\subsection{Throughput Optimality of Queue Nonemptiness-based Scheduling in Fully Connected Graphs}\label{AppendixTOFullyConnected}
The proof of Thm.~\ref{lemPropertyPMeansTO}, i.e., throughput optimality of policies satisfying property $\mathcal{P}$, proceeds via a Lyapunov argument. The sole purpose of this this subsection is to provide some intuition to the reader about how we came to construct the Lyapunov function used therein. This subsection, therefore, is not necessary to understand the proof and may be skipped without loss of continuity. 

Consider stabilizing a collocated network of $N$ queues described by a \emph{Fully Connected} interference graph. Here the capacity region consists of all rate vectors $\boldsymbol{\lambda}\in\mathbb{R}_+^N$ that satisfy $\sum_{i=1}^N\lambda_i<1$. Any policy that schedules a nonempty queue in every slot (if there exists one) is T.O. To see this, define $Q(t):=\sum_{i=1}^NQ_i(t)$, $A(t+1)=\sum_{i=1}^NA_i(t+1)$ and $D(t):=\sum_{i=1}^ND_i(t)$ and consider 
the Lyapunov Function 
\begin{equation}
L(\mathbf{Q}(t)):=\left(\sum_{i=1}^NQ_i(t)\right)^2=Q^2(t).
\label{eqnLyapunovFullyConnected}
\end{equation}
Using the fact that for any three non negative reals $x,y,z$, $((x-y)^++z)^2\leq x^2+y^2+z^2-2x(y-z)$, we see that
\begin{equation*}
Q^2(t+1)\leq Q^2(t)+D^2(t)+A^2(t)-2Q(t)\left(D(t)-A(t)\right).
\end{equation*}
Hence, the expected single slot drift becomes
\begin{eqnarray}
&&\mathbb{E}\bigg[L(\mathbf{Q}(t+1))-L(\mathbf{Q}(t))\mid\mathbf{Q}(t)=\mathbf{q}\bigg]\nonumber\\
&&\stackrel{\dagger1}{\leq} \mathbb{E}\bigg[D^2(t)+A^2(t)\mid \mathbf{q}\bigg]- 2q\mathbb{E}\bigg[D(t)- A(t)\mid \mathbf{q}\bigg]\nonumber\\
&&\stackrel{\dagger2}{=} 1+N^2-2q\left(\mathbb{E}\bigg[D(t)\mid \mathbf{q}\bigg]- \sum_{i=1}^N\lambda_i\right),
\label{eqnExpectedDriftFullyConnected}
\end{eqnarray}
where in $\dagger1$, $\mathbf{q}=[q_1,\cdots,q_N]$, and $q=\sum_{i=1}^Nq_i$. In $\dagger2$ we have used the fact that $D(t)\leq 1$ since at most one queue can be served per slot, $A(t)\leq N$ since at most one packet can arrive in each of the $N$ queues in a slot, and because arrivals are independent of the current system state, $\mathbb{E}\bigg[A(t)\mid q\bigg]=\mathbb{E}A(t)=\sum_{i=1}^N\lambda_i$. Since the policy schedules a non empty queue in every slot,
\begin{eqnarray}
\mathbb{E}\bigg[D(t)\mid q\bigg]&=&\mathbb{I}_{\{Q(t)>0\}},\text{ i.e., }\nonumber\\
\mathbb{E}\bigg[\sum_{i=1}^ND_i(t)\mid q\bigg]&=&\mathbb{I}_{\{\sum_{i=1}^NQ_i(t)>0\}}
\label{eqnConditionalExpectationToIndicator}
\end{eqnarray}
Taking expectations on both sides of \eqref{eqnExpectedDriftFullyConnected}, we see that 
\begin{eqnarray*}
\mathbb{E}Q^2(t+1)-\mathbb{E}Q^2(t)\leq 1+N^2-2\mathbb{E}\left(Q(t)\mathbb{I}_{\{Q(t)>0\}}\right)\\
-\mathbb{E}Q(t)\left(\sum_{i=1}^N\lambda_i\right),
\end{eqnarray*}
 and since $Q(t)\geq0,~\forall t,$ $\mathbb{E}\left(Q(t)\mathbb{I}_{\{Q(t)>0\}}\right)=\mathbb{E}Q(t)$. Setting $\epsilon=1-\sum_{i=1}^N\lambda_i$ ($\epsilon>0$ by stability considerations) we get
 \begin{eqnarray}
 \mathbb{E}Q^2(t+1)-\mathbb{E}Q^2(t)\leq 1+N^2-2\epsilon\mathbb{E}Q(t).
 \end{eqnarray}
Summing the above over $t=0,1,\cdots,T-1,$ we get
\begin{eqnarray}
 \mathbb{E}Q^2(T)-\mathbb{E}Q^2(0)\leq T(1+N^2)-2\epsilon\sum_{t=0}^{T-1}\mathbb{E}Q(t).
\end{eqnarray}
A little bit of algebra shows that 
\begin{eqnarray}
\frac{1}{T}\sum_{t=0}^{T-1}\mathbb{E}Q(t)&\leq& \frac{1+N^2}{2\epsilon} + \frac{\mathbb{E}Q^2(0)}{2\epsilon T}\nonumber\\
\Rightarrow \limsup_{T\rightarrow\infty}\frac{1}{T}\sum_{t=0}^{T-1}\mathbb{E}Q(t)&<&\infty,\nonumber
\end{eqnarray} 
which implies strong stability. In the sequel, we will call this rather standard technique \cite{neely10stochastic-network-optimization-book} of showing strong stability, the \emph{telescoping sum method.} As a precursor to the proof of Lem.~\ref{lemPropertyPMeansTO}, observe that this class of policies satisfies property $\mathcal{P}$ in the lemma, i.e., \eqref{eqnPropertyP} since $\mathbb{E}\bigg[D(t)\mid q\bigg]=\mathbb{I}_{\{Q(t)>0\}}$ means that $\sum_{i=1}^ND_i(t)=0\iff\sum_{i=1}^NQ_i(t)=0$. 
\qed
\subsection{Proof of Thm.~\ref{thmPi3_1IsTO}}\label{AppendixProofOfPi3_1IsTO}
Queues 1 and 2 form a priority queue and $\pi^{(3)}_{TD}$ serves the pair of queues 1 and 2 whenever either of them is nonempty. So, $\pi^{(3)}_{TD}$ satisfies Property $\mathcal{P}$ for $i=1$ (specifically, Eqn.~\eqref{eqnPropertyP}), which means that the process $\left\lbrace[Q_1(t),Q_2(t)],~t\geq0\right\rbrace$, is strongly stable. Further, since Queue 1 receives the highest priority, as soon as a packet arrives it is served and leaves the system at the end of the slot. Consequently, Queue 1 behaves like a $Geo(\lambda_1)/D/1$ queue, with service time being exactly one slot. This also means that by starting out with $Q_1(0)\leq 1$, in any time slot, Queue 1 has at most 1 packet, which is the arrival during that slot, i.e., $Q_i(t)=A_i(t),~\forall t\geq1.$ Also, $P\{Q_i(t)>0\}=\lambda_1.$

Queues~1 and 2 form a priority queueing system. This means that the packet at the Head of Line (HOL) position in Queue~2 is served whenever $Q_1(t)=0$. Since $Q_1(t)=A_1(t),$ $P\{Q_1(t)=0\}=1-\lambda_1$ independently of $Q_2(t)$. Moreover, the arrivals to Queue~1 are Bernoulli with mean $\lambda_1$, which means that 
the service time $B_2$ of every packet in Queue 2 is IID and geometrically distributed with mean $\frac{1}{1-\lambda_1}.$ Specifically, $B_2\sim Geo(\frac{1}{1-\lambda_1})$. This means that Queue 2 behaves like a (refer glossary \ref{glossaryOfNotationForPart2} for an explanation of this notation) $Geo(\lambda_1)/Geo(\frac{1}{1-\lambda_1})/1$ queue, and 
since $\lambda_2<(1-\lambda_1),$ Queue 2 is stable. Furthermore, $\{[Q_1(t),Q_2(t)],t\geq0\}$ forms an aperiodic, irreducible positive recurrent DTMC.
This means that there exists a steady state probability measure on Queue 2's backlog such that, 
\begin{eqnarray}
\lim_{t\rightarrow\infty}P\{Q_2(t)=0\}&=&1-\lambda_2\mathbb{E}B_2\nonumber\\
&=&1-\frac{\lambda_2}{1-\lambda_1}.
\label{eqnSteadyStateProbabilityQ2IsEmpty}
\end{eqnarray}
From the definition of $\pi^{(3)}_{TD}$ we see that Queue 3 is scheduled for service whenever either Queue 1 is nonempty or when \emph{both} Queue 1 and Queue 2 are empty. Specifically, the choice of the activation set is completely governed by the backlogs of queues 1 and 2 and does not depend on Queue 3 at all. This means the service given to Queue 3 in every slot is independent of its backlog in that slot. 
Suppose we begin both Queue 1 and Queue 2 in their steady state distributions, 
\begin{small}
\begin{eqnarray}
\hspace{-0.750cm}
P\{S_3(t)=1\}&=& P\{Q_1(t)>0\}+P\{Q_1(t)=0, Q_2(t)=0\}\nonumber\\
&=&\lambda_1+P\{Q_2(t)=0\}P\{Q_1(t)=0\mid Q_2(t)=0\}\nonumber\\
&\stackrel{*1}{=}& \lambda_1+\left(1-\frac{\lambda_2}{1-\lambda_1}\right)P\{A_1(t)=0\mid Q_2(t)=0\}\nonumber\\
&\stackrel{*2}{=}& \lambda_1+\left(1-\frac{\lambda_2}{1-\lambda_1}\right)(1-\lambda_1)\nonumber\\
&=& 1-\lambda_2
>\lambda_3.
\end{eqnarray}
\end{small}
Equality $*1$ uses Eqn.~\eqref{eqnSteadyStateProbabilityQ2IsEmpty} and $*2$ uses the fact that external arrivals to Queue 2 in a slot are independent of the backlog of Queue 2 in that slot. To show that Queue 3 is strongly stable, define $L:\mathbb{N}\rightarrow\mathbb{R}_+$ as
$L(Q_3(t))=Q^2_3(t).$

\begin{small}
\begin{eqnarray}
\hspace{-0.80cm}
\mathbb{E}\bigg[L(t+1)-L(t)\bigg]&=&\mathbb{E}[Q^2_3(t+1)-Q^2_3(t)]\nonumber\\
&=&\mathbb{E}[\left(\left(Q_3(t)-S_3(t)\right)^++A_3(t+1)\right)^2-Q^2_3(t)]\nonumber\\
&\stackrel{\star3}{\leq}& 2-2\mathbb{E}Q_3(t)\mathbb{E}[S_3(t)-A_3(t+1)]\nonumber\\
&=&2-2\mathbb{E}Q_3(t)\left(1-\lambda_2-\lambda_3\right)\nonumber\\
&\stackrel{\star4}{=}&2-2\delta\mathbb{E}Q_3(t)
\end{eqnarray}
\end{small}
In $\star3$, we have once again used the fact that for any four non negative reals $w,x,y\text{ and }z,$ with $w\leq(x-y)^++z,$ $w^2\leq x^2+y^2+z^2-2x(y-z)$, with $w=Q_3(t+1), x=Q_3(t), y=S_3(t)\text{ and }z=A_3(t+1)$. Further, $S_3(t)\leq 1$ and $A_3(t+1)\leq 1.$ Finally, in $\star4,$ $\delta=1-\lambda_2-\lambda_3>0$ by capacity constraints. This shows that Queue 3 is also strongly stable, and since 
\begin{eqnarray*}
&&\limsup_{T\rightarrow\infty}\frac{1}{T}\sum_{t=0}^{t-1}\sum_{i=1}^3\mathbb{E}_{\pi^{(3)}_{TD}} Q_i(t)\\
&&\leq \limsup_{T\rightarrow\infty}\frac{1}{T}\sum_{t=0}^{t-1}\mathbb{E}_{\pi^{(3)}_{TD}} Q_1(t) +\limsup_{T\rightarrow\infty}\frac{1}{T}\sum_{t=0}^{t-1}\mathbb{E}_{\pi^{(3)}_{TD}} Q_2(t)\\
&&+\limsup_{T\rightarrow\infty}\frac{1}{T}\sum_{t=0}^{t-1}\mathbb{E}_{\pi^{(3)}_{TD}}Q_3(t),
\end{eqnarray*}
the system is also strongly stable under this policy. The other policy $\pi^{(3)}_{BU}$ simply swaps the priorities of Queues 1 and 3 in the enumeration \ref{enumerateDecisionTreeP3_1}, and its proof proceeds as before, mutatis mutandis. 
\qed
\subsection{Proof of Thm.~\ref{thmPi3_3IsTO}}\label{AppendixProofOfPi3_3IsTO}
Since, by definition, for any Queue $i,$ $D_i(t)=S_i(t)\mathbb{I}_{\{Q_i(t)>0\}}$, $Q_1(t)+Q_2(t)=0$ always means $D_1(t)+D_2(t)=0$. To show the converse, we consider several cases 
\begin{itemize}
\item $Q_1(t)>0$ and $Q_2(t)>0$ means that either in step 2 or 3 of MSM, on of these queues will get scheduled and either $D_1(t)=1$ or $D_2(t)=1$.
\item $Q_1(t)=0$ and $Q_2(t)>0$ means that MSM schedules Queue 2 in step 2, and $D_2(t)=1$. 
\item $Q_2(t)=0$ and $Q_1(t)>0$ means that MSM schedules Queue 1 in either step 1 or step 3, depending on the length of Queue $3$, whereby $D_1(t)=1$. 
\end{itemize} 
Following the same logic, we state a similar result for $D_2(t)+D_3(t)$. This means that $\tilde{\pi}^{(3)}_{IQ}$ satisfies property $\mathcal{P}$, and from Lem.~\ref{lemPropertyPMeansTO}, $\tilde{\pi}^{(3)}_{IQ}$ is T.O.
\qed
\subsection{Proof of Thm.~\ref{Chapter03thmPi33isDelayOptimal}}\label{AppendixProofOfPi33isDelayOptimal}
We adapt the technique used in Lem.~4.2 of \cite{tassiulas-ephremides94dynamic-scheduling-tandem-parallel} to prove this result\footnote{The reader should note that there is a \emph{typo} in \cite{tassiulas-ephremides94dynamic-scheduling-tandem-parallel} that labels two results as Lem~4.1. Here, we refer to the latter as 4.2 to avoid confusion.}. The main idea is to construct a sequence $\{\pi'_k,k\geq0\}$ of intermediate policies such that the backlog in every queue converges sample path-wise to that of $\tilde{\pi}^{(3)}_{IQ}$ which means that for every $t\geq0,$
\begin{equation}
\lim_{k\rightarrow\infty}Q^{\pi'_{k}}_l(t)=Q^{\tilde{\pi}^{(3)}_{IQ}}_l(t),~1\leq l\leq3,
\end{equation}
\emph{over every sample path}. Each policy in the sequence $\pi'_k$ is designed to provide smaller sum queue length than its predecessor $\pi'_{k-1}$ and the chosen policy $\pi.$ Towards this end, we first couple arrivals to the systems on which $\tilde{\pi}^{(3)}_{IQ}$, $\left\lbrace \pi'_k,~k\geq0\right\rbrace$ and $\pi$ act (by assumption, the initial conditions are equal, i.e., on every sample path $\mathbf{Q}^\pi(0)= \mathbf{Q}^{\tilde{\pi}^{(3)}_{IQ}}(0)= \mathbf{Q}^{{\pi}'_k}(0)=\mathbf{Q},~\forall k\geq0$, where $\mathbf{Q}\in\mathbb{N}^3$ is some generic queue length vector). We then define $\pi'_0$ as follows. 
In slot 0, $\pi'_0$ follows $\tilde{\pi}^{(3)}_{IQ}$ which means the activation vectors chosen by the two policies are the same. In other words, $\mathbf{s}^{{\pi}'_0}(0)=\mathbf{s}^{\tilde{\pi}^{(3)}_{IQ}}(0),$ with the superscripts denoting the policy. We now show that at $t=1,$ the following conditions are satisfied by $\mathbf{Q}^{\pi'_0}(1)$ and $\mathbf{Q}^\pi(1)$. In Condition~\ref{Chapter03condnAllQueuesNonempty} below, the indices $j_k$ are as  defined in Lem.~\ref{Chapter03lemSufficientForActivationMSM}.
\begin{enumerate}\label{condnPi33DelayOptimalInductionHypothesis}
\item $Q^{\pi'_0}_l(t)\leq Q^\pi_l(t)+1$ for $1\leq l\leq N.$ Here, $N=3.$
\item\label{condnDONonEmptyLessThanN} If $Q^{\pi'_0}_l(t)=Q^\pi_l(t)+1$ and $l<N$ then $Q^{\pi'_0}_{l+1}(t)=Q^\pi_{l+1}(t)-1$.
\item\label{condnDONonEmptyMoreThan1} If $Q^{\pi'_0}_l(t)=Q^\pi_l(t)+1$ and $l>1$ then $Q^{\pi'_0}_{l-1}(t)=Q^\pi_{l-1}(t)-1$.
\item\label{Chapter03condnAllQueuesNonempty} If $j_1=1$, $j_2=N$ and $N$ is odd, then $Q^{\pi'_0}_1(t)\leq Q^\pi_1(t)$ and $Q^{\pi'_0}_3(t)\leq Q^\pi_3(t)$.
\end{enumerate}
The first condition is obvious since at most one packet can depart from a queue in a slot and since arrivals to the two systems are coupled. For the second condition observe that since arrivals are coupled, $Q^{\pi'_0}_l(1)=Q^\pi_l(1)+1\Rightarrow Q^{\pi'_0}_l(0)=Q^\pi_l(0)=Q_l\neq0,$ and $s^\pi_l(0)=1$ while $s^{\pi'_0}_l(0)=0$, i.e., Queue $l$ was not empty at 0, and $\pi$ served it while $\pi'_0$ did not. This is because if either the queue was empty at time 0 or both the policies served it, the $1$ packet mismatch would never have occurred. We now have two cases.
\begin{itemize}
\item $l=1\Rightarrow$ $Q_2(0)\neq0\text{, and, }Q_3(0)=0$, since by the definition, $\tilde{\pi}^{(3)}_{IQ}$ ignores the extreme queues when they are nonempty only when Queue 2 is nonempty and $Q_1(t)\cdot Q_3(t)=0.$ Hence, $\pi'_0$ serves Queue 2 in slot 0 while $\pi$ does not, resulting in $Q^{\pi'_1}_2(0)=Q^\pi_2(0)-1.$
\item $l=2\Rightarrow$ $Q_1(0)\neq0$ and $Q_3(0)\neq0,$ since by definition, $\tilde{\pi}^{(3)}_{IQ}$ ignores Queue 2 queues when it is nonempty only when both Queue 1 and Queue 3 are nonempty. In this case $\pi'_0$ serves Queue 1 and Queue 3 in slot 0 while $\pi$ does not, resulting in $Q^{\pi'_1}_1(1)=Q^\pi_1(1)-1$ and $Q^{\pi'_0}_3(1)=Q^\pi_3(1)-1.$
\end{itemize}
The third condition is explained in a similar manner and follows easily from symmetry. When $j_1=1$ and $j_2=3,$ all three queues are nonempty and $\pi'$ serves both. This proves the fourth condition. When these 4 conditions hold, the sum backlog with $\pi'$ is not larger than with $\pi$ due to the following reason. When all queues are initially nonempty (meaning $Q_l(0)>0,1\leq l\leq 3)$, this is true from condition~\ref{Chapter03condnAllQueuesNonempty}. When only two adjacent queues are nonempty, conditions~\ref{condnDONonEmptyLessThanN} and \ref{condnDONonEmptyMoreThan1}, as the case may be, ensure this. When only queues 1 and 3 are nonempty, $\tilde{\pi}^{(3)}_{IQ}$ and hence, $\pi'_0$ serve both of them. The case with only one nonempty queue at $t=0$ is trivial. Thus, 
\begin{equation}
\sum_{i=1}^3Q^{\pi'_{0}}_i(1)\leq \sum_{i=1}^3Q^{\pi}_i(1).\nonumber
\end{equation}

For $t\geq1,$ the definition of $\pi'_0$ and the rest of the proof of how the above inequality is ensured at every $t\geq 0$ is the same as in \cite{tassiulas-ephremides94dynamic-scheduling-tandem-parallel} and will not be repeated. 
For every $k\geq0,$ $\pi'_{k+1}$ is defined as the policy that follows $\pi^{(3)}_{OQ}$ over slots $0,1,\dots,k$ and over $k+1,\dots,$ is defined as in \cite{tassiulas-ephremides94dynamic-scheduling-tandem-parallel} so as to satisfy 
\begin{equation}
\sum_{i=1}^3Q^{\pi_{k+1}}_i(t)\leq \sum_{i=1}^3Q^{\pi_{k}}_i(t),~\forall t\geq0,~\forall k\geq1.
\label{eqnDOofPi3TildePoliciesProgressivelyBetter}
\end{equation}
Again, by construction, it is clear that 
\begin{equation}
\lim_{k\rightarrow\infty}Q^{\pi'_{k}}_i(t)\stackrel{s}{=}Q^{\tilde{\pi}^{(3)}_{IQ}}_i(t),~\forall t\geq0,~1\leq i\leq N,
\label{eqnDOofPi3TildeSureConvergenceOfEveryQueue}
\end{equation}
where $\stackrel{s}{=}$ means over every sample path (and is stronger than \enquote{a.s.}). Eqn.~\ref{eqnDOofPi3TildePoliciesProgressivelyBetter} together with Eqn.~\ref{eqnDOofPi3TildeSureConvergenceOfEveryQueue} give us \ref{eqnPi33isDelayOptimal}. This completes the proof.
\qed
\subsection{Proof of Prop.~\ref{thmPi3_4IsNotTO}}\label{AppendixPi3_4IsNotTO}
Consider real numbers $\epsilon$ and $\delta$, such that $\delta>0,0<\epsilon<0.5$ and $\epsilon+\delta\leq0.5$. Let $\boldsymbol{\lambda}= [0.5-\epsilon-\delta,0.5+\epsilon,0.5-\epsilon-\delta]$. Clearly, $\boldsymbol{\lambda}\in\Lambda^o$. At time $t$, define the event $S_2:=\{\text{Queue 2 is served in slot }t\}$ and let $\mathbf{Q}(t)=[q_1, q_2, q_3].$
\begin{eqnarray}
P\{S_2=1\} &=& P\{S_2=1\mid q_1+q_3>0\}P\{q_1+q_3>0\}\nonumber\\
&&+P\{S_2=1\mid q_1+q_3=0\}P\{q_1+q_3=0\}.\nonumber\\
&\leq & 0\cdot P\{q_1+q_3>0\} + 1\cdot P\{q_1+q_3=0\}\nonumber\\
&=& 1-P\{q_1+q_3>0\}\nonumber\\
&=& 1-P\{q_1>0\text{ or }q_3>0\}\nonumber\\
&\leq& 1-P\{A_1(t)>0\text{ or }A_3(t)>0\}\nonumber\\
&=&P\{A_1(t)=A_3(t)=0\}=\left(1-\left(0.5-\epsilon-\delta\right)\right)^2\nonumber\\
&=&\left(0.5+\epsilon+\delta\right)^2
\end{eqnarray}
If $x:=(0.5+\epsilon)=\lambda_2,$ to prove the instability of $\pi^{(3)}_{OQ}$ one only needs to solve the nonlinear program \eqref{eqnNonlinearProgramPolicyUnstableProof} below. That will establish that $P\{S_2(t)=1\}<\lambda_2,$ and hence prove that Queue 2 is unstable. 
\begin{eqnarray}
 Find~~~~~(x+\delta)^2 &<& x,\nonumber\\
s.t.~~~~~\delta &>& 0,\nonumber\\
x &>& 0.5,\nonumber\\
x &<& 1,\nonumber\\
 x+\delta & \leq & 1.
 \label{eqnNonlinearProgramPolicyUnstableProof}
\end{eqnarray}
The problem above is easily solved, for example, by $x=0.75,$ from which we conclude that $\pi^{(3)}_{OQ}$ is unstable.
\qed


\subsection{Proof of Lem.~\ref{Chapter04SecQLenAgnosticTOSplicingThmGenericSplicedPolicyIsAdmissible}}
\label{Chapter04SecQLenAgnosticTOSplicingAppendixPfOfThmGenericSplicedPolicyIsAdmissible}
We know that $\pi^{(N)}_{TD}$ and $\pi^{(N)}_{BU}$ they produce a single activation vector $\mathbf{s}(t)\in\{0,1\}^N$ for every $\boldsymbol{\zeta}(t)\in\{0,1\}^N$. We have also seen that $\pi^{(2N-1)}_{SP}$ induces $\pi^{(N)}_{BU}$ on Queues 1 through $N$, and $\pi^{(N)}_{TD}$ on Queues $N$ through $2N-1$. Given any occupancy vector for the new system $\boldsymbol{\zeta}(t)\in\{0,1\}^{2N-1}$ notice that $\pi^{(N)}_{BU}$ maps coordinates 1 through $N$ to a single activation vector and $\pi^{(N)}_{TD}$ maps coordinates $N$ through $2N-1$ to a single activation vector, with a non-conflicting overlap at Queue~$N.$ Thus, every $\boldsymbol{\zeta}(t)\in\{0,1\}^{2N-1}$ gets mapped to a unique activation vector $\boldsymbol{s}(t)\in\{0,1\}^{2N-1}$, resulting in an admissible policy.
\qed
%
%
\subsection{Proof of Prop.~\ref{Chapter04SecAgnosticSplicingPropPi4_TDIsTO}}\label{AppendixProofOfAgnosticSplicingPropPi4_TDIsTO}
Before we prove the stability of $\pi^{(4)}_{TD}$, we will need the following two lemmas.
\begin{lem}\label{Chapter04SecAgnosticSplicingLemExpectationOfTruncatedDifferencesOfIndependentRVs}
Let $A$ and $B$ be two {independent} random variables, with $A$ taking values in $\left\lbrace0,1,2,\cdots\right\rbrace$ and $B$ taking values in $\left\lbrace0,1\right\rbrace$. Define $Z:=(A-B)^+.$ Then,
\begin{equation}
\mathbb{E}Z=\mathbb{E}A-P\left\lbrace B=1\right\rbrace\left(1-P\left\lbrace A=0\right\rbrace\right).
\end{equation}
\end{lem}
\begin{pf}
Let $p_k=P\left\lbrace A=k\right\rbrace,k\geq0,$ and $q=P\left\lbrace B=1\right\rbrace.$ \\
Then, $P\left\lbrace Z = 0\right\rbrace=P\left\lbrace A=0\right\rbrace+P\left\lbrace A=1, B=1\right\rbrace =p_0+p_1q$, and for all $k\geq1$,
\begin{eqnarray}
P\left\lbrace Z = k\right\rbrace &=& P\left\lbrace A=k, B=0\right\rbrace+P\left\lbrace A=k+1, B=1\right\rbrace \nonumber\\
&=& p_k(1-q)+p_{k+1}q.
\end{eqnarray}
Hence,
\begin{eqnarray}
\mathbb{E}Z &=& \sum_{k=1}^\infty kP\left\lbrace Z = k\right\rbrace \nonumber\\
&=& \sum_{k=1}^\infty k\left(p_k(1-q)+p_{k+1}q\right)\nonumber\\
&=& \left(1-q\right)\sum_{k=1}^\infty k p_k+q\sum_{k=1}^\infty \left(k+1~-1\right) p_{k+1}\nonumber\\
&=& \left(1-q\right)\mathbb{E}A+q\sum_{k=1}^\infty \left(k+1\right) p_{k+1}-q\sum_{k=1}^\infty p_{k+1}\nonumber\\
&=& \left(1-q\right)\mathbb{E}A+q\left(\mathbb{E}A-p_1\right)-q\left(1-p_1-p_0\right)\nonumber\\
&=& \mathbb{E}A-q\left(1-p_0\right).
\end{eqnarray}
\end{pf}

\begin{lem}\label{Chapter04SecAgnosticSplicingLemProbabilityOfSystemEmptyUnderPi3_1}
Under the policy $\pi^{(3)}_{TD},$
\begin{eqnarray}
\lim_{t\rightarrow\infty} P\left\lbrace Q_1(t) = 0, Q_2(t) = 0, Q_3(t) = 0\right\rbrace = \nonumber\\
\left(1-\lambda_1\right) \left(1-\frac{\lambda_2}{1-\lambda_1}\right) \left(1-\frac{\lambda_3}{1-\lambda_2}\right).
\end{eqnarray}
\end{lem}
\begin{pf}
We have already shown that the policy $\pi^{(3)}_{TD}$ is throughput optimal. From this we see that 
\begin{itemize}
\item the vector-valued process $\left\lbrace\left[Q_1(t),Q_2(t),Q_3(t)\right],t\geq0\right\rbrace$ is strongly stable, under $\pi^{(3)}_{TD}$ and hence, also a positive recurrent DTMC. 
\item Recall that while $s_i(t)$ is used to indicate if service is \enquote{offered} to Queue $i$ at time $t$, $D_i(t)$ indicates if a packet actually leaves Queue $i$ at the end of that slot, i.e., $D_i(t)=s_i(t)\mathbb{I}_{\{Q_i(t)>0\}}$.
The proof of throughput optimality of $\pi^{(3)}_{TD}$ (Thm.~\ref{thmPi3_1IsTO}) already showed that when Queues 1 and 2 are started out in their steady state distributions, 
\begin{equation}\label{eqnProbabilityQ3IsOfferedService}
P\left\lbrace s_3(t)=1\right\rbrace=1-\lambda_2.
\end{equation}
\end{itemize}
So, assume queues 1, 2 and 3 are started out in their steady state distributions. Since $Q_3(t+1)=\left(Q_3(t)-s_3(t)\right)^++A_3(t+1),$ using the fact that Queue 3 is in steady state and Lem.~\ref{Chapter04SecAgnosticSplicingLemExpectationOfTruncatedDifferencesOfIndependentRVs}, we see that
\begin{eqnarray}
\mathbb{E}Q_3(t+1)&=&\mathbb{E}Q_3(t)\nonumber\\
&&-P\left\lbrace s_3(t)=1\right\rbrace\left(1-P\left\lbrace Q_3(t)=0\right\rbrace\right)+\lambda_3,\nonumber\\
\Rightarrow 1-P\left\lbrace Q_3(t)=0\right\rbrace &=& \frac{\lambda_3}{P\left\lbrace s_3(t)=1\right\rbrace},\nonumber\\
\Rightarrow P\left\lbrace Q_3(t)=0\right\rbrace &=& 1-\frac{\lambda_3}{1-\lambda_2},~\text{in the \emph{steady state.}}
\label{eqnSteadyStateProbQueue3Empty}
\end{eqnarray}

Next, note that under $\pi^{(3)}_{TD},$ the system\footnote{In this proof, by \enquote{system}, we mean the 3 queues system.} transmits a packet whenever it is nonempty, i.e., it never so happens that the system is nonempty in a slot and none of the queues is served in that slot. Secondly, the system transmits \emph{two} packets in a slot iff both queues 1 and 3 are nonempty. 

Now, define, for all $t\geq0$, $Q(t):=Q_1(t)+Q_2(t)+Q_3(t),$ and $A(t):=A_1(t)+A_2(t)+A_3(t).$ Then, following the above argument,
\begin{equation}
Q(t+1)=Q(t)-\mathbb{I}_{\left\lbrace Q(t)>0\right\rbrace}-\mathbb{I}_{\left\lbrace Q_1(t)>0,Q_3(t)>0\right\rbrace}+A(t+1).
\end{equation}
Note that the mean arrival rate to the three queue subsystem is $\mathbb{E}A(t)=\lambda_1+\lambda_2+\lambda_3.$
Taking expectation on both sides of the above equation and letting $t\rightarrow\infty,$ we get
\begin{eqnarray*}
\lim_{t\rightarrow\infty}\mathbb{E}Q(t+1)&=&\lim_{t\rightarrow\infty}\mathbb{E}Q(t)-\lim_{t\rightarrow\infty}P\left\lbrace Q(t)>0\right\rbrace\\
&&-\lim_{t\rightarrow\infty}P\left\lbrace Q_1(t)>0,Q_3(t)>0\right\rbrace\\
&&+\lambda_1+\lambda_2+\lambda_3.\nonumber\\
\Rightarrow\lim_{t\rightarrow\infty}P\left\lbrace Q(t)>0\right\rbrace&=&-\lim_{t\rightarrow\infty}P\left\lbrace Q_1(t)>0,Q_3(t)>0\right\rbrace\\
&&+\lambda_1+\lambda_2+\lambda_3.\nonumber\\
&\stackrel{\dagger}{=}&-\lim_{t\rightarrow\infty}P\left\lbrace A_1(t)=1\right\rbrace P\left\lbrace Q_3(t)>0\right\rbrace\\
&&+\lambda_1+\lambda_2+\lambda_3\nonumber\\
&=&-\frac{\lambda_1\lambda_3}{1-\lambda_2}+\lambda_1+\lambda_2+\lambda_3\nonumber\\
&=&1-\left(\left(1-\lambda_1\right) \left(1-\frac{\lambda_2}{1-\lambda_1}\right)\right.\\
&&\left.\times\left(1-\frac{\lambda_3}{1-\lambda_2}\right)\right).
\end{eqnarray*}

\end{pf}

We now continue with the proof of Prop.~\ref{Chapter04SecAgnosticSplicingPropPi4_TDIsTO}. From the definition of $\pi^{(4)}_{TD},$ we see that Queue $4$ is offered service under one of the following conditions.
\begin{itemize}
\item Queue 1 is nonempty and Queue 3 is empty
\item Queue 1 is empty and Queue 2 is nonempty
\item Queues 1, 2 and 3 are all empty.
\end{itemize}
Let us now compute the probability that Queue 4 is offered service in a slot, under the assumption that queues 1, 2 and 3 are started out in stationarity. 
\begin{eqnarray}
P\left\lbrace s_4(t)=1\right\rbrace &=& P\left\lbrace Q_1(t)>0, Q_3=0\right\rbrace + P\left\lbrace Q_1(t)=0, Q_2(t)>0\right\rbrace\nonumber \\
&+& P\left\lbrace Q_1(t) = 0, Q_2(t) = 0, Q_3(t) = 0\right\rbrace \nonumber\\
&\stackrel{\star}{=}& \lambda_1\left(1-\frac{\lambda_3}{1-\lambda_2}\right)+ \left(\left(1-\lambda_1\right)\frac{\lambda_3}{1-\lambda_2}\right) \nonumber\\
&& + \left(\left(1-\lambda_1\right) \left(1-\frac{\lambda_2}{1-\lambda_1}\right) \left(1-\frac{\lambda_3}{1-\lambda_2}\right)\right)\nonumber\\
&=& 1-\lambda_3\nonumber\\
&>& \lambda_4\nonumber
\label{eqnQueue4IsOfferredServiceInPi4_3}
\end{eqnarray}
In equality $\star,$ we used the result of Lem.~\ref{Chapter04SecAgnosticSplicingLemProbabilityOfSystemEmptyUnderPi3_1}. Now, using the same drift argument as in the proof of throughput optimality of $\pi^{(3)}_{TD}$ on $Q_4(t)$, we see that the policy is throughput optimal. 
\qed

\subsection{Analyzing the priority policies in greater detail}\label{AppendixAll3IndicatorFunctionsAreIndependent}
We will now make a few more observations about $\pi^{(3)}_{TD}$ and $\pi^{(4)}_{TD}$. In what follows, we will drop the time index and represent $Q_i(t)$ by $Q_i$ to simplify notation.
\begin{eqnarray}
P\left\lbrace Q_3>0,Q_1=0,Q_2>0\right\rbrace &=& P\left\lbrace Q_3>0,Q_2>0\right\rbrace \nonumber\\
&& \times P\left\lbrace Q_1=0|Q_3>0,Q_2>0\right\rbrace\nonumber\\
&=& (1-\lambda_1)P\left\lbrace Q_3>0,Q_2>0\right\rbrace\nonumber\\
\end{eqnarray}
Next, recall that under $\pi^{(4)}_{TD}$, Queue 4 is \emph{offered} service (i.e., $s_4=1$) whenever either Queue 3 is empty, or when Queue 3 is nonempty, but Queue 1 is empty \emph{and} Queue 2 is non empty. Additionally, from Eqn.~\ref{eqnQueue4IsOfferredServiceInPi4_3} we gather  that $P\left\lbrace s_4(t)=1\right\rbrace=1-\lambda_3.$ 
Hence, 
\begin{eqnarray}
P\left\lbrace s_4(t)=1\right\rbrace &=& 1-\lambda_3\nonumber\\
&=&P\left\lbrace Q_3=0\right\rbrace \nonumber\\
&&+P\left\lbrace Q_3>0,Q_1=0,Q_2>0\right\rbrace\nonumber\\
&\stackrel{\dagger a}{=}& \left(1-\frac{\lambda_3}{1-\lambda_2}\right)\nonumber\\
&&+(1-\lambda_1)P\left\lbrace Q_3>0,Q_2>0\right\rbrace\nonumber\\
\Rightarrow P\left\lbrace Q_3>0,Q_2>0\right\rbrace &=& \frac{1}{1-\lambda_1} \left(1-\lambda_3-\left(1-\frac{\lambda_3}{1-\lambda_2}\right)\right)\nonumber\\
&=& \frac{1}{1-\lambda_1} \lambda_3\left(\frac{1}{1-\lambda_2}-1\right)\nonumber\\
&=& \frac{\lambda_2}{1-\lambda_1}\cdot \frac{\lambda_3}{1-\lambda_2}\nonumber\\
&=& P\left\lbrace Q_2>0\right\rbrace\cdot P\left\lbrace Q_3>0\right\rbrace,
\label{eqnProbabilityQ2Geq0Q3Geq0UnderPriorityPolicies}
\end{eqnarray}
where equality $\dagger a$ above comes from Eqn.~\eqref{eqnSteadyStateProbQueue3Empty}. Next, 
\begin{eqnarray}
(1-\lambda_1)\left(1-\frac{\lambda_2}{1-\lambda_2}\right)&=& P\left\lbrace Q_1=0, Q_2=0\right\rbrace\nonumber\\
&=& P\left\lbrace Q_1=0, Q_2=0, Q_3>0\right\rbrace \nonumber\\
&& +P\left\lbrace Q_1=0,Q_2=0,Q_3=0\right\rbrace \nonumber\\
&\stackrel{\dagger b}{=}& (1-\lambda_1)P\left\lbrace Q_2=0, Q_3>0\right\rbrace \nonumber\\
&& + (1-\lambda_1)\left(1-\frac{\lambda_2}{1-\lambda_1}\right) \left(1-\frac{\lambda_3}{1-\lambda_2}\right)\nonumber\\
\Rightarrow P\left\lbrace Q_2=0, Q_3>0\right\rbrace &=& \left(1-\frac{\lambda_2}{1-\lambda_1}\right)\frac{\lambda_3}{1-\lambda_2}\nonumber\\
&=& P\left\lbrace Q_2=0\right\rbrace\cdot P\left\lbrace Q_3>0\right\rbrace,\nonumber\\
\label{eqnProbabilityQ2Eq0Q3Geq0UnderPriorityPolicies}
\end{eqnarray}
where, in equality $\dagger b$ we have made use of Lem.~\ref{Chapter04SecAgnosticSplicingLemProbabilityOfSystemEmptyUnderPi3_1}. Next,
\begin{eqnarray}
P\left\lbrace Q_2=0, Q_3=0\right\rbrace &=&P\left\lbrace Q_1>0, Q_2=0, Q_3=0\right\rbrace \nonumber\\
&& +P\left\lbrace Q_1=0,Q_2=0,Q_3=0\right\rbrace \nonumber\\
&=& \lambda_1 P\left\lbrace Q_2=0, Q_3=0\right\rbrace \nonumber\\
&& + (1-\lambda_1)\left(1-\frac{\lambda_2}{1-\lambda_1}\right) \left(1-\frac{\lambda_3}{1-\lambda_2}\right),\nonumber\\
\end{eqnarray}
which means that
\begin{eqnarray}
\hspace{-2.00cm}
(1-\lambda_1) P\left\lbrace Q_2=0, Q_3=0\right\rbrace &=& (1-\lambda_1) \left(1-\frac{\lambda_2}{1-\lambda_1}\right) \left(1-\frac{\lambda_3}{1-\lambda_2}\right)\nonumber\\
\Rightarrow P\left\lbrace Q_2=0, Q_3=0\right\rbrace &=&  \left(1-\frac{\lambda_2}{1-\lambda_1}\right) \left(1-\frac{\lambda_3}{1-\lambda_2}\right)\nonumber\\
&=&P\left\lbrace Q_2=0\right\rbrace\cdot P\left\lbrace Q_3=0\right\rbrace.
\label{eqnProbabilityQ2Eq0Q3Eq0UnderPriorityPolicies}
\end{eqnarray}
Finally, 
\begin{eqnarray}
\hspace{-1.00cm}
P\left\lbrace Q_2>0, Q_3=0\right\rbrace &=& 1-\left( P\left\lbrace Q_2>0, Q_3>0\right\rbrace \right.\nonumber\\
&&\left.+P\left\lbrace Q_2=0, Q_3>0\right\rbrace \right) +P\left\lbrace Q_2=0, Q_3=0\right\rbrace\nonumber\\
&=& 1-\left(\frac{\lambda_2}{1-\lambda_1}\cdot \frac{\lambda_3}{1-\lambda_2} + \left(1-\frac{\lambda_2}{1-\lambda_1}\right)\cdot \frac{\lambda_3}{1-\lambda_2}\right. \nonumber\\
&+& \left.\left(1-\frac{\lambda_2}{1-\lambda_1}\right) \left(1-\frac{\lambda_3}{1-\lambda_2}\right) \right) \nonumber\\
&=& 1-\left(\frac{\lambda_3}{1-\lambda_2} + 1- \frac{\lambda_2}{1-\lambda_1} - \frac{\lambda_3}{1-\lambda_2} \right.\nonumber\\
&& \left. + \frac{\lambda_2}{1-\lambda_1}\frac{\lambda_3}{1-\lambda_2}\right)\nonumber\\
&=& \frac{\lambda_2}{1-\lambda_1}\left(1-\frac{\lambda_2}{1-\lambda_1}\right)\nonumber\\
&=&P\left\lbrace Q_2>0\right\rbrace\cdot P\left\lbrace Q_3=0\right\rbrace.
\label{eqnProbabilityQ2Geq0Q3Eq0UnderPriorityPolicies}
\end{eqnarray}
From Eqn.~\eqref{eqnProbabilityQ2Geq0Q3Geq0UnderPriorityPolicies}, Eqn.~\eqref{eqnProbabilityQ2Eq0Q3Geq0UnderPriorityPolicies},  Eqn.~\eqref{eqnProbabilityQ2Eq0Q3Eq0UnderPriorityPolicies} and Eqn.~\eqref{eqnProbabilityQ2Geq0Q3Eq0UnderPriorityPolicies} and  we see that the random variables $\mathbb{I}_{\lbrace Q_1>0\rbrace}$, $\mathbb{I}_{\lbrace Q_2>0\rbrace}$ and $\mathbb{I}_{\lbrace Q_3>0\rbrace}$ are \textbf{independent}, for every $t\geq 0$ under the condition that the initial queue length vector $[Q_1(0),Q_2(0),Q_3(0)]$ follows the steady state distribution, which always exists since $\pi^{(3)}_{TD}$ is throughput optimal.
\begin{rem}
Since the top-down priority policies $\pi^{(3+k)}_{TD}$, for all $k\geq0$ induce $\pi^{(3)}_{TD}$ on queues 1, 2 and 3,  this independence is \emph{always true} in steady state. 
\end{rem}

\subsection{Proof of Prop.~\ref{Chapter04SecAgnosticSplicingPropPi5_TDIsTO}}\label{AppendixProofOfAgnosticSplicingPropPi5_TDIsTO}
Recall that under the top-down policies, Queue 1 receives highest priority and is served whenever it is non empty, followed by Queue 2 and so on. $\pi^{(5)}_{TD}$ offers service to Queue 5 (i.e., $s_5(t)=1$) \textbf{iff} the following conditions are satisfied
\textsl{
\begin{itemize}
\item $Q_1>0\text{ and }Q_3>0$
\item $Q_1>0,~Q_3=0,\text{ and }Q_4=0$,
\item $Q_1=0,~Q_2>0,\text{ and }Q_4=0$,
\item $Q_1=0,~Q_2=0,\text{ and }Q_3>0$ and
\item $Q_1=0,~Q_2=0,~Q_3=0\text{ and }Q_4=0$.
\end{itemize} 
}
So, 
\begin{eqnarray}
\hspace{-0.75cm}
P\left\lbrace s_5(t)=1\right\rbrace &=& P\left\lbrace Q_1>0,~Q_3>0\right\rbrace + P\left\lbrace Q_1=0,~Q_2=0,~Q_3>0\right\rbrace \nonumber\\
&& + P\left\lbrace Q_1>0,~Q_3=0,~Q_4=0\right\rbrace \nonumber\\
&& +P\left\lbrace Q_1=0,~Q_2>0,~Q_4=0\right\rbrace \nonumber\\
&& + P\left\lbrace Q_1=0,~Q_2=0,~Q_3=0,~Q_4=0\right\rbrace\nonumber\\
&\stackrel{\dagger c}{=}& \lambda_1\frac{\lambda_3}{1-\lambda_2}+(1-\lambda_1)\left(1-\frac{\lambda_2}{1-\lambda_1}\right)\frac{\lambda_3}{1-\lambda_2}\nonumber\\
&& +P\left\lbrace Q_1>0,~Q_3=0,~Q_4=0\right\rbrace \nonumber\\
&& +P\left\lbrace Q_1=0,~Q_2>0,~Q_4=0\right\rbrace \nonumber\\
&& + P\left\lbrace Q_1=0,~Q_2=0,~Q_3=0,~Q_4=0\right\rbrace, 
\label{eqnIntermediateResultOfferedServiceToQueue5}
\end{eqnarray}
where in equality $c$ we have used the independence results of Sec.\ref{AppendixAll3IndicatorFunctionsAreIndependent}. Now, consider the subsystem comprising queues 1, 2 and 4 under this policy and call this subsystem $\mathcal{Q}_{124}$. Since $\pi^{(5)}_{TD}$ restricted to the first 4 queues is essentially $\pi^{(4)}_{TD}$, the top-down policy for the 4 queue system, and since Thm.~\ref{Chapter04SecAgnosticSplicingPropPi4_TDIsTO} already showed that $\pi^{(4)}_{TD}$ is throughput-optimal, $\mathcal{Q}_{124}$ is a stable subsystem and \emph{has a steady state distribution} which is simply a \emph{marginal} of the distribution of the 4 queue system with Queue 3's coordinate summed out. 

The arrival rate to $\mathcal{Q}_{124}$ is $\lambda_1+\lambda_2+\lambda_4$. Also, under $\pi^{(4)}_{TD}$ and hence $\pi^{(5)}_{TD}$, the $\mathcal{Q}_{124}$ transmits 
\begin{itemize}
\item At least 1 packet in slots with
\begin{itemize}
\item $Q_1>0$, or
\item $Q_1=0,~Q_2>0$, or 
\item $Q_1=0,~Q_2=0,~Q_3=0\text{ and }Q_4>0$, and
\end{itemize}
\item 2 packets in slots with 
\begin{itemize}
\item $Q_1>0,~Q_3=0,\text{ and }Q_4>0$, or
\item $Q_1=0,~Q_2>0,\text{ and }Q_4>0$.
\end{itemize}
\end{itemize}
Assume $\mathcal{Q}_{124}$ is started out in its steady state. Let $Q(t)=\sum_{i\in\mathcal{Q}_{124}}Q_i(t),$ and $A(t)=\sum_{i\in\mathcal{Q}_{124}}A_i(t),$ for all $t\geq0.$ Consequently, 
\begin{eqnarray*}
\hspace{-2.00cm}
Q(t+1) &=& Q(t)-\mathbb{I}_{\left\lbrace Q_1(t)>0\right\rbrace}-\mathbb{I}_{\left\lbrace Q_1(t)=0, Q_2(t)>0\right\rbrace} \\
&& -\mathbb{I}_{\left\lbrace Q_1(t)=0, Q_2(t)=0, Q_3(t)=0, Q_4(t)>0\right\rbrace}\nonumber\\ 
&& -\mathbb{I}_{\left\lbrace Q_1(t)>0, Q_3(t)=0, Q_4(t)>0\right\rbrace}\\
&& -\mathbb{I}_{\left\lbrace Q_1(t)=0, Q_2(t)>0, Q_4(t)>0\right\rbrace} +A(t+1)\nonumber\\
\mathbb{E}Q(t+1)-\mathbb{E}Q(t) &=& -P{\left\lbrace Q_1(t)>0\right\rbrace}\\
&& -P{\left\lbrace Q_1(t)=0, Q_2(t)>0\right\rbrace} \\
&& -P\left\lbrace Q_1(t)=0, Q_2(t)=0, \right.\\
&& \left. Q_3(t)=0, Q_4(t)>0\right\rbrace\nonumber\\ 
&& -P{\left\lbrace Q_1(t)>0, Q_3(t)=0, Q_4(t)>0\right\rbrace}\\
&& -P{\left\lbrace Q_1(t)=0, Q_2(t)>0, Q_4(t)>0\right\rbrace} \\
&&+\mathbb{E}A(t+1),\nonumber
\end{eqnarray*}
which, using the fact that $\mathbb{E}Q(t+1)-\mathbb{E}Q(t)=0$ in the steady state and that $\mathbb{E}A(t+1)=\lambda_1+\lambda_2+\lambda_4$, gives us
\begin{eqnarray}
&& P{\left\lbrace Q_1(t)=0, Q_2(t)=0, Q_3(t)=0, Q_4(t)>0\right\rbrace} \nonumber\\
&& + P{\left\lbrace Q_1(t)>0, Q_3(t)=0, Q_4(t)>0\right\rbrace} \nonumber\\
&& + P{\left\lbrace Q_1(t)=0, Q_2(t)>0, Q_4(t)>0\right\rbrace}\nonumber\\
&& =\lambda_1+\lambda_2+\lambda_4-P{\left\lbrace Q_1(t)>0\right\rbrace}-P{\left\lbrace Q_1(t)=0, Q_2(t)>0\right\rbrace}\nonumber\\
&& =\lambda_1+\lambda_2+\lambda_4-\lambda_1-(1-\lambda_1)\frac{\lambda_2}{1-\lambda_1}\nonumber\\
&& = \lambda_4.
\label{eqnIntermediateResultAnalysisOfQ124SubsystemLambda4RHS}
\end{eqnarray}
Notice that
\begin{eqnarray}
&& P{\left\lbrace Q_1(t)=0, Q_2(t)=0, Q_3(t)=0, {\color{blue}Q_4(t)>0}\right\rbrace} \nonumber\\
&& +P{\left\lbrace Q_1(t)>0, Q_3(t)=0, {\color{blue} Q_4(t)>0}\right\rbrace} \nonumber\\
&& + P{\left\lbrace Q_1(t)=0, Q_2(t)>0,{\color{blue}Q_4(t)>0}\right\rbrace}\nonumber\\
&& + P{\left\lbrace Q_1(t)=0, Q_2(t)=0, Q_3(t)=0, {\color{blue}Q_4(t)=0}\right\rbrace}\nonumber\\
&& + P{\left\lbrace Q_1(t)>0, Q_3(t)=0, {\color{blue}Q_4(t)=0}\right\rbrace} \nonumber\\
&& + P{\left\lbrace Q_1(t)=0, Q_2(t)>0, {\color{blue}Q_4(t)=0}\right\rbrace}\nonumber\\
&& = P{\left\lbrace Q_1(t)=0, Q_2(t)=0, Q_3(t)=0\right\rbrace}\nonumber\\
&& + P{\left\lbrace Q_1(t)>0, Q_3(t)=0\right\rbrace} \nonumber\\
&& + P{\left\lbrace Q_1(t)=0, Q_2(t)>0\right\rbrace}.\nonumber
\end{eqnarray}
Using Eqn.~\ref{eqnIntermediateResultAnalysisOfQ124SubsystemLambda4RHS} on the first three terms on the LHS of the above equation and invoking the independence results in Sec.~\ref{AppendixAll3IndicatorFunctionsAreIndependent} on the RHS, we get
\begin{eqnarray}
&& \lambda_4+P{\left\lbrace Q_1(t)=0, Q_2(t)=0, Q_3(t)=0,{\color{blue}Q_4(t)=0}\right\rbrace}\nonumber\\
&& + P{\left\lbrace Q_1(t)>0, Q_3(t)=0, {\color{blue}Q_4(t)=0}\right\rbrace} \nonumber\\
&& + P{\left\lbrace Q_1(t)=0, Q_2(t)>0, {\color{blue}Q_4(t)=0}\right\rbrace}\nonumber\\
&& = (1-\lambda_1) \left(1-\frac{\lambda_2}{1-\lambda_1}\right) \left(1-\frac{\lambda_3}{1-\lambda_2}\right) + \lambda_1\left(1-\frac{\lambda_3}{1-\lambda_2}\right) \nonumber\\
&& + (1-\lambda_1)\frac{\lambda_2}{1-\lambda_1},\nonumber
\end{eqnarray}
which means that,
\begin{eqnarray*}
&& P{\left\lbrace Q_1(t)=0, Q_2(t)=0, Q_3(t)=0, {Q_4(t)=0}\right\rbrace}\\
&& +P{\left\lbrace Q_1(t)>0, Q_3(t)=0, {Q_4(t)=0}\right\rbrace} \nonumber\\
&& + P{\left\lbrace Q_1(t)=0, Q_2(t)>0, {Q_4(t)=0}\right\rbrace}\nonumber\\
&& = (1-\lambda_1) \left(1-\frac{\lambda_2}{1-\lambda_1}\right) \left(1-\frac{\lambda_3}{1-\lambda_2}\right) + \lambda_1\left(1-\frac{\lambda_3}{1-\lambda_2}\right) \\
&& + (1-\lambda_1)\frac{\lambda_2}{1-\lambda_1}-\lambda_4.
\end{eqnarray*}

Substituting this on the RHS of Eqn.~\ref{eqnIntermediateResultOfferedServiceToQueue5}, we get 
\begin{eqnarray*}
P\left\lbrace S_5(t)=1\right\rbrace &=&  \lambda_1\frac{\lambda_3}{1-\lambda_2}+(1-\lambda_1)\left(1-\frac{\lambda_2}{1-\lambda_1}\right)\frac{\lambda_3}{1-\lambda_2}\nonumber\\
&+&  \lambda_1\left(1-\frac{\lambda_3}{1-\lambda_2}\right) + (1-\lambda_1) \left(1-\frac{\lambda_2}{1-\lambda_1}\right) \left(1-\frac{\lambda_3}{1-\lambda_2}\right)\\
&+& (1-\lambda_1)\frac{\lambda_2}{1-\lambda_1}-\lambda_4\nonumber\\
&=& \lambda_1 + (1-\lambda_1)\left(1-\frac{\lambda_2}{1-\lambda_1}\right) + (1-\lambda_1)\frac{\lambda_2}{1-\lambda_1}-\lambda_4\nonumber\\
&=& 1-\lambda_4\nonumber\\
&>& \lambda_5.
\end{eqnarray*}
Thus, both $\pi^{(5)}_{TD}$ and $\pi^{(5)}_{BU}$, the top-down and bottom-up policies are stable.
\qed

\qed
\subsection{Proof of Lem.~\ref{lemSplicedPolicy4QueuesIsTO}}\label{AppendixTOofSplicedPolicy4Queues}
The proof of throughput-optimality of $\pi^{(4)}_{TI}$ is along the lines of the proof of Thm.~\ref{Chapter04SecQLenAgnosticTOSplicingThmGenericSplicedPolicyIsTO}. By that we mean we shall show that the policy is a splicing of two throughput-optimal scheduling policies for the 3-queue system. Recall from our earlier discussion that Policy $\pi^{(4)}_{TI}$ is defined as follows.

\textsl{At time $t:$
\begin{enumerate}
\item If $Q_2(t)>0,$ $\mathbf{S}(t)=[0,1,0,1]$.
\item Else, if $Q_3(t)>0,~\mathbf{S}(t)=[1,0,1,0]$.
\item Else, $\mathbf{S}(t)=[1,0,0,1]$.
\end{enumerate}
}
Notice that the above definition can be split into two parts, the first acting on queues 1, 2 and 3 and the second on queues 2, 3 and 4, as shown below.

\textsl{At time $t:$
\begin{enumerate}
\item\label{condn4QueuesOnlyFirst3Appendix} Check $[\zeta_1(t), \zeta_2(t), \zeta_3(t)]$
\begin{enumerate}
\item If $i_2(t)=1,~\mathbf{S}(t)=[0,1,0,1].$
\item Else if $i_3(t)=1,~\mathbf{S}(t)=[1,0,1,0].$
\item Else $\mathbf{S}(t)=[1,0,0,1].$
\end{enumerate}
\item\label{condn4QueuesOnlyLast3Appendix} Check $[\zeta_2(t), \zeta_3(t), \zeta_4(t)]$
\begin{enumerate}
\item If $i_2(t)=1,~\mathbf{S}(t)=[0,1,0,1].$
\item Else if $i_3(t)=1,~\mathbf{S}(t)=[1,0,1,0].$
\item Else $\mathbf{S}(t)=[1,0,0,1].$
\end{enumerate}
\end{enumerate}
}
The first portion of the policy, i.e., part~\ref{condn4QueuesOnlyFirst3Appendix} can be seen as a very simple variant of policy $\pi^{(3)}_{IQ}$, which was  the non MSM policy defined in Sec.~\ref{secNonMSMStablePolicy}. Instead of always scheduling queues 1 and 3 when $Q_2(t)=0,$ here, we schedule queues 1 and 3 when $[\zeta_2(t), \zeta_3(t)]=[0,1]$ and queues 1 and 4 when $Q_2(t)=Q_3(t)=0.$ This amounts to choosing $\mathbf{S}(t)=[1,0,1]$ when $Q_2(t)=0,\text{ but }Q_3(t)>0$, and $\mathbf{S}(t)=[1,0,0]$ when $Q_2(t)=Q_3(t)=0.$ This is permissible since it does not violate the definition of $\pi^{(3)}_{IQ}$ in any way. We already know that $\pi^{(3)}_{IQ}$ is stabilizing for all $[\lambda_1,\lambda_2,\lambda_3]\in\Lambda^o_3,$ which is the interior of the capacity region for 3-queue path-graph interference networks defined in \eqref{Chapter04SecQLenAgnosticTOSplicingEqnCapacityPathGraphs}.

The second portion of the above policy, i.e., part~\ref{condn4QueuesOnlyLast3Appendix} can be seen as a similar variant of policy $\pi^{(3)}_{TD}$ which is defined on queues 2, 3 and 4. From the analysis of the 3-queue system we already know that $\pi^{(3)}_{TD}$ is stabilizing for all $[\lambda_2,\lambda_3,\lambda_4]\in\Lambda^o_3.$ Since any $\boldsymbol{\lambda}\in\Lambda^o_4$ satisfies $\lambda_i+\lambda_{i+1}<1,~1\leq i\leq 3$, it automatically means that $[\lambda_1,\lambda_2,\lambda_3]\in\Lambda^o_3,$ and $[\lambda_2,\lambda_3,\lambda_4]\in\Lambda^o_3.$ The analysis of $\pi^{(4)}_{TI}$ so far implies that the rate region stabilized by it ($\Lambda_{\pi^{(4)}_{TI}}$) satisfies 
\begin{equation}
\Lambda^o_4\subseteq\Lambda^o_{\pi^{(4)}_{TI}},
\label{eqnT&E4QueueCapRegionSubsetOfNonMSMPolicy}
\end{equation}
where $\Lambda^o_4$ is the interior of the capacity region of the system with $N$ queues. But since rate vectors outside $\Lambda^o_4$ cannot be stabilized anyway, $\pi^{(4)}_{TI}$ is indeed throughput-optimal.
\qed
\subsection{Proof of Prop.~\ref{Chapter04propPiTilde4Other2PoliciesAlsoStable}}\label{AppendixProofOfPropPiTilde4Other2PoliciesAlsoStable}
From Table.~\ref{Chapter04tableAll4QueuePolicies}, we see that $\tilde{\pi}^{(4)}_3$ differs from $\tilde{\pi}^{(4)}_1$ at exactly one system occupancy vector, viz, $\boldsymbol{\zeta}(t)=[1,1,1,1]$. 
When $\boldsymbol{\zeta}(t)=[1,1,1,1]$, both activate the same number of queues, but while $\tilde{\pi}^{(4)}_3$ serves queue 1 and 3 while $\tilde{\pi}^{(4)}_1$ serves queues 2 and 4. This means that 
all four conditions in the proof of delay optimality of the policy $\tilde{\pi}^{(3)}_{IQ}$ (see Sec.~
\ref{AppendixProofOfPi33isDelayOptimal}) are satisfied. Note that  condition~\ref{Chapter03condnAllQueuesNonempty} therein is vacuously satisfied in this case, since $N=4$ is not an odd number. Hence, using the same technique as in Sec.~\ref{AppendixProofOfPi33isDelayOptimal}, we see that 
\begin{equation}
\sum_{i=1}^4Q^{\tilde{\pi}^{(4)}_3}_i(t)\stackrel{st}{\leq}\sum_{i=1}^4Q^{\tilde{\pi}^{(4)}_1}_i(t),~\forall t\geq0.\nonumber
\end{equation}
But by switching the roles of $\tilde{\pi}^{(4)}_3$ and $\tilde{\pi}^{(4)}_1$ and going through the same argument, we also see that 
\begin{equation}
\sum_{i=1}^4Q^{\tilde{\pi}^{(4)}_1}_i(t)\stackrel{st}{\leq}\sum_{i=1}^4Q^{\tilde{\pi}^{(4)}_3}_i(t),~\forall t\geq0,\nonumber
\end{equation}
which leads us to the conclusion that when the initial conditions are identical, both the policies have the same total system occupancy distributions, i.e.,
\begin{equation}
\sum_{i=1}^4Q^{\tilde{\pi}^{(4)}_3}_i(t)\stackrel{st}{=}\sum_{i=1}^4Q^{\tilde{\pi}^{(4)}_1}_i(t),~\forall t\geq0.
\label{eqnSystemOccupancyOf1and3AreTheSame}
\end{equation}
Since $\tilde{\pi}^{(4)}_1$ is strongly stable, arguments following \eqref{Chapter04eqnStabilityThroughProjections}, show us that $\tilde{\pi}^{(4)}_3$ is strongly stable as well. Finally, a similar proof with $\tilde{\pi}^{(4)}_2$  and $\tilde{\pi}^{(4)}_4$  shows that 
\begin{equation}
\sum_{i=1}^4Q^{\tilde{\pi}^{(4)}_4}_i(t)\stackrel{st}{=}\sum_{i=1}^4Q^{\tilde{\pi}^{(4)}_2}_i(t),~\forall t\geq0.
\label{eqnSystemOccupancyOf2and4AreTheSame}
\end{equation}
and hence, that $\tilde{\pi}^{(4)}_4$ is throughput optimal. This concludes the proof.
\qed
\subsection{Proof of Prop.~\ref{Chapter04Prop2LambdasForNonDOinPi4}}\label{AppendixProofOf2LambdasForNonDOinPi4}
 
Consider rate vectors of the form $\boldsymbol{\lambda}_1=(\lambda_1,\lambda_2,\lambda_3,0)$ and $\boldsymbol{\lambda}_2=(0,\lambda_2,\lambda_3,\lambda_4)$ in $\Lambda^o$. First consider $\boldsymbol{\lambda}_1$ and let $Q_4(0)=0,$ under this rate vector. By the Borel Cantelli Lemma applied to the Bernoulli arrival process, a zero arrival rate implies that, w.p.1, on sample paths of $A_4(t),$ there will be finitely many arrivals. This means that, w.p.1, along sample paths the policy $\tilde{\pi}^{(4)}_2$ restricted to Queues~1, 2 and 3 reduces 
 to the policy described below. Note that the fourth coordinate of $\mathbf{s}(t)$ is not shown since Queue 4 is always empty. 

At time $t,$
\begin{enumerate}
\item If $Q_3(t)>0,$ $\mathbf{s}(t)=[1,0,1]$.
\item Else, if $Q_2(t)>0,~\mathbf{s}(t)=[0,1,0]$.
\item Else, $\mathbf{s}(t)=[1,0,0]$.
\end{enumerate}
This is simply $\pi^{(3)}_{BU}.$ It follows that the distribution of sum queue length converges to that of the sum queue length of $\pi^{(3)}_{BU}.$ In the same way, $\tilde{\pi}^{(4)}_1$ now reduces to

At time $t,$
\begin{enumerate}
\item If $\boldsymbol{\zeta}(t)=[1,1,1]$, $\mathbf{s}(t)=[1,0,1]$.
\item Else, if $Q_2(t)>0,$ $\mathbf{s}(t)=[0,1,0]$.
\item Else, if $Q_3(t)>0,~\mathbf{s}(t)=[1,0,1]$.
\item Else, $\mathbf{s}(t)=[1,0,0]$,
\end{enumerate}
This is simply $\tilde{\pi}^{(3)}_{IQ},$ which, as we have already seen, is delay optimal (in the stochastic ordering sense) for the system with 3 queues. In Sec.~\ref{Chapter05SimulationResults} we show arrival rate vectors for which strict inequality holds. Since $\tilde{\pi}^{(4)}_1$ and $\tilde{\pi}^{(4)}_2$ differ from $\tilde{\pi}^{(4)}_3$ and $\tilde{\pi}^{(4)}_4$ respectively only on $\boldsymbol{\zeta}(t)=[1,1,1,1]$, and since this occupancy never occurs (Queue 4 is always empty), the same $\boldsymbol{\lambda}_1$ continues to work. This proves Equations~\eqref{eqnLamForWhich1and3BetterThan2and4}. A similar argument can be made with $\boldsymbol{\lambda}_2$ to prove Eqns.~\eqref{eqnLamForWhich2and4BetterThan1and3}. This concludes the proof.
\qed
\subsection{Proof of Prop.~\ref{propPoliciesInPi4TildeAreBetterThanPi4M}} \label{AppendixProofOfPoliciesInPi4TildeAreBetterThanPi4M}

\begin{table}[tb]
\caption{Comparison of $\mathbf{S}(t)$ under policies in $\Pi^{(4)}_M$ and $\tilde{\Pi}^{(4)}.$ Note here, that $s_1=[1,0,1,0],s_2=[0,1,0,1]\text{ and }s_3=[1,0,0,1]$.}
\centering 
\begin{tabular}{c c c} 
\hline\hline 
$[i_4(t),i_3(t),i_2(t),i_1(t)]$ & $\Pi^{(4)}_M$ & $\tilde{\Pi}^{(4)}$\\ [0.5ex] 
\hline 
0000 &    1001 & 1001   \\ 
0001 &    1001 & 1001   \\
0010 &    1010 & 1010   \\
$\mathbf{0011}$ &  $\{s_1,s_2\}$ & 1010 \\ 

0100 &    0101 & 0101  \\
0101 &    0101 & 0101  \\
$\mathbf{0110}$ &  $\{s_1,s_2\}$ & $\{s_1,s_2\}$ \\
0111 &    1010 & 1010 \\

1000 &    1001 & 1001 \\
1001 &    1001 & 1001 \\
1010 &    1010 & 1010 \\
$\mathbf{1011}$ &  $\{s_1,s_3\}$ &1010 \\ 
 
$\mathbf{1100}$ & $\{s_1,s_2\}$ & 0101 \\
$\mathbf{1101}$ &  $\{s_2,s_3\}$ & 0101 \\
1110 &    1010 & 1010 \\
$\mathbf{1111}$ & $\{s_1,s_2,s_3\}$  & $\{s_1,s_2\}$ \\[1ex] \\
\hline 
\end{tabular}
\label{Chapter04TablePoliciesInPi4TildeAreBetterThanPi4M} 
\end{table} 

 Denote by $s_1,s_2\text{ and }s_3$ the three activation vectors $[1,0,1,0],[0,1,0,1]\text{ and }[1,0,0,1]$. Any MSM policy in $\Pi^{(4)}_M$ will only choose from among these three vectors in any slot. Consider once again, the set of all 16 possible values that $\boldsymbol{\zeta}(t)$ can take, as given in Table.~\ref{Chapter04TablePoliciesInPi4TildeAreBetterThanPi4M}. As the table shows, among the occupancy vectors shaded red, $\boldsymbol{\zeta}(t)=[0110]$ is the only one  for which policies in both sets completely agree on the set of activation vectors from which to choose. We will, therefore, focus on the other five in the remainder of the proof. Recall that the condition which separates $\Pi^{(N)}_M$ from $\tilde{\Pi}^{(N)}$ is condition~\ref{Chapter03conditionMSMDelaySmall} in Lem.~\ref{Chapter03lemSufficientForActivationMSM}. Since $\pi\in\Pi^{(4)}_M\setminus\tilde{\Pi}^{(4)}$ there exists at least one occupancy vector among the five, for which $\pi$ chooses an activation vector that violates this condition. 

Consider for example, $\boldsymbol{\zeta}(t)=[1,1,0,0],$ which means that $j_1=1$ and $j_2=2$ and suppose, under this occupancy vector, that $\pi$ chooses $s_2$, i.e., serves queues 1 and 3. We have already seen that Condition~\ref{Chapter03conditionMSMDelaySmall} in  Lem.~\ref{Chapter03lemSufficientForActivationMSM} (that gave the conditions for an activation vector to be MSM) demands that a policy schedule queues 2 and 4, i.e., choose $s_1=[0,1,0,1].$  Let $\pi'$ be the policy that $\boldsymbol{\zeta}(t)=[1,1,0,0]\mapsto s_1=[0,1,0,1]$ and is identical to $\pi$ for all other occupancy vectors. The objective is to now prove that $\pi'$ produces stochastically smaller system backlog than $\pi$ by using the procedure in the proof of delay optimality of $\tilde{\pi}^{(3)}_{IQ}$ (Sec.~\ref{AppendixProofOfPi33isDelayOptimal}).

Specifically, consider condition~\ref{condnDONonEmptyLessThanN} in the proof. From the preceding discussion we gather that when $\boldsymbol{\zeta}(t)=[1,1,0,0]$, there exists $l<4$ such that $\pi'$ does not serve Queue~$l$ while $\pi$ does serve it. This is obviously Queue~1. But in this case, $\pi'$ does serve Queue~$l+1$ in the same slot, while $\pi$ does not, which means that condition~\ref{condnDONonEmptyLessThanN} is satisfied. Note that since $N$ here is an even number and $\pi$ and $\pi'$ are identical on all other occupancy vectors, conditions Conditions~3 and 4 are vacuously satisfied. Hence, following the procedure outlined in the rest of the proof of delay optimality of $\tilde{\pi}^{(3)}_{IQ}$,
we see that $\pi'$ results in stochastically smaller sum queue lengths than $\pi.$

The same argument can now be given for any combination of the shaded occupancy vectors (in Table.~\ref{Chapter04TablePoliciesInPi4TildeAreBetterThanPi4M}) for which policies in $\Pi^{(4)}_M\setminus\tilde{\Pi}^{(4)}$ violate condition~\ref{Chapter03conditionMSMDelaySmall} in Lem.~\ref{Chapter03lemSufficientForActivationMSM}. This proves the result.

 \qed


\subsection{Proof of Prop.~\ref{Chapter05PropPhiNTildeIsTO}}\label{AppendixProofOfPhiNTildeIsTO}
 We extend our initial idea of Property~$\mathcal{P}$ to prove this proposition as follows. For all $1\leq m\leq N,$ define $Q_m(t):=\sum_{j\in\mathcal{C}_m}Q_{m,j}(t),$ and $D_m(t):=\sum_{j\in\mathcal{C}_m}D_{m,j}(t).$
As in the proof of sufficiency of Property~$\mathcal{P}$, the idea is to prove that $\tilde{\phi}$ satisfies the following version of the property, which immediately leads to strong stability. For all $2\leq m\leq N,$
\begin{equation}
D_1(t)+D_m(t)=0\iff Q_1(t)+Q_m(t)=0,~\forall t\geq0.
\label{eqnPropertyPForNonPathGraphs}
\end{equation}
Let $m\geq2$ in the sequel. By the definition of the departure processes $\left\lbrace D_i(t),i\geq1\right\rbrace$, in every slot $t\geq0,$ $Q_1(t)+Q_m(t)=0$ always means $D_1(t)+D_m(t)=0$. To show the converse, we consider several cases 
\begin{itemize}
\item $Q_1(t)>0$ and $Q_m(t)>0$ means that in one of the 3 steps of the definition of $\tilde{\phi}$, one of the queues in either of these cliques will get scheduled and either $D_1(t)=1$ or $D_m(t)=1$.
\item $Q_1(t)>0$ and $Q_m(t)=0$ means that $\tilde{\phi}$ schedules a nonempty queue in $\mathcal{C}_1$ in step 2, and $D_1(t)=1$. 
\item $Q_1(t)=0$ and $Q_m(t)>0$ means that $\tilde{\phi}$ schedules a nonempty queue in $\mathcal{C}_m$ in either step 1 or step 3, depending on whether the other cliques have nonempty queues. In either case, $D_m(t)=1$. 
\end{itemize} 
The proof of sufficiency of Property~$\mathcal{P}$ can now be extended using the Lyapunov function defined below to show that $\tilde{\phi}$ is indeed throughput-optimal. 
\begin{eqnarray}
L(\mathbf{Q}(t)):=\sum_{m=2}^{N}(Q_1(t)+Q_{m}(t))^2
\label{Chapter05EqnLyapunovFunctionForPropertyPNewVersion}
\end{eqnarray}
\qed

\subsection{Throughput-optimality of $\Phi^{(S)}_{IC}(T)$}\label{Chapter05AppendixProofOfTOOfPhiT_1}
The proof consists of two parts. We will first prove that $\Phi^{(S)}_{IC}(T)$ specialized to a single collocated network, i.e., a single clique is throughput-optimal and then use a new version of property $\mathcal{P}$ to complete the proof for our \enquote{star of cliques} interference graphs (of the type shown in Fig.~\ref{Chapter05FigExtendingPi3PoliciesToOtherInterferenceGraphs}).\\
\indent So consider once again, a \emph{collocated} system of $N$ queues described by a fully connected interference graph. As is the case in Sec.~\ref{Chapter05SecPhiWithPeriodicOccupancyInformation}, suppose the system only knows $\boldsymbol{\zeta}(t)\in\left\lbrace0,1\right\rbrace^{N}$, at times $t=0,T,2T,\cdots$. Following Sec.~\ref{Chapter05SecPhiWithPeriodicOccupancyInformation}, arrivals in the $k^{th}$ frame are \emph{not} served in the $k^{th}$ frame. We denote by $\psi_T,$ the scheduling policy that, during $kT,kT+1,\dots,KT+T-1$, serves every queue known to be nonempty at $kT$ until either 
\begin{enumerate}
\item The next frame, i.e., $k+1$ begins, or
\item All packets queued in the system until the beginning of slot $kT$ have been served. In this case the system obviously idles until the next frame begins.
\end{enumerate}
Since only one queue can be served in any slot, the capacity region of this system is $\left\lbrace\boldsymbol{\lambda}\in\mathbb{R}^N_+\mid \sum_{i=1}^N\lambda_i<1\right\rbrace$. 
In what follows, we will analyse the process $\left\lbrace\mathbf{q}(k),k\geq0\right\rbrace$, where $\mathbf{q}(k):=\mathbf{Q}(kT).$
\begin{lem}\label{Chapter05LemAppendixProofOfPsiTIsTO}
Under $\psi_T$, for any $\boldsymbol{\lambda}$ inside the capacity region, 
\begin{itemize}
\item the process $\left\lbrace\mathbf{q}(k),k\geq0,\right\rbrace$ is strongly stable, i.e., $\psi_T$ is throughput-optimal, and 
\item mean packet delay under $\psi_T$ is \emph{linear} in $T$ which means that there exists an $\alpha\in\mathbb{R}_+$, such that
\begin{equation}
\mathbb{E}_{\psi_T}\sum_{i=1}^NQ_i(kT)\leq\alpha T,~\forall k\geq0.
\end{equation}
\end{itemize}

\end{lem}
\begin{pf}
Let $A_i[x,y]$ denote the number of arrivals to Queue $i$ over the slots $x,x+1,\dots,y$. Since the arrivals are all Bernoulli, $A_i[x,y]$ is a Binomial$\left(y-x+1,\Lambda_i\right)$ random variable. Denote the total system backlog at $kT$ by  $q(k):=\sum_{i=1}^N q_i(k)$ and total arrival to the system during the $k^{th}$ frame by $A(k+1):=\sum_{i=1}^N A_i[kT+1,k(T+1)]$. It is then easy to see that
\begin{equation}
q(k+1)=\left(q(k)-T\right)^++A(k+1),
\end{equation}
since $\psi_T$ serves the network until all $q(k)$ packets leave, if $q(k)<T$, or exactly $T$ packets depart (this happens when the $k^{th}$ frame begins with at least $T$ packets in the network). With this 
we get, 
\begin{eqnarray*}
\mathbb{E}_{\psi_T}\left[q^2(k+1)-q^2(k)\mid q(k)=q\right]&\leq& q^2+(N^2+1)T^2\\&-&2qT\left(1-\sum_{i=1}^N\Lambda_i\right),\\
&=& q^2+(N^2+1)T^2-2\epsilon q,
\end{eqnarray*}
where $\epsilon:=\left(1-\sum_{i=1}^N\Lambda_i\right)>0.$ Taking expectations on both sides of the above equation, we get
\begin{eqnarray}
\mathbb{E}_{\psi_T}\left[q^2(k+1)-q^2(k)\right]&\leq&\mathbb{E}_{\psi_T}q^2(k)+(N^2+1)T\nonumber\\
&-&2\epsilon\mathbb{E}_{\psi_T}q(k),\nonumber\\
\mathbb{E}_{\psi_T}\sum_{i=1}^NQ_i(kT)&\stackrel{\star}{\leq}&\frac{(N^2+1)}{2\epsilon}T,\label{eqnAppendixDelayLinearInT}\\
\Rightarrow \limsup_{k\rightarrow\infty}\frac{1}{kT}\sum_{l=0}^{k-1}\mathbb{E}_{\psi_T}\sum_{i=1}^NQ_i(lT)&<&\infty,\label{eqnAppendixPsi_TIsTO}
\end{eqnarray}
In inequality $\star$, we have used the fact that $\mathbb{E}_{\psi_T}q^2(k+1)\geq0$. In particular, Eqn.~\ref{eqnAppendixPsi_TIsTO} shows that the system is strongly stable under $\psi_T$ and setting $\alpha=\frac{(N^2+1)}{2\epsilon},$ and using Little's theorem along with  Eqn.~\eqref{eqnAppendixDelayLinearInT} we see that mean packet delays are linear in $T.$
\end{pf}

The proof of throughput-optimality of $\Phi^{(T)}$ follows by using the above lemma with the $(N-1)$ queue lengths $\left[\left(\sum_{j\in\mathcal{C}_1}Q_j(t)+\sum_{j\in\mathcal{C}_k}Q_j(t)\right),~2\leq k\leq N\right].$ It also means that delay with $\Phi^{(T)}$ increases linearly in $T.$
\qed
\subsection{Proof of Prop.~\ref{Chapter05PropPhi3_TildeIsBetterThanPhi3_1}}\label{AppendixProofOfPropPhi3_TildeIsBetterThanPhi3_1}
This proof proceeds along the same lines as the proof of delay optimality of Policy~$\tilde{\pi}^{(3)}_{IQ}$ that we presented in Sec.~\ref{AppendixProofOfPi33isDelayOptimal}.$\phi^{(S)}_{IC}$ and $\tilde{\phi}$ differ only when every peripheral clique has a nonempty queue and behave identically otherwise. Verifying the conditions required to establish sample pathwise and hence, stochastic ordering are very similar to our proof of delay-optimality of $\tilde{\pi}^{(3)}_{IQ}$ and will not be repeated. 
\qed

\subsection{Proof of Prop.~\ref{Chapter05PropPhi3TDIsTO}}\label{AppendixProofOfPropPhi3TDIsTO}
Let $Q_i(t):=\sum_{j=1}^{\mathcal{N}_i}Q_{i,j}(t)$ be the total backlog of Clique~i, i.e., $\mathcal{C}_i$, at the beginning of time slot $t$ and let the total arrival rate to $\mathcal{C}_i$ be denoted by $\lambda_i:=\sum_{j\in\mathcal{C}_i}\lambda_{i,j}=\sum_{j=1}^{\mathcal{N}_i}\lambda_{i,j}.$
Define $Q_{1,2}(t):=Q_1(t)+Q_2(t).$ Notice that Clique~1 is scheduled for service in every slot in which any queue in it has a packet and whenever $\mathcal{C}_1$ is not scheduled, $\mathcal{C}_2$ is scheduled provided it is non empty. So, we have that 
\begin{eqnarray}
Q_{1,2}(t+1)=Q_{1,2}(t)-\mathbb{I}_{\left\lbrace Q_{1,2}(t)>0\right\rbrace}+A_{1,2}(t+1),
\label{eqnEvolutionOfQ12Under3CliquesTopDownPolicy}
\end{eqnarray}
where $A_{1,2}(t+1),~t\geq0$ is the total number of arrivals to $\mathcal{C}_1$ and $\mathcal{C}_2$ at the beginning of slot $t+1,$ and $\mathbb{E}A_{1,2}(t+1)=\lambda_1+\lambda_2.$ It is easy to show that Cliques 1 and 2 are stable under this policy (a simple sum of queue length squares Lyapunov function suffices), which means that there exists a stationary distribution for the process $\lbrace Q_{1,2}(t),~t\geq0\rbrace$. Now, from \eqref{Chapter05EqnCapacityRegionOfLAoCModel} with $N=3$ cliques, we know that $\lambda_1+\lambda_2<1$. Taking expectation on both sides of the equation \emph{in steady state}, we get
\begin{eqnarray}
\mathbb{E}Q_{1,2}(t+1)&=&\mathbb{E}Q_{1,2}(t)-P\left\lbrace Q_{1,2}(t)>0\right\rbrace+\lambda_1+\lambda_2,\nonumber\\
\Rightarrow P\left\lbrace Q_{1,2}(t)=0\right\rbrace &=& 1-\lambda_1-\lambda_2
\label{eqnProbabilityThatQ12IsEmptyUnderTopDownCliquePolicy}
\end{eqnarray}
So, since a non empty queue in $\mathcal{C}_3$ is served in every slot in which either
\begin{itemize}
\item  there is a non empty queue in $\mathcal{C}_1$, or
\item  there are no non empty queues in $\mathcal{C}_1$, or $\mathcal{C}_2.$
\end{itemize}
With this we see that the \emph{offered service process} to Clique~3, i.e., $\lbrace S_{\mathcal{C}_3}(t),~t\geq0\rbrace$ satisfies 
\begin{eqnarray}
P\lbrace S_{\mathcal{C}_3}(t)=1\rbrace &=& \lambda_1+(1-\lambda_1-\lambda_2)\nonumber\\
&=& 1-\lambda_2>\lambda_3.
\end{eqnarray}
Notice that $P\lbrace S_{\mathcal{C}_3}(t)=1\rbrace$ is independent of $Q_3(t)$. Hence, repeating the drift argument from Thm.~\ref{thmPi3_1IsTO} that showed the throughput optimality of $\pi^{(3)}_{TD}$ on $Q_{3}(t)$, we see that $\mathcal{C}_3$ is also stable which means that $\theta^{(3L)}_{TD}$ is throughput optimal. 
\qed
\subsection{Proof of Prop.~\ref{Chapter05PropPhi3_5IsTO}}\label{AppendixProofOfPhi3_5IsTO}
We first show that $\phi^{(S)}_{IC}$ satisfies Eqn.~\ref{eqnPropertyPForNonPathGraphs} at every $t\geq0.$ Thereafter, the analysis in the proof of $\tilde{\phi}^{(S)}_{IC}$ using the same Lyapunov function as in Eqn.~\eqref{Chapter05EqnLyapunovFunctionForPropertyPNewVersion} can be used to establish strong stability.
Let $m\geq2$ in the sequel. By the definition of the departure processes $\left\lbrace D_i(t),i\geq1\right\rbrace$, in every slot $t\geq0,$ $Q_1(t)+Q_m(t)=0$ always means $D_1(t)+D_m(t)=0$. To show the converse, we consider several cases 
\begin{itemize}
\item $Q_1(t)>0\Rightarrow D_1(t)=1$.
\item $Q_1(t)=0$ and $Q_m(t)>0$ means that $\phi^{(S)}_{IC}$ schedules a nonempty queue in $\mathcal{C}_m$ in step 1 or step 2 ensuring $D_m(t)=1$. 
\end{itemize} 
The proof now uses the same Lyapunov function as \ref{AppendixProofOfPhiNTildeIsTO} and proceeds along the same lines.
\qed

\subsection{Proof of Thm.~\ref{Chapter05ThmGammaForQZSoCIsTO}}\label{AppendixProofOfThmGammaForQZSoCIsTO}
\begin{figure*}[tb]
\centering
\includegraphics[height=3.50cm, width=11.0cm]{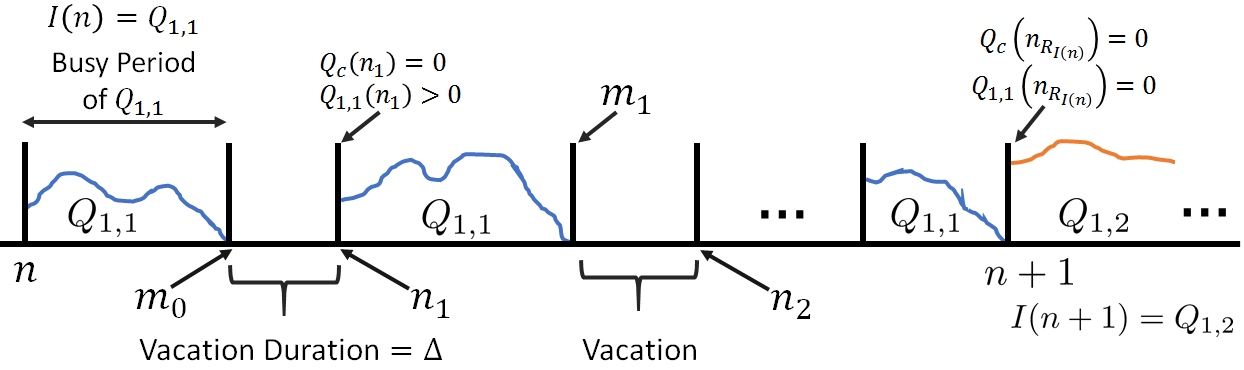}
\caption{A sample path illustrating the service process to one of the peripheral queues, here, $Q_{1,1}.$ Even with \emph{one} queue in the central clique the resulting service process to peripheral queues is found to be quite complex. Note that here, $I(n)=Q_{1,1}$, \emph{in Clique~1,} while $I(n+1)=Q_{1,2}$ since it was chosen in Step~\ref{Chapter05ProtocolGammaNoPowerInMinislot2MeansArgMaxVi} in the definition of $\phi^{(S)}_{CS}.$}
\label{Chapter05FigServiceProcessForPeripheralQUnderGamma}
\end{figure*}

We begin by first analysing the service processes to different queues under $\phi^{(S)}_{CS}$. This will yield important insights into how the stability proof should proceed. 
\paragraph{Service Processes under $\phi^{(S)}_{CS}$ :} For the purposes of this proof, we relabel the queue in the central clique as $Q_c(t)$ and assume that the peripheral cliques are numbered $\mathcal{C}_1,\cdots,\mathcal{C}_N.$ Clearly, since the central clique gets maximum priority and Queue~$c$, the only queue in this clique,  is served whenever it is nonempty, it behaves as a $Geo/D/1$ queue with service time equal to $1$ slot. We now move on to queues in the peripheral cliques. WLOG we consider clique $\mathcal{C}_1$ and $Q_{1,1}$  in it and note that every clique is running an \emph{exhaustive service policy} locally. Fig.~\ref{Chapter05FigServiceProcessForPeripheralQUnderGamma} shows a sample path of the queue length-evolution process $Q_{1,1}(t)$. 
Note that Queue~$c$ is served in every slot in which there is an arrival to it and that due to the Bernoulli nature of the arrival processes, the interarrival duration is Geometric with mean $\frac{1}{1-\lambda_c}.$ A packet in $Q_{1,1}$ that reaches the Head-of-Line (HOL) position therefore sees a service duration that is Geometric with mean $\frac{1}{1-\lambda_c}.$ This, of course, is true for every queue in the peripheral cliques. 
Once $Q_{1,1}$ becomes empty, the rest of the clique does not necessarily obtain this knowledge in that same slot. It depends on whether the central clique is empty or not. For example, in slot $m_0$ in Fig.~\ref{Chapter05FigServiceProcessForPeripheralQUnderGamma}, $Q_{1,1}$ has become empty, but Queue~$c$ has received an arrival which means that it is Queue~$c$ that is served, power is sensed in minislot 1 itself, and the protocol $\phi^{(S)}_{CS}$ never enters Step~\ref{Chapter05ProtocolGammaNoPowerInMinislot2MeansArgMaxVi}. The other queues in $\mathcal{C}_1$ don't know if $Q_{1,1}$ is empty and so, in the next slot in which Queue~$c$ is empty, it is $Q_{1,1}$ that is given channel access and if, by then it has received any arrivals, it begins another busy period. This is what happens in Fig.~\ref{Chapter05FigServiceProcessForPeripheralQUnderGamma} until instant $m_1$ and this entire process repeats resulting in a random number of busy periods of $Q_{1,1}$ until the instant when \emph{both} Queue~$c$ \emph{and} $Q_{1,1}$ are found empty. This happens in slot $n_{R_{I(n)}}$ in the figure. At this instant, $\phi^{(S)}_{CS}$ enters Step.~\ref{Chapter05ProtocolGammaNoPowerInMinislot2MeansArgMaxVi} and the queue with the largest $V_i$ takes over. Notice that there are several portions labelled \enquote{Vacation} in the figure. A \emph{Vacation} is the duration between a peripheral queue becoming empty and the first time since then that it is granted channel access since the central queue is empty. The durations of these vacations are also distributed Geometric with mean $\frac{1}{1-\lambda_c}.$
To summarize, the service to a peripheral queue under $\phi^{(S)}_{CS}$ consists of a random number of (Busy Period + Vacation) durations. 

The proof will focus on analysing $\phi^{(S)}_{CS}$ restricted to a single clique and this analysis will be extended later to show that the entire star-of-cliques system is stable. For now, WLOG, we focus on clique $\mathcal{C}_1.$ Clearly, the queue length vector process $\mathbf{Q}(t)$ under $\phi^{(S)}_{CS}$ is a DTMC. We prove the throughput optimality of $\phi^{(S)}_{CS}$ using a drift argument for the queue length vector process $\mathbf{Q}(n)$ which is $\mathbf{Q}(t)$ sampled at instants $n\geq0$ when a new peripheral queue is granted channel access (see Fig.~\ref{Chapter05FigServiceProcessForPeripheralQUnderGamma}). 
We define the following quantities
\begin{itemize}\label{Chapter05ItemizedListOfQuantitiesProofOfTOOfPhiS_CS}
\item $n:$ the slot in which Queue~$I(n)$ (here $Q_{1,1}$) begins transmission.
\item $I(n):$ the queue having channel access 
\item $n+1:$ the slot in which channel access is granted to the next peripheral queue in $\mathcal{C}_1.$ 
\item $m_l:$ instant at which the $l$\textsuperscript{th} (Vacation + Busy Period) begins.
\item $R_{I(n)}:$ the number of (Vacation + Busy Period)'s for Queue~$I(n)$, here $Q_{1,1}$.
\item $\lambda_c:$ packet arrival rate at Queue~$c.$
\item $n_i,~i\geq0:$ instants at which busy periods of Queue~$I(n)$ begin. Clearly, $n_0=n$.
\end{itemize}

\begin{pf}
Denote by $\rho_j$ the load on Queue~$j$, i.e., $\rho_j=\lambda_j\mathbb{E}B=\lambda_j\frac{1}{1-\lambda_c}$, and by $\rho=\sum_{j=1}^{\mathcal{N}_1}\rho_j$ the total load on the clique. The interference constraints dictate that $\rho<1.$
Let $b=\mathbb{E}B=\frac{1}{\lambda_c}$
Define $\Delta=\inf\left\lbrace m>m_0|A_c(m)=0\right\rbrace-m_0$, i.e., the duration after $m_0$ until the slot without any arrival to Queue~$c$ (the central queue). Note that the vacation may be of length $0$ slots as well, since when the busy period of Queue~$I(n)$ ends, Queue~$c$ could be empty. Therefore, $\forall k\geq 0$ $P\{\Delta=k\}=\lambda_c^k\left(1-\lambda_c\right)$, and the mean of $\Delta$ is
\begin{eqnarray}
\mathbb{E}\Delta=\sum_{k=0}^\infty k\lambda_c^k\left(1-\lambda_c\right) &=& \frac{\lambda_c}{1-\lambda_c}
\label{Chapter05EqnMeanDurationVacation}\\
\mathbb{E}A_j(\Delta) &=& \mathbb{E}\left(\mathbb{E}\left[A_j(\Delta)\mid \Delta\right]\right) = \mathbb{E}\left(\lambda_j\mathbb{E}\Delta\right) \nonumber\\
&=& \lambda_c\frac{\lambda_j}{1-\lambda_c} = \lambda_c\rho_j 
\label{Chapter05EqnMeanNumberArrivalsDuringVacation}
\end{eqnarray} 
Now, notice that
\begin{equation}
Q_j(n_1)=
\begin{cases}
Q_j(n)+\underbrace{A_j\left(G_{I(n)}\left(Q_{I(n)}(n)\right)\right)}_{\text{arrivals to Queue}~j\text{ during busy period of Queue } I(n)}\\
\underbrace{+A_j(\Delta)}_{\text{arrivals during the subsequent vacation}},~~\text{ for }j\neq I(n)\\
A_j(\Delta),~~\text{ for }j = I(n).\\
\end{cases}
\label{Chapter05EqnQLenEvolutionUpToN1}
\end{equation}
Using \eqref{Chapter05EqnMeanNumberArrivalsDuringVacation} and \eqref{Chapter05EqnQLenEvolutionUpToN1}, we get
\begin{eqnarray}
b\mathbb{E}\left[Q_j(n_1)|\mathbf{Q}(n)\right] &\leq& bQ_j(n) + bQ_{I(n)}(n)\frac{\rho_j}{1-\rho_{I(n)}}+b\lambda_c\rho_j\nonumber\\
\mathbb{E}\left[\sum_{j=1}^{\mathcal{N}_1}bQ_{j}(n_1)|\mathbf{Q}(n)\right] &\leq& \sum_{j=1}^{\mathcal{N}_1}bQ_j(n)+\left(\frac{\rho-\rho_j}{1-\rho_{I(n)}}-1\right)bQ_{I(n)}(n)\nonumber\\
&& +b\lambda_c\rho \nonumber\\
&\stackrel{\star}{=}& \sum_{j=1}^{\mathcal{N}_1}bQ_j(n)+\underbrace{h_{I(n)}\left(\rho-1\right)bQ_{I(n)}(n)}_{\text{strictly negative}}\nonumber\\
&& +b\lambda_c\rho,
\label{Chapter05EqnConditionalMeanSumQLenAtN1}
\end{eqnarray}
where in equality $\star$ we have defined $h_{I(n)}=\frac{1}{1-\rho_{I(n)}}$. Also observe the fact that on the R.H.S of \eqref{Chapter05EqnConditionalMeanSumQLenAtN1}, $(\rho-1)$ is \emph{strictly negative,} by capacity constraints. 
 Similarly,
\begin{eqnarray*}
\mathbb{E}\left[\sum_{j=1}^{\mathcal{N}_1}bQ_{j}(n_2)|\mathbf{Q}(n)\right] &=& \mathbb{E}\left[\underbrace{\mathbb{E}\left[\sum_{j=1}^{\mathcal{N}_1}bQ_{j}(n_2)\bigg|\mathbf{Q}(n_1),\mathbf{Q}(n)\right]}_{\sigma\left(\mathbf{Q}(n)\right)\subset\sigma\left(\mathbf{Q}(n_1),\mathbf{Q}(n)\right)}\bigg|\mathbf{Q}(n)\right] \nonumber\\
&\stackrel{\dagger}{=}& \mathbb{E}\left[\mathbb{E}\left[\sum_{j=1}^{\mathcal{N}_1}bQ_{j}(n_2)\bigg|\mathbf{Q}(n_1)\right]\bigg|\mathbf{Q}(n)\right] \nonumber\\
\mathbb{E}\left[\sum_{j=1}^{\mathcal{N}_1}bQ_{j}(n_2)|\mathbf{Q}(n)\right] &\stackrel{\star1}{\leq}& \mathbb{E}\left[\sum_{j=1}^{\mathcal{N}_1} bQ_j(n_1)\bigg|\mathbf{Q}(n)\right]\nonumber\\
&& +h_{I(n)}\left(\rho-1\right)b\mathbb{E}\left[Q_{I(n)}(n_1)\bigg|\mathbf{Q}(n)\right]+b\lambda_c\rho \nonumber\\
&=& \left(\sum_{j=1}^{\mathcal{N}_1}bQ_j(n)\right.\nonumber\\
&& +h_{I(n)}\left(\rho-1\right)bQ_{I(n)}(n) +b\lambda_c\rho\bigg)\nonumber\\
&& +\left(h_{I(n)}\left(\rho-1\right)b\lambda_c\rho_{I(n)}\right)+b\lambda_c\rho,\nonumber\\
&=& \sum_{j=1}^{\mathcal{N}_1}bQ_j(n)+ 2b\lambda_c\rho\nonumber\\
&& +\underbrace{\left(\rho-1\right)bh_{I(n)}\left(Q_{I(n)}+\rho_{I(n)}\right)}_{\text{strictly negative}}.
\end{eqnarray*}
Proceeding similarly,
\begin{eqnarray}
\mathbb{E}\left[\sum_{j=1}^{\mathcal{N}_1}bQ_{j}(n_k)|\mathbf{Q}(n)\right]&=& \sum_{j=1}^{\mathcal{N}_1}bQ_j(n)+ kb\lambda_c\rho\nonumber\\
&& +\left(\rho-1\right)bh_{I(n)}\left(Q_{I(n)}\right.\nonumber\\
&& \left. +(k-1)\rho_{I(n)}\right),~\forall k\geq1.
\label{Chapter05EqnConditionalMeanSumQLenForGeneralNk}
\end{eqnarray}
Equality $\dagger$ follows from the Markovian nature of the evolution of the queue-length vector, and we have used \eqref{Chapter05EqnConditionalMeanSumQLenAtN1} and the fact that $I(n_1)=I(n)$ in inequality $\star1$. Let us now compute the mean number of secondary busy periods which will inform the choice of $k$ in \eqref{Chapter05EqnConditionalMeanSumQLenForGeneralNk} while computing the conditional drift between instants $n$ and $n+1.$ The visit to Queue~$I(n)$ ends when it receives $0$ arrivals during a vacation. Let the number of vacations during a visit to Queue~$I(n)$ be $R_{I(n)}.$
\begin{eqnarray}
P\left(R_{I(n)}=k\right) &=& \prod_{l=1}^k P\left(A^{(l)}_{I(n)}(\Delta)>0\right)\nonumber\\
&& \times P\left(A^{(k+1)}_{I(n)}(\Delta)=0\right),~\forall k\geq0.\nonumber\\
\end{eqnarray}
But all the $A^{(l)}_{I(n)}(\Delta)$ are iid and 
\begin{eqnarray}
P\left(A^{(l)}_{I(n)}(\Delta)=0\right) &=& \sum_{k=0}^\infty P\left(A^{(l)}_{I(n)}(\Delta)=0\bigg|\Delta=k\right)\nonumber\\
&=& \sum_{k=0}^\infty (1-\lambda_{I(n)})^k\lambda_{c}(1-\lambda_{c})^k\nonumber\\
&=& \frac{1-\lambda_c}{1-(1-\lambda_{I(n)})\lambda_c},
\label{Chapter05EqnProbOfZeroVacationsForQueueIOfN}
\end{eqnarray}
Which gives us 
\begin{equation}
\mathbb{E}R_{I(n)}=\frac{\lambda_c\lambda_{I(n)}}{1-\lambda_c}+1
\label{Chapter05EqnMeanNumberOfVacationsForQueueIOfN}
\end{equation}
Now, using \eqref{Chapter05EqnMeanNumberOfVacationsForQueueIOfN} in \eqref{Chapter05EqnConditionalMeanSumQLenForGeneralNk}, we get 
\begin{eqnarray*}
\mathbb{E}\left[\sum_{j=1}^{\mathcal{N}_1}bQ_{j}(n_{R_{I(n)}})|\mathbf{Q}(n)\right] &=& \sum_{j=1}^{\mathcal{N}_1}bQ_j(n)+ \left(\frac{\lambda_c\lambda_{I(n)}}{1-\lambda_c}+1\right)b\lambda_c\rho \\
&& +\left(\rho-1\right)bh_{I(n)}\\
&& \times\left(Q_{I(n)}+\left(\frac{\lambda_c\lambda_{I(n)}}{1-\lambda_c}\right)\rho_{I(n)}\right), \nonumber
\end{eqnarray*}
from which we get the conditional drift as
\begin{eqnarray*}
&&\mathbb{E}\left[\sum_{j=1}^{\mathcal{N}_1}bQ_{j}(n+1)-\sum_{j=1}^{\mathcal{N}_1}bQ_{j}(n)|\mathbf{Q}(n)\right] \\
&& =\mathbb{E}\left[\sum_{j=1}^{\mathcal{N}_1}bQ_{j}(n_{R_{I(n)}})-\sum_{j=1}^{\mathcal{N}_1}bQ_{j}(n)|\mathbf{Q}(n)\right] \nonumber\\
&&\leq\left(\frac{\lambda_c\lambda_{I(n)}}{1-\lambda_c}+1\right)b\lambda_c\rho \\
&& + \left(\rho-1\right)bh_{I(n)}\left(Q_{I(n)}+\left(\frac{\lambda_c\lambda_{I(n)}}{1-\lambda_c}\right)\rho_{I(n)}\right)\nonumber\\
&& <-\epsilon,\nonumber
\end{eqnarray*}
for large enough $Q_{I(n)}$. Invoking the Foster-Lyapunov theorem \cite{fayolle-etal95constructive-theory-markov-chains} we see that the chain $\mathbf{Q}(n)$ is positive recurrent. This process can be used repeatedly to show that each of the ${\mathcal{N}_1}$ DTMCs $\left\lbrace\mathbf{Q}_{n{\mathcal{N}_1}+K}\right\rbrace_{n=0}^\infty$ is positive recurrent for $K=0,1,\cdots,{\mathcal{N}_1}-1.$ Finally, this same procedure can be repeated for each peripheral clique $\mathcal{C}_i,1\leq i\leq N,$ to show that $\phi^{(S)}_{CS}$ is throughput optimal. 
\end{pf}


\subsection{Proof of Prop.~\ref{Chapter05PropPsi3_1TIsTO}}\label{AppendixProofOfPropPsi3_1TIsTO}
Let $A_j(s,t]:=\sum_{k=s+1}^tA_j(t)$ be the total number of arrivals to $\mathcal{C}_j$ in $\lbrace s+1,s+2,\cdots,t\rbrace$. Recall that the total backlog of clique $j$ at the beginning of slot $t$ is denoted by $Q_j(t)=\sum_{k=1}^{\mathcal{N}_j}Q_{j,k}(t)$ .
We will prove that the process $\lbrace\mathbf{Q}(kT):=\left[Q_1(kT),Q_2(kT),Q_3(kT)\right],~t\geq0\rbrace$ is stable. For this, we first look at a policy that is clearly suboptimal (in terms of delay) compared to $\psi^{(3)}_{1,T}$. This policy only serves, during the $k^{th}$ frame, the packets that have already arrived to a scheduled queue at or before the beginning of the frame, i.e., it does not serve the packets arriving at the queue over $(kT,KT+(T-1)]$.

Firstly, since $\mathcal{C}_1$ is served whenever it is non empty, $Q_1(t)$ evolves as $Q_1\left((k+1)T\right)=Q_1\left((t)\right)-\mathbb{I}_{\lbrace \zeta_1(kT)>0\rbrace} + A_1(kT,(k+1)T]$. Also notice that if either  $\mathcal{C}_1$ or $\mathcal{C}_2$ is non empty at the beginning of a frame, one of these two cliques always gets served, i.e.,
\begin{eqnarray}
Q_{1,2}\left((k+1)T\right)&=& Q_{1,2}\left(kT\right)-T\mathbb{I}_{\lbrace Q_{1,2}(kT)>0\rbrace}\nonumber\\
&& +A_{1,2}\left(kT, kT+(T-1)\right]
\label{eqnEvolutionOfQ12UnderPsi3_1T}
\end{eqnarray}
Using the modified Property $\mathcal{P}$ for batch arrivals and departures, we see that $\lbrace Q_{1,2}(kT),k=0,1,2,\cdots\rbrace$ is strongly stable, and, also being an aperiodic, irreducible DTMC, it is also positive recurrent. This proves the existence of a stationary measure $\mu_{1,2}$ for the chain. Using arguments similar to the one in \eqref{eqnProbabilityThatQ12IsEmptyUnderTopDownCliquePolicy}, we see that, in the steady state, $P\lbrace Q_{1,2}(kT)=0\rbrace=\mu_{1,2}(\lbrace0\rbrace)=T(1-\lambda_1-\lambda_2)$. Now, coming to $\mathcal{C}_3,$ let $S_{\mathcal{C}_3}\left(kT, kT+(T-1)\right]$ be the total number of slots in the $k^{th}$ frame during which clique 3 is offered service. Then,
\begin{eqnarray}
Q_3\left((k+1)T\right)&=&\bigg(Q_3\left(kT\right)-S_{\mathcal{C}_3}\left(kT, kT+(T-1)\right]\bigg)^{+}\nonumber\\
&& +A_3\left(kT, kT+(T-1)\right]
\end{eqnarray}
Assume $Q_{1,2}(0)\sim\mu_{1,2}$, then 
\begin{eqnarray}
P\lbrace S_{\mathcal{C}_3}(t) &=& 1\rbrace=P\left\lbrace Q_1(kT)=1\right\rbrace+P\lbrace Q_{1,2}(kT)=0\rbrace\nonumber\\
&=& \lambda_1T+T(1-\lambda_1-\lambda_2)\nonumber\\
&=& (1-\lambda_2)T\nonumber\\
&>&\lambda_3T=\mathbb{E}A_3\left(kT, kT+(T-1)\right].
\end{eqnarray}
Hence, Clique 3 is also stable and this proves the claim. 
\qed

\subsection{Simulation Details}\label{AppendixSimulationDetails}
Here we provide the arrival rate vectors for:
\begin{itemize}
\item Path graph network simulations in Table~\ref{table5QueuesSimulationsPiTilde5MWAndLOfMWAlpha}
\begin{enumerate}
\item $N=4$ queues: $\boldsymbol{\lambda}=[0.49, 0.49, 0.49, 0.49]$
\item $N=5$ queues: $\boldsymbol{\lambda}=[0.15, 0.049, 0.95, 0.049, 0.15]$
\item $N=15$ queues: \\$\boldsymbol{\lambda}=[0.80,0.15,0.15,0.15,0.15,0.8,0.049,\\0.95,0.049,0.8,0.15,0.15,0.15,0.15,0.80]$
\end{enumerate}
\item Cluster of Cliques simulations in Table~\ref{tableNonPathGraphSimulationsPhi3TildePhi3_1MWAndLOfMW}
\begin{enumerate}
\item Star-of-Cliques network: \\$\boldsymbol{\lambda}=[0.3, 0.3, 0.3, 0.09, 0.9, 0.9]$
\item Linear-Array-of-Cliques: \\$\boldsymbol{\lambda}=[0.1,0.1,0.1, 0.049, 0.65,0.3, 0.049,0.0,0.0]$
\end{enumerate}
\end{itemize}


\bibliographystyle{IEEEtran}
\bibliography{IEEEabrv,techreport}
\begin{IEEEbiographynophoto}{Avinash Mohan (S.M.'16-M'17)} 
obtained his M.Tech. and PhD degrees from the Indian Institute of Technology (IIT) Madras and the Indian Institute of Science (IISc) Bangalore, respectively. He is currently a postdoctoral fellow with the Reinforcement Learning Research Laboratory ($(RL)^2$ Lab) at the Technion, Israel Institute of Technology. His research interests include analysis of pricing in deregulated electricity markets, stochastic control and reinforcement learning and resource allocation in wireless communication networks.
\end{IEEEbiographynophoto}
\begin{IEEEbiographynophoto}{Aditya Gopalan}
Aditya Gopalan is an Assistant Professor and INSPIRE Faculty Fellow at the Indian Institute of Science, Electrical Communication Engineering. He received the Ph.D. degree in electrical engineering from The University of Texas at Austin, and the B.Tech. and M.Tech. degrees in electrical engineering from the Indian Institute of Technology Madras. He was an Andrew and Erna Viterbi Post-Doctoral Fellow at the Technion-Israel Institute of Technology. His research interests include machine learning and statistical inference, control, and algorithms for resource allocation in communication networks.
\end{IEEEbiographynophoto}
\begin{IEEEbiographynophoto}{Anurag Kumar}
(B.Tech., Indian Institute of Technology (IIT)
Kanpur; PhD, Cornell University, both in Electrical Engineering) was
with Bell Labs, Holmdel, N.J., for over 6 years. Since then he has
been on the faculty of the ECE Department at the Indian Institute of
Science (IISc), Bangalore; he is at present the Director of the
Institute.  His area of research is communication networking, and he
has recently focused primarily on wireless networking. He is a Fellow
of the IEEE, the Indian National Science Academy (INSA), the Indian
National Academy of Engineering (INAE), and the Indian Academy of
Sciences (IASc).  He was an associate editor of IEEE Transactions on
Networking, and of IEEE Communications Surveys and Tutorials.
\end{IEEEbiographynophoto}







%

%
%
%
%
%

\ifCLASSOPTIONcaptionsoff
  \newpage
\fi

\end{document}